\documentclass[aps,a4paper,superscriptaddress,showpacs,preprintnumbers,amsmath,amssymb]{revtex4}

\usepackage{ulem}

\usepackage{psfrag}
\usepackage{graphicx} 
\usepackage{dcolumn} 
\usepackage{color}
\usepackage{latexsym,amsfonts} 
\usepackage{bm} 
\usepackage{amssymb}
\begin{document}

\title{Deficit of reactor antineutrinos\\ at distances smaller than
  $100\,{\rm m}$ and inverse $\beta$--decay}

\author{A. N. Ivanov}\email{ivanov@kph.tuwien.ac.at}
\affiliation{Atominstitut, Technische Universit\"at Wien, Stadionalle
  2, A-1020 Wien, Austria}
\author{R. H\"ollwieser}\affiliation{Atominstitut, Technische
  Universit\"at Wien, Stadionalle 2, A-1020 Wien, Austria}
\affiliation{Department of Physics, New Mexico State University, Las
  Cruces, NM 88003, USA} \author{N. I. Troitskaya} \affiliation{State
  Polytechnic University of St. Petersburg, Polytechnicheskaya 29,
  195251, Russian Federation}
\author{M. Wellenzohn}\affiliation{Atominstitut, Technische
  Universit\"at Wien, Stadionalle 2, A-1020 Wien, Austria}
\author{O. M. Zherebtsov}\affiliation{Petersburg Nuclear Physics
  Institute, 188300 Gatchina, Orlova roscha 1, Russian
  Federation}\author{A. P. Serebrov}\affiliation{Petersburg Nuclear
  Physics Institute, 188300 Gatchina, Orlova roscha 1, Russian
  Federation}

\date{\today}

\begin{abstract}
We analyse a change in a deficit of reactor antineutrinos at distances
smaller than $100\,{\rm m}$ by changing the lifetime of the neutron
from $\tau_n = 885.7\,{\rm s}$ to $\tau_n = 879.6\,{\rm s}$,
calculated for the axial coupling constants $\lambda = - 1.2694$ and
$\lambda = - 1.2750$, respectively, in order to get a result
corresponding the new world average value $\tau_n = 880.1(1.1)\,{\rm
  s}$. We calculate the angular distribution and cross section for the
inverse $\beta$--decay, taking into account the contributions of the
``weak magnetism'' and the neutron recoil to next--to-leading order in
the large baryon mass expansion and the radiative corrections of order
$\alpha/\pi \sim 10^{-3}$, calculated to leading order in the large
baryon mass expansion. We obtain an increase of a deficit of reactor
antineutrinos of about $0.734\,\%$. We discuss a universality of
radiative corrections to order $\alpha$ to the neutrino (antineutrino)
reactions induced by weak charged currents, pointed out by Kurylov,
Ramsey-Musolf and Vogel (Phys. Rev. D {\bf 67}, 035502 (2003)), and
calculate the antineutrino--energy spectrum of the neutron
$\beta^-$--decay to order $\alpha/\pi$ and taking into account the
contributions of the ``weak magnetism'' and the proton recoil.
\end{abstract}

\maketitle

\section{Introduction}
\label{sec:introduction}

In this paper we analyse a deficit of the reactor antineutrinos at
distances smaller than $100\,{\rm m}$ from reactors. As has been
pointed out in Ref.\cite{RNA1} the ratio of the observed event rate of
antineutrinos, emitted by reactor and detected at distances smaller
than $100\,{\rm m}$, to the predicted rate is of about
$0.943(23)$. This implies an existence of a deficit of antineutrinos
of about $5.7\,\%$. Such a deficit of antineutrinos may be, for
example, explained by the electron--sterile antineutrino oscillations,
where a mass of sterile antineutrinos is of about $m_s \sim 1\,{\rm
  eV}$ \cite{RNA1} (see also Ref.\cite{PDG12}). The experiment on the
observation of the sterile antineutrinos from reactors at distances $6
- 13\,{\rm m}$ has been recently proposed by in Ref.\cite{Serebrov2012}.

The yield of reactor antineutrinos is being detected by the inverse
$\beta$--decay $\bar{\nu}_e + p \to n + e^+$ in terms of the yield of
the positrons, produced by antineutrinos in the energy region $2\,{\rm
  MeV} \le E_{\bar{\nu}} \le 8\,{\rm MeV}$ \cite{RNA1}. The
calculation of the angular distribution and the cross section of the
inverse $\beta$--decay $\bar{\nu}_e + p \to n + e^+$, induced by
reactor antineutrinos, was calculated in Ref.\cite{IBDa} in the
non--relativistic approximation and without radiative
corrections. Then, in Refs.\cite{IBDb,IBDc} the obtained results were
applied to the experimental analysis of the limits on the parameters
of the electron antineutrino oscillations. The account for the
contributions of the ``weak magnetism'' and the neutron recoil,
calculated to next--to--leading order in the large baryon mass
expansion, and the radiative corrections, calculated to leading order
in the large baryon mass expansion, has been carried out in
Refs.\cite{IBD1}--\cite{IBD6}. A comparison of the radiative and
recoil corrections in the neutron $\beta^-$--decay $n \to p + e^- +
\bar{\nu}_e$ to the inverse $\beta$--decay $\bar{\nu}_e + p \to n +
e^+$ has been performed in Ref.\cite{IBD7} within the heavy--baryon chiral
perturbation theory (HB$\chi$PT). The same approach has been also used
in \cite{IBD4}--\cite{IBD7}. The authors of the papers
\cite{IBD1}--\cite{IBD6} discussed the results on the inverse
$\beta$--decay in connection with measurements of the $\theta_{13}$
mixing angle of the antineutrino mass eigenstates and
electron--sterile antineutrino oscillations \cite{PDG12}.

A reactor antineutrino deficit of about $5.7\,\%$ has been observed in
\cite{RNA1} at the level of $98.6\,\%\,{\rm C.L.}$ by using the
theoretical cross section for the inverse $\beta^-$--decay, calculated
for the lifetime of the neutron $\tau_n = 885.7\,{\rm s}$ or the axial
coupling constant $\lambda = - 1.2694$. It is important to emphasise
that the lifetime of the neutron $\tau_n = 885.7\,{\rm s}$ disagrees
with recent world average value $\tau_n = 880.1(1.1)\,{\rm s}$
\cite{PDG12}. Thus one may expect \cite{Serebrov2012} that a reduction
of the lifetime of the neutron from $\tau_n = 885.7\,{\rm s}$ to
$\tau_n = 880.1(1.1)\,{\rm s}$ might lead to an increase of a deficit
of reactor antineutrinos of about $0.7\,\%$. Since the theoretical
value $\tau_n = 879.6\,{\rm s}$ of the lifetime of the neutron,
calculated in Ref.\cite{Ivanov2013}, agrees well with the world
average one $\tau_n = 880.1(1.1)\,{\rm s}$ \cite{PDG12}, below we
follow Ref.\cite{Ivanov2013} and calculate the inverse $\beta$--decay.

Thus, in connection with an analysis of a deficit of reactor
antineutrinos we calculate the angular distribution and cross section
for the inverse $\beta$--decay by taking into account the
contributions of the ``weak magnetism'' and the neutron recoil to
next--to--leading order in the large baryon mass expansion and the
radiative corrections of order $\alpha/\pi$, calculated to leading
order in the large baryon mass expansion, where $\alpha = 1/137.036$
is the fine--structure constant \cite{PDG12}.

The paper is organised as follows. In section~\ref{sec:yield} we give
a numerical analysis of the cross section for the inverse
$\beta$--decay, calculated in Appendices A, B and C. We discuss the
yield of positrons, induced by reactor antineutrinos, in connection
with a deficit of reactor antineutrinos, observed in \cite{RNA1}.  In
section~\ref{sec:asymmetry} we analyse the asymmetry of the inverse
$\beta$--decay as an alternative method for measurements of the axial
coupling constant $\lambda$ and the determination of the correlation
coefficient $a_0$, describing in the neutron $\beta^-$--decay
correlations between the 3-momenta of the electron and antineutrino
(see \cite{Abele1,Konrad} and \cite{Ivanov2013}). We calculate also
the average value $\langle \cos\theta_{e\bar{\nu}}\rangle$ as a
function of the antineutrino energy. In section~\ref{sec:universality}
we confirm a universality of the radiative corrections, calculated to
order $\alpha/\pi \sim 10^{-3}$, to the neutrino (antineutrino)
reactions with the electron (positron) in the final state, pointed out
by Kurylov, Ramsey-Musolf and Vogel \cite{RamseyM2003,RamseyM2002}. We
show that the radiative corrections to the cross sections for the
reactions of the neutrino (antineutrino) disintegration of the
deuteron, calculated in \cite{RamseyM2003,RamseyM2002}, coincide with
the radiative corrections to the cross section for the inverse
$\beta$--decay. In addition we have calculated the
antineutrino--energy spectrum of the neutron $\beta^-$--decay by
taking into account the radiative corrections to order $\alpha/\pi$
and the contributions of the ``weak magnetism'' and the proton recoil
to next--to--leading order in the proton mass expansion. In
section~\ref{sec:conclusion} we summarise the obtained results and
discuss their connection to possible existence of sterile
neutrinos. In Appendices A, B and C we give detailed calculations i)
of the correlation coefficients $A(E_{\bar{\nu}})$, $B(E_{\bar{\nu}})$
and $C(E_{\bar{\nu}})$ (see Eq.(\ref{label1})) to next--to--leading
order in the large baryon mass expansion, caused by the ``weak
magnetism'' and the neutron recoil, ii) of the radiative corrections
to the correlation coefficient $A(E_{\bar{\nu}})$ (see
Eq.(\ref{label1})) or to the cross section for the inverse
$\beta$--decay (see Eq.(\ref{label4})) and iii) of the radiative
corrections to the correlation coefficient $B(E_{\bar{\nu}})$ (see
Eq.(\ref{label1})) or to the asymmetry $B_{\exp}(E_{\bar{\nu}})$ (see
Eq.(\ref{label7})) and $\langle \cos\theta_{e\bar{\nu}}\rangle$ (see
Eq.(\ref{label8})). In Appendix D we calculate the cross section for
the radiative inverse $\beta$--decay by taking into account the
contributions of the proton--photon interaction to leading order in
the large proton mass expansion. Such contributions are important for
a gauge invariance of the amplitude and the final expression for the
cross section for the radiative inverse $\beta$--decay.

\section{Deficit of reactor antineutrinos}
\label{sec:yield}

The angular distribution or the differential cross section for the
inverse $\beta$--decay can be written in the following general from
(see Appendix A and Eq.(\ref{labelA.70}))
\begin{eqnarray}\label{label1}
\frac{d\sigma(E_{\bar{\nu}},\cos\theta_{e\bar{\nu}})}{d\cos\theta_{e\bar{\nu}}}
= \frac{1}{2}\,\sigma_0\,\Big(A(E_{\bar{\nu}})\,\Big(1 +
\frac{\alpha}{\pi}\,f_A(\bar{E})\Big) + B(E_{\bar{\nu}})\Big(1 +
\frac{\alpha}{\pi}\,f_B(\bar{E})\Big)\,\bar{\beta}\,\cos\theta_{e\bar{\nu}} +
C(E_{\bar{\nu}})\,\bar{\beta}^2\,\cos^2\theta_{e\bar{\nu}}\Big)\,\bar{k}\bar{E},
\end{eqnarray}
where $\theta_{e\bar{\nu}}$ is an antineutrino--positron correlation
angle, $\bar{E} = E_{\nu} - \Delta$, $\bar{k} = \sqrt{(E_{\bar{\nu}} -
  \Delta)^2 - m^2_e}$ and $\bar{\beta} = \bar{k}/\bar{E}$ are the
energy, momentum and velocity of the positron with $\Delta = m_n - m_p
= 1.2934\,{\rm MeV}$, calculated at $m_n = 939.5654\,{\rm MeV}$ and
$m_p = 938.2720\,{\rm MeV}$ \cite{PDG12}. The correlation coefficients
$A(E_{\bar{\nu}})$, $B(E_{\bar{\nu}})$ and $C(E_{\bar{\nu}})$ are
calculated to next--to--leading order in the large baryon mass or
large $M$ expansion, caused by the ``weak magnetism'' and the neutron
recoil (see Appendix A), where $M = (m_n + m_p)/2$ is the average
nucleon mass \cite{Ivanov2013}. To leading order in the large $M$
expansion the correlation coefficients are equal to $A(E_{\bar{\nu}})
= 1$, $B(E_{\bar{\nu}}) = a_0$ and $C(E_{\bar{\nu}}) = 0$, where $a_0$
is the correlation coefficient of correlations between the 3--momenta
of the electron and antineutrino in the neutron $\beta^-$--decay
\cite{Abele1}. The constant $\sigma_0$ is equal to
\begin{eqnarray}\label{label2}
\hspace{-0.3in}\sigma_0 = (1 +
3\lambda^2)\,\frac{G^2_F|V_{ud}|^2}{\pi}\,(1 + \Delta_R) =
\frac{2\pi^2}{\tau_n f(E_0, Z = 1)}\,\frac{1}{(1 + \delta_R)},
\end{eqnarray}
where $\lambda$ is the axial coupling constant \cite{Abele1}, $G_F =
1.1664\times 10^{-11}\,{\rm MeV^{-2}}$ and $V_{ud} = 0.97427(15)$ are
the Fermi weak coupling constant and the Cabibbo-Kobayashi-Maskawa
(CKM) quark mixing matrix element \cite{PDG12}, $\tau_n$ is the
lifetime of the neutron, $f(E_0, Z = 1)$ and $E_0 = (m^2_n - m^2_p +
m^2 _e)/2 m_n = 1.2927\,{\rm MeV}$ are the Fermi integral and the
end--point energy of the electron--energy spectrum of the neutron
$\beta^-$--decay \cite{Ivanov2013}, respectively. The Fermi integral,
calculated in \cite{Ivanov2013} by taking into account the
contributions of the ``weak magnetism'' and the proton recoil to
next--to--leading order in the large $M$ expansion, is related to the
phase--space factor $f$, defined by $f(E_0, Z = 1) = m^5_e f$.  The
phase--space factor $f$ depends slightly on the axial coupling
constant $\lambda$ and for $\lambda = - 1.2694$ \cite{RNA1} and for
$\lambda = - 1.2750$ \cite{Abele1} it is equal to $f = 1.6894$
\cite{Ivanov2013}. In turn, the lifetime of the neutron $\tau_n$,
being inversely proportional to the factor $(1 + 3 \lambda^2)$, for
$\lambda = - 1.2694$ \cite{RNA1} and for $\lambda = - 1.2750$
\cite{Abele1} changes from $\tau_n = 885.7\,{\rm s}$ to $\tau_n =
879.6\,{\rm s}$ \cite{Ivanov2013}, respectively. This leads to an
increase of $\sigma_0$ in of about $0.69\,\%$. Such a change of
$\sigma_0$ should lead to an increase of a deficit of reactor
antineutrinos \cite{RNA1,Serebrov2012}. The later is important for the
planning experiments on the detection of sterile antineutrinos at
distances $(6 - 13)\,{\rm m}$ from the reactor by Serebrov {\it et
  al.}  \cite{Serebrov2012}.  Then, $(1 + \delta_R)(1 + \Delta_R) = 1
+ {\rm RC}$ are the radiative corrections \cite{RC8}--\cite{RC17}
integrated over the electron--energy spectrum of the neutron
$\beta^-$--decay (see also \cite{Ivanov2013}), where $\delta_R =
0.01505$ is defined by one--photon exchanges only \cite{RC1,RC17} and
$\Delta_R = 0.02381$ is caused by electroweak boson exchanges and QCD
corrections \cite{RC1,RC17}. The numerical values of $\delta_R$ and
$\Delta_R$, obtained in \cite{RC1,RC17} and discussed in
\cite{Abele1}, have been also confirmed in \cite{Ivanov2013} (see
discussion below Eq.(68) of Ref.\cite{Ivanov2013}). The functions
$(\alpha/\pi)\,f_A(\bar{E})$ and $(\alpha/\pi)\,f_B(\bar{E})$ define
the radiative corrections to the correlation coefficients
$A(E_{\bar{\nu}})$ and $B(E_{\bar{\nu}})$, respectively, calculated to
leading order in the large $M$ expansion (see Appendices A, B and
C). Since the radiative corrections $(\alpha/\pi)\,f_A(\bar{E})$ and
$(\alpha/\pi)\,f_B(\bar{E})$ do not depend on the axial coupling
constant $\lambda$, they do not influence on the change of a deficit
of reactor antineutrinos by changing the axial coupling constant
$\lambda$ from $\lambda = - 1.2694$ to $\lambda = - 1.2750$.

A deficit of reactor antineutrinos may be observed by measuring the
yield of positrons in the inverse $\beta$--decay, defined by
\cite{IBDb,IBDc,IBD1}
\begin{eqnarray}\label{label3}
\hspace{-0.3in}Y_{e^+} = \int^{(E_{\bar{\nu}})_{\rm
    max}}_{(E_{\bar{\nu}})_{\rm min}}dE_{\bar{\nu}}\,\sigma(E_{\bar{\nu}})\,n(E_{\bar{\nu}}),
\end{eqnarray}
where $n(E_{\bar{\nu}})$ is the reactor antineutrino flux
\cite{IBDb,IBDc,IBD1} (see also \cite{RNA1}) for the antineutrino
energy region $(E_{\bar{\nu}})_{\rm min} = 2\,{\rm MeV} \le
E_{\bar{\nu}} \le (E_{\bar{\nu}})_{\rm max} = 8\,{\rm MeV}$ and
$\sigma(E_{\bar{\nu}})$ is the cross section for the inverse
$\beta$--decay. Integrating the angular distribution Eq.(\ref{label1})
over $\cos\theta_{e\bar{\nu}}$ in the limits $-1 \le
\cos\theta_{e\bar{\nu}} \le + 1$ (see Appendix A) we obtain the cross
section for the inverse $\beta$--decay
\begin{eqnarray}\label{label4}
\hspace{-0.3in}\sigma(E_{\bar{\nu}}) = \sigma_0\,\Big(A(E_{\bar{\nu}})
+ \frac{1}{3}\,C(E_{\bar{\nu}})\,\bar{\beta}^2\Big)\,\Big(1 +
\frac{\alpha}{\pi}\,f_A(\bar{E})\Big)\,\bar{k}\bar{E}.
\end{eqnarray}
For the derivation of Eq.(\ref{label4}) we have
neglected the contributions of the terms, which are of order
$(\alpha/\pi)(E_{\bar{\nu}}/M) \sim 10^{-5}$ in the antineutrino
energy region $2\,{\rm MeV} \le E_{\bar{\nu}} \le 8\,{\rm MeV}$.

For the numerical analysis of the cross section for the inverse
$\beta$--decay Eq.(\ref{label4}) we use analytical expressions for the
correlation coefficients $A(E_{\bar{\nu}})$ and $C(E_{\bar{\nu}})$,
given in Appendix A (see Eq.(\ref{labelA.24}) and
Eq.(\ref{labelA.75})), and the function $(\alpha/\pi)\,f_A(\bar{E})$,
calculated in detail in Appendices A and B. The correlation
coefficients $A(E_{\bar{\nu}})$ and $C(E_{\bar{\nu}})$ and the
radiative corrections $(\alpha/\pi)\,f_A(\bar{E})$ are in analytical
agreement with the results, obtained in \cite{IBD1}--\cite{IBD5}. For
the antineutrino energy region $2\,{\rm MeV} \le E_{\bar{\nu}} \le
8\,{\rm MeV}$ the function $(\alpha/\pi)\,f_A(\bar{E})$ is plotted in
Fig.\,1.
\begin{figure}
\centering \includegraphics[height=0.18\textheight]{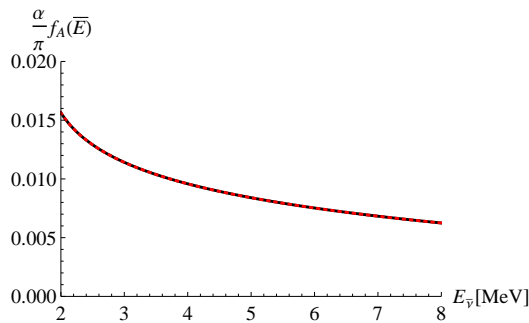}
\caption{(Color online) The radiative corrections $(\alpha/\pi)\,f_A(\bar{E})$ to the
  total cross section for the inverse $\beta$--decay in the
  antineutrino energy region $2\,{\rm MeV} \le E_{\bar{\nu}} \le
  8\,{\rm MeV}$ are given by the dotted line, defined by
  Eq.(\ref{labelA.71}) with the function $f^{(\gamma)}_A(\bar{E})$ in
  Eq.(\ref{labelA.64}), and the solid line, defined by the analytical
  expression for the function $f_A(\bar{E})$ in Eq.(\ref{labelB.20}).}
\end{figure}

A numerical analysis of a deficit of the reactor antineutrinos has
been carried out in \cite{RNA1} for the lifetime of the neutron
$\tau_n = 885.7\,{\rm s}$, corresponding the axial coupling constant
$\lambda = - 1.2694$. This lifetime of the neutron does not agree with
the recent world average value $\tau_n = 880.1(1.1)\,{\rm s}$
\cite{PDG12}, whereas the lifetime of the neutron $\tau_n =
879.6(1.1)\,{\rm s}$, calculated for the axial coupling constant
$\lambda = - 1,2750(9)$ \cite{Ivanov2013}, agrees well with the world
average value $\tau_n = 880.1(1.1)\,{\rm s}$. This is the basis of our
revision of a deficit of reactor antineutrinos, observed in
\cite{RNA1}.

The cross sections for the inverse $\beta$--decay, calculated for
$\lambda = - 1.2750$ and $\lambda = - 1.2694$, defined in the
antineutrino energy region $2\,{\rm MeV} \le E_{\bar{\nu}} \le 8\,{\rm
  MeV}$, are shown in Fig.\,2 (left).
\begin{figure}
\centering \includegraphics[height=0.18\textheight]{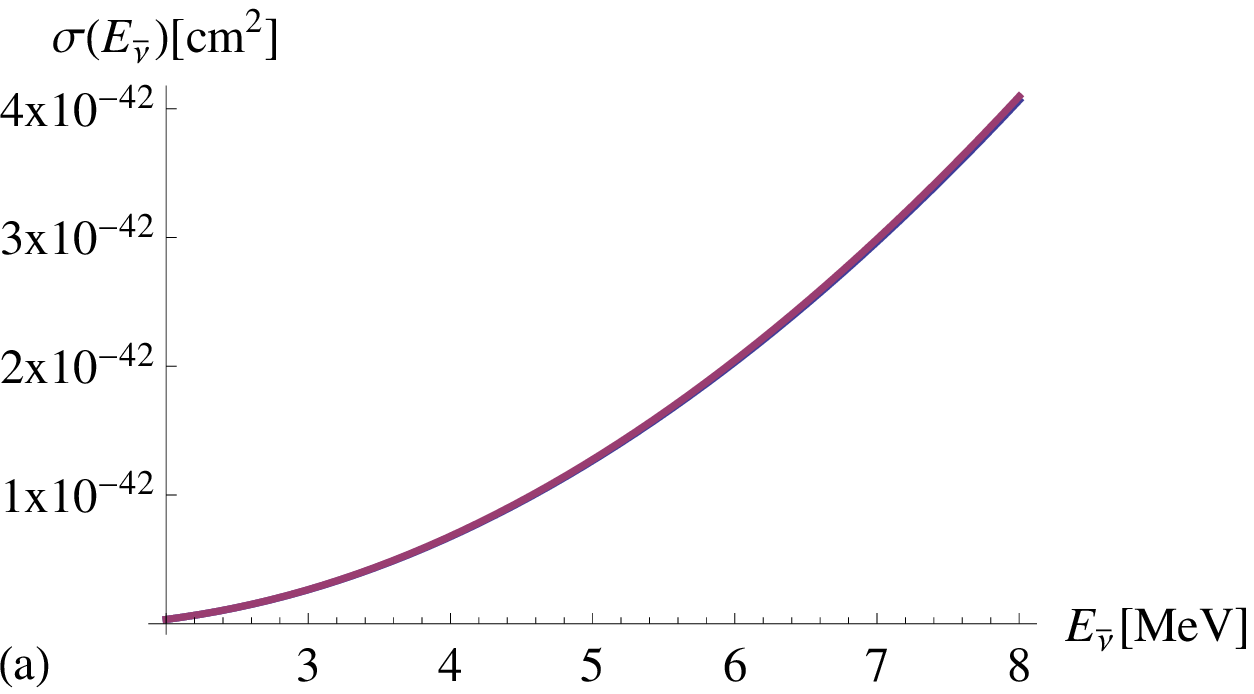}
\centering \includegraphics[height=0.18\textheight]{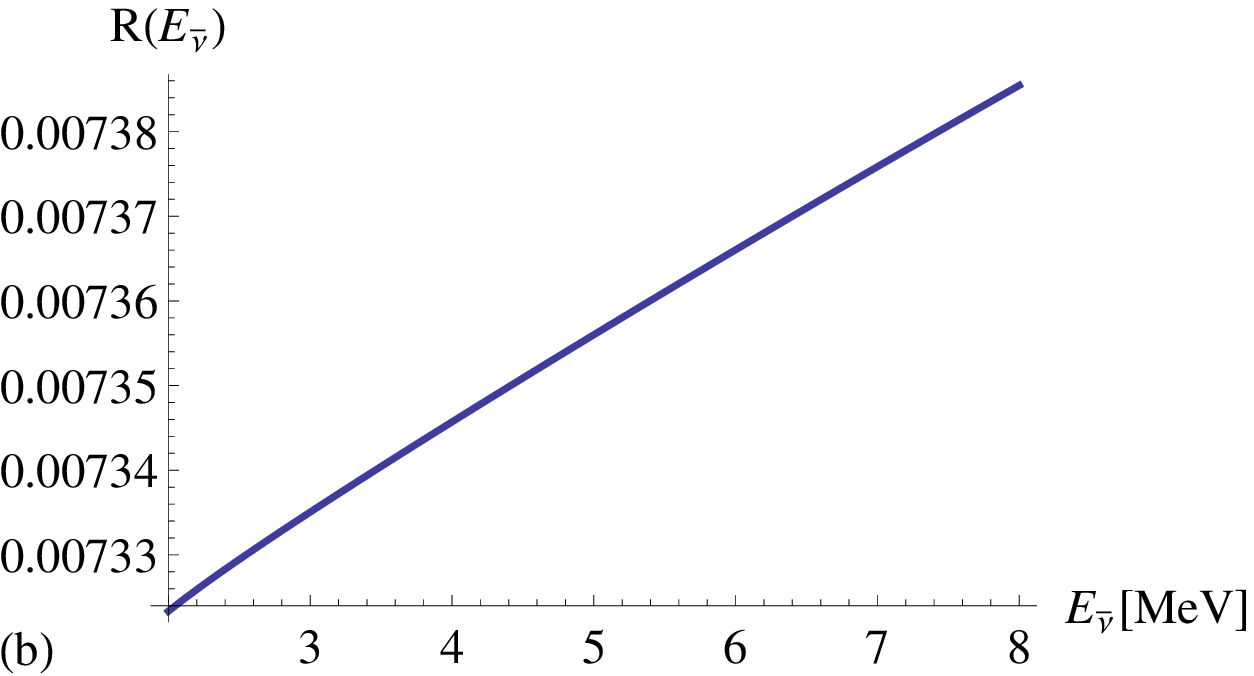}
\caption{(Color online) The cross sections for the inverse $\beta$--decay (a),
  calculated for $\lambda = - 1.2750$ and $\lambda = - 1.2694$, and
  the ratio $R(E_{\bar{\nu}}) = \Delta
  \sigma(E_{\bar{\nu}})/\sigma(E_{\bar{\nu}})$ (b) in the
  antineutrino energy region $2\,{\rm MeV} \le E_{\bar{\nu}} \le
  8\,{\rm MeV}$.}
\end{figure}
Since the cross sections for the inverse $\beta$--decay, calculated
for $\lambda = - 1.2750$ and $\lambda = - 1.2694$, are practically
indistinguishable, in Fig.\,2 (right) we plot the ratio $R(E_{\bar{\nu}}) =
\Delta\sigma(E_{\bar{\nu}})/\sigma(E_{\bar{\nu}})$, given by
\begin{eqnarray}\label{label5}
\hspace{-0.3in}R(E_{\bar{\nu}}) &=& \frac{6\lambda \Delta \lambda}{1 +
  3\lambda^2}\Big\{1 + \frac{1}{M}\,\frac{2}{3}\Big[\Big(1 -
  \frac{\kappa + 1}{2\lambda}\Big)\,\Delta + \Big(- \frac{1}{2} +
  \frac{\kappa + 1}{\lambda}\Big)\,E_{\bar{\nu}} + \Big(-
  \frac{1}{2}\, - \frac{\kappa + 1}{2
    \lambda}\Big)\,\frac{m^2_e}{\bar{E}}\Big]\Big\}\nonumber\\
\hspace{-0.3in}&&\times\,\Big[1 - \frac{1}{M}\,\frac{1}{1 +
    3\lambda^2}\Big(2(\lambda^2 - (\kappa + 1)\lambda)\Delta +
  4(\kappa + 1)\lambda E_{\bar{\nu}} - (\lambda^2 + 2(\kappa +
  1)\lambda + 1)\,\frac{m^2_e}{\bar{E}}\Big) -
  a_0\,\frac{E_{\bar{\nu}}}{M}\Big].
\end{eqnarray}
The ratio $R(E_{\bar{\nu}})$ defines a relative deviation of the cross
section for the inverse $\beta$--decay, calculated at $\lambda = -
1.2694$, from the cross section, calculated at $\lambda = - 1.2750$.
Since in the energy region $2\,{\rm MeV} \le E_{\bar{\nu}} \le 8\,{\rm
  MeV}$ the ratio $R(E_{\bar{\nu}})$ depends slightly on the
antineutrino energy and a maximum of the reactor antineutrino--energy
spectrum is smeared over the region $2\,{\rm MeV} \le E_{\bar{\nu}}
\le 4\,{\rm MeV}$ (see Fig.\,12 of Rev.\cite{RNA1}), one may set that
on average a variation of the axial coupling constant $\lambda$ from
$\lambda = - 1.2694$ to $\lambda = - 1.2750$ changes the value of the
cross section for the inverse $\beta$--decay in of about
$0.734\,\%$. Such a change of the cross section agrees well with a
change of the lifetime of the neutron in $0.69\,\%$ from $\tau_n =
885.7\,{\rm s}$ \cite{RNA1} to $\tau_n = 879.6\,{\rm s}$
\cite{Ivanov2013}.

Since in the antineutrino energy region $2\,{\rm MeV} \le
E_{\bar{\nu}} \le 8\,{\rm MeV}$ the ratio $R(E_{\bar{\nu}})$ depends
slightly on the antineutrino energy, the proposed analysis of the
cross section for the inverse $\beta$--decay implies that a deficit
$\Delta Y_{\bar{\nu}} = 5.7\,\%$ of the reactor antineutrinos,
observed in \cite{RNA1} for the lifetime of the neutron $\tau_n =
885.7\,{\rm s}$ and the axial coupling constant $\lambda = - 1.2694$,
should be increased up to $\Delta Y_{\bar{\nu}} \simeq 6.434\,\%$
including $0.734\,\%$, as has been pointed out in \cite{Serebrov2012}.
We would like to emphasise that the obtained increase of a deficit of
antineutrinos does not depend on the radiative corrections to the
inverse $\beta$--decay, which are cancelled in the ratio
$R(E_{\bar{\nu}})$ due to their independence of the axial coupling
constant.

\section{Asymmetry of inverse $\beta$--decay as tool for
 measurement of correlation coefficient $a_0$}
\label{sec:asymmetry}

In the neutron $\beta^-$--decay the correlation coefficient $a_0$ is a
quantitative characteristic of correlations between 3--momenta of the
electron and antineutrino, calculated to leading order in the large
proton mass expansion \cite{Abele1} (see also
\cite{Ivanov2013}). Since the antineutrino in the final state of the
neutron $\beta^-$--decay is hard to detect, for the experimental
determination of $a_0$ one should measure correlations of the
3--momenta of the decay proton and electron \cite{Ivanov2013}. The
inverse $\beta$--decay may be a good laboratory for a measurement of
the correlation coefficient $a_0$. In terms of the angular
distribution of the cross section for the inverse $\beta$--decay we
may define the asymmetry $B_{\exp}(E_{\bar{\nu}})$ as
\begin{eqnarray}\label{label6}
\hspace{-0.3in}B_{\exp}(E_{\bar{\nu}}) = \frac{\displaystyle
  \frac{d\sigma(E_{\bar{\nu}},\cos\theta_{e\bar{\nu}})}{
    d\cos\theta_{e\bar{\nu}}}\Big|_{\theta_{e\bar{\nu}} = 0} -
  \frac{d\sigma(E_{\bar{\nu}},\cos\theta_{e\bar{\nu}})}{
    d\cos\theta_{e\bar{\nu}}}\Big|_{\theta_{e\bar{\nu}} =
    \pi}}{\displaystyle
  \frac{d\sigma(E_{\bar{\nu}},\cos\theta_{e\bar{\nu}})}{
    d\cos\theta_{e\bar{\nu}}}\Big|_{\theta_{e\bar{\nu}} = 0} +
  \frac{d\sigma(E_{\bar{\nu}},\cos\theta_{e\bar{\nu}})}{
    d\cos\theta_{e\bar{\nu}}}\Big|_{\theta_{e\bar{\nu}} = \pi} }.
\end{eqnarray}
Substituting Eq.(\ref{label1}) into Eq.(\ref{label6}) we obtain
\begin{eqnarray}\label{label7}
\hspace{-0.3in}B_{\exp}(E_{\bar{\nu}}) =
\frac{B(E_{\bar{\nu}})\bar{\beta}}{A(E_{\bar{\nu}}) +
  C(E_{\bar{\nu}})\,\bar{\beta}^2}\,\Big(1 +
\frac{\alpha}{\pi}\,(f_B(\bar{E}) - f_A(\bar{E}))\Big).
\end{eqnarray}
Since the experimental data on the asymmetry $B_{\exp}(E_{\bar{\nu}})$
may be fitted by only one parameter $\lambda$ and at the neglect of
the $1/M$ corrections the asymmetry $B_{\exp}(E_{\bar{\nu}})$ is equal
to $B_{\exp}(E_{\bar{\nu}}) = a_0\bar{\beta}$, the asymmetry
$B_{\exp}(E_{\bar{\nu}})$ may be a good tool for an experimental
determination of the coefficient $a_0$ alternative to the experiments,
discussed in \cite{Ivanov2013}. In Fig.\,3 we plot
$B_{\exp}(E_{\bar{\nu}})$ in the antineutrino energy region $2\,{\rm
  MeV} \le E_{\bar{\nu}} \le 8\,{\rm MeV}$.

\begin{figure}
\centering \includegraphics[height=0.20\textheight]{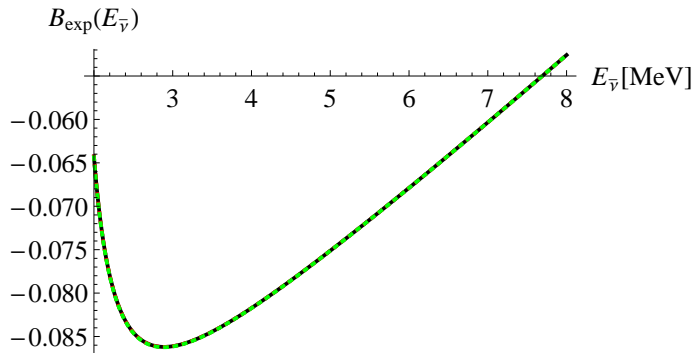}
\caption{(Color online) The asymmetry $B_{\exp}(E_{\bar{\nu}})$ of the inverse
  $\beta$--decay in the antineutrino energy region $2\,{\rm MeV} \le
  E_{\bar{\nu}} \le 8\,{\rm MeV}$. The dotted line corresponds the
  calculation of the radiative corrections
  $(\alpha/\pi)\,f_A(\bar{E})$ and $(\alpha/\pi)\,f_B(\bar{E})$,
  defined by Eq.(\ref{labelA.71}) and Eq.(\ref{labelA.72}) with
  $(\alpha/\pi)\,f^{(\gamma)}_A(\bar{E})$ and
  $(\alpha/\pi)\,f^{(\gamma)}_B(\bar{E})$, given by the integrals in
  Eq.(\ref{labelA.64}) and Eq.(\ref{labelA.65}), respectively. The
  green line is calculated for the radiative corrections, given by the
  analytical expressions in Eq.(\ref{labelB.20}) and
  Eq.(\ref{labelC.16}), respectively.}
\end{figure}
\begin{figure}
\centering \includegraphics[height=0.20\textheight]{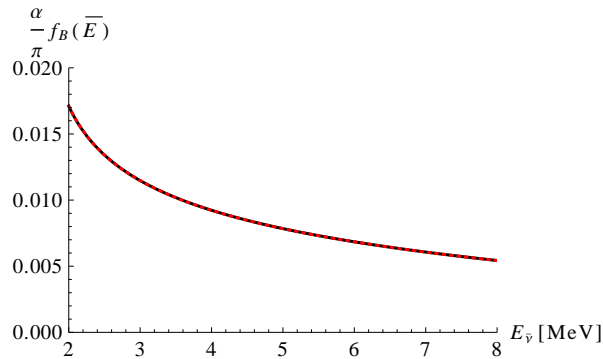}
\caption{(Color online) The radiative correction
  $(\alpha/\pi)\,f_B(\bar{E})$. The dotted line corresponds the
  calculation of the radiative corrections
  $(\alpha/\pi)\,f_B(\bar{E})$, defined by Eq.(\ref{labelA.72}) with
  $(\alpha/\pi)\,f^{(\gamma)}_B(\bar{E})$, given by the integrals over
  the positron energy in Eq.(\ref{labelA.65}). The red line
  corresponds the radiative corrections, given by the analytical
  expressions Eq.(\ref{labelC.16}).}
\end{figure}

The radiative corrections to the asymmetry $B_{\exp}(E_{\bar{\nu}})$
are defined by the functions $(\alpha/\pi)\,f_A(\bar{E})$ and
$(\alpha/\pi)\,f_B(\bar{E})$. The radiative corrections
$(\alpha/\pi)\,f_A(\bar{E})$ to the correlation coefficient
$A(E_{\bar{\nu}}))$ or that is the same to the cross section for the
inverse $\beta$--decay are calculated in analytical agreement with the
results, obtained by Vogel \cite{IBD1}, Fayans \cite{IBD3} and Raha,
Myhrer and Kudobera \cite{IBD5}. In turn, the radiative corrections
$(\alpha/\pi)\,f_B(\bar{E})$ to the correlation coefficient
$B(E_{\bar{\nu}}))$ were calculated by Fukugita and Kubota \cite{IBD4}
and recently by Raha, Myhrer and Kudobera \cite{IBD5}. The function
$(\alpha/\pi)\,f_B(\bar{E})$ agrees well with results, obtained by
Fukugita and Kubota \cite{IBD4} and by Raha, Myhrer and Kudobera
\cite{IBD5}.

We would like to note that the asymmetry $B_{\exp}(E_{\bar{\nu}})$
acquires a maximal absolute value in the region of the antineutrino
energies $2\,{\rm MeV} \le E_{\bar{\nu}} \le 4\,{\rm MeV}$ in the
vicinity of the maximum of the antineutrino--energy spectrum. Such a
property of the asymmetry $B_{\exp}(E_{\bar{\nu}})$ makes meaningful a
measurement of the axial coupling constant $\lambda$ and the
determination of the correlation coefficient $a_0$ from the inverse
$\beta$--decay as a method of the determination of the correlation
coefficient $a_0$ alternative to the electron--proton energy
distribution and the proton--energy spectrum, discussed in
\cite{Ivanov2013}.

Then we follow \cite{IBD3} and calculate the average value of
$\cos\theta_{e\bar{\nu}}$. We obtain
\begin{eqnarray}\label{label8}
\hspace{-0.3in}\langle \cos\theta_{e\bar{\nu}}\rangle =
\frac{\displaystyle \int^{+1}_{-1}\cos\theta_{e\bar{\nu}}\, \frac{
    d\sigma(E_{\bar{\nu}},\cos\theta_{e\bar{\nu}})}{d\cos\theta_{e\bar{\nu}}}\,d\cos\theta_{e\bar{\nu}}}
     {\displaystyle \int^{+1}_{-1} \frac{
         d\sigma(E_{\bar{\nu}},\cos\theta_{e\bar{\nu}})}{d\cos\theta_{e\bar{\nu}}}\,
       d\cos\theta_{e\bar{\nu}}} = \frac{B(E_{\bar{\nu}})\bar{\beta}}{3 A(E_{\bar{\nu}}) +
  C(E_{\bar{\nu}})\,\bar{\beta}^2}\,\Big(1 +
\frac{\alpha}{\pi}\,(f_B(\bar{E}) - f_A(\bar{E}))\Big).
\end{eqnarray}
In Fig.\,5 we plot $\langle \cos\theta_{e\bar{\nu}}\rangle$ in the
region of the antineutrino energy $2\,{\rm MeV} \le E_{\bar{\nu}} \le
8\,{\rm MeV}$. In comparison with \cite{IBD3}, the average value
$\langle \cos\theta_{e\bar{\nu}}\rangle$, given by Eq.(\ref{label8}),
is improved by the contributions of the radiative corrections.

\begin{figure}
\centering \includegraphics[height=0.20\textheight]{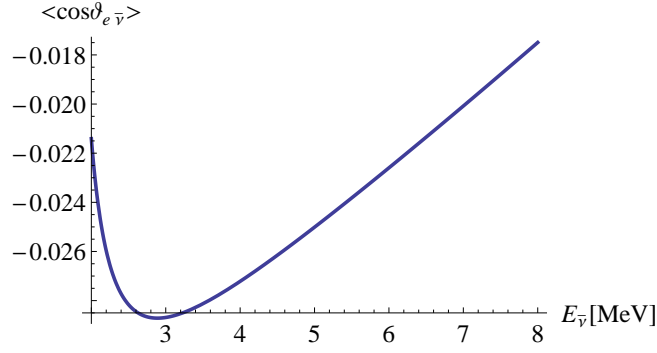}
\caption{(Color online) Average value $\langle \cos\theta_{e\bar{\nu}}\rangle$,
  plotted in the antineutrino energy region $2\,{\rm MeV} \le
  E_{\bar{\nu}} \le 8\,{\rm MeV}$.}
\end{figure}

\section{Universality of radiative corrections to order $\alpha$ to  
neutrino (antineutrino) reactions, induced by weak charged currents, and
antineutrino energy spectrum of neutron $\beta^-$--decay}
\label{sec:universality}

As has been pointed out by Kurylov, Ramsey-Musolf and Vogel
\cite{RamseyM2003,RamseyM2002}, the radiative corrections to order
$\alpha/\pi \sim 10^{-3}$, calculated to the cross sections for a
class of nuclear reactions, induced by weak charged currents and
involving neutrino (antineutrino) in the initial state with electron
(positron) in the final state, are universal. Such a universality has
been proved in \cite{RamseyM2003,RamseyM2002} for the reactions of the
neutrino (antineutrino) disintegration of the deuteron, caused by
charged weak currents. As has been found in
\cite{RamseyM2003,RamseyM2002}, the cross sections for the neutrino
(antineutrino) reactions, induced by charged weak currents, can be
written in the following form
\begin{eqnarray}\label{label9}
\hspace{-0.3in}\sigma_{\rm CC}(E) = \sigma^{(\rm tree)}_{\rm
  CC}(E)\,\Big(1 + \frac{\alpha}{\pi}\,g(E)\Big),
\end{eqnarray}
where $\sigma^{(\rm tree)}_{\rm CC}(E)$ and $\sigma_{\rm CC}(E)$ are
the cross sections for the reaction under consideration, calculated to
leading and to next--to--leading order in $\alpha/\pi$, respectively,
$E$ is the energy observed in the detector. The function $g(E)$ is
given by a sum of one--virtual photon exchanges, bremsstrahlung,
electroweak boson exchanges and QCD corrections. Using
Eq.(\ref{labelB.7}) the function $g(E)$, calculated in
\cite{RamseyM2003}, can be given by
\begin{eqnarray}\label{label10}
\hspace{-0.3in}&&g(E) = \frac{3}{2}\,{\ell
  n}\Big(\frac{m_p}{m_e}\Big) + \frac{3}{4} - \frac{1}{2}\,{\ell
  n}\Big(\frac{1 + \sqrt{1 - \beta^2}}{1 - \sqrt{1 -
    \beta^2}}\Big)\,\Big[\frac{1}{\beta}\,{\ell n}\Big(\frac{1 +
    \beta}{1 - \beta}\Big) - 2\Big] + {\ell n}\Big(\frac{2\beta}{1 +
  \beta}\Big)\,\Big[\frac{1}{\beta}\,{\ell n}\Big(\frac{1 + \beta}{1 -
    \beta}\Big) - 2\Big] + \frac{2}{\beta}\,L\Big(\frac{2\beta}{1 +
  \beta}\Big) \nonumber\\
\hspace{-0.3in}&& + \frac{(1 - \beta)^2}{2\beta}\,{\ell n}\Big(\frac{1 +
  \beta}{1 - \beta}\Big)+ \frac{1}{2\beta E^2}\int^{E}_{m_e}dx\,(E - x)\,{\ell
  n}\Big(\frac{1 + \beta(x)}{1 - \beta(x)}\Big) + \frac{1}{\beta
  E}\int^{E}_{m_e}\frac{dx}{E - x}\,\Big\{x\beta(x)\Big[\frac{1}{\beta(x)}\,{\ell
    n}\Big(\frac{1 + \beta(x)}{1 - \beta(x)}\Big) - 2\Big]\nonumber\\
\hspace{-0.3in}&& - E \beta
\Big[\frac{1}{\beta}\,{\ell n}\Big(\frac{1 + \beta}{1 - \beta}\Big) -
  2\Big]\Big\} + C_R,
\end{eqnarray}
where the constant $C_R$ is defined by the contributions of
electroweak boson exchanges and QCD corrections. The calculation of
the integrals one can perform analytically by using the procedure,
expounded in Appendices B and C. As a result we arrive at the
analytical expressions, adduced for these integrals in the Appendix to
Ref.\cite{RamseyM2003}. Using these expressions we may transcribe the
function $g(E)$ into the form
\begin{eqnarray}\label{label13}
\hspace{-0.3in}g(E) &=& \frac{3}{2}\,{\ell
  n}\Big(\frac{m_p}{m_e}\Big) + \frac{23}{8} + 2 {\ell
  n}\Big(\frac{2\beta}{1 + \beta}\Big)\,\Big[\frac{1}{\beta}\,{\ell
    n}\Big(\frac{1 + \beta}{1 - \beta}\Big) - 2\Big] +
\frac{4}{\beta}\,L\Big(\frac{2\beta}{1 + \beta}\Big) +
\frac{3}{8}\,\Big(\beta^2 + \frac{7}{3}\big)\,\frac{1}{\beta}\,{\ell
  n}\Big(\frac{1 + \beta}{1 - \beta}\Big)\nonumber\\
\hspace{-0.3in}&-& 2\,{\ell n}\Big(\frac{1 + \beta}{1 - \beta}\Big) +
C'_{\rm WZ} = f_A(E) + C'_{\rm WZ},
\end{eqnarray}
where we have redefined the contributions of the electroweak boson
exchanges and QCD corrections. The function $f_A(E)$, given by
Eq.(\ref{labelA.76}) (see also Eq.(\ref{labelB.20})), defines the
radiative corrections to the cross section for the inverse
$\beta$--decay Eq.(\ref{label4}), induced by the charged weak
currents. Thus we may rewrite Eq.(\ref{label9}) as follows
\begin{eqnarray}\label{label14}
\hspace{-0.3in}\sigma_{\rm CC}(E) = \sigma^{(\rm tree)}_{\rm
  CC}(E)\,(1 + \Delta'_R)\,\Big(1 + \frac{\alpha}{\pi}\,f_A(E)\Big),
\end{eqnarray}
where $\Delta'_R = (\alpha/\pi)\,C'_{\rm WZ}$. According to
\cite{RamseyM2003}, the constant $\Delta'_R$ may contain also the
contributions, depending on a nuclear structure of a target.

As has been pointed by Sirlin \cite{Sirlin2011} (see also
\cite{RamseyM2003,RamseyM2002}), the radiative corrections to the
antineutrino (neutrino) energy spectra of the $\beta$--decays are
described by the function $f_A(\bar{E})$. Indeed, one may show that
the antineutrino energy spectrum of the neutron $\beta^-$--decay takes
the form
\begin{eqnarray}\label{label15}
\hspace{-0.3in}\frac{d\lambda_n(E_{\bar{\nu}})}{dE_{\bar{\nu}}} = (1 +
3\lambda^2)\,\frac{G^2_F|V_{ud}|^2}{2\pi^3}\,(1 + \Delta_R)\,\bar{k}_e
\bar{E}_e F(\bar{E}_e, Z =
1)\,E^2_{\bar{\nu}}\,\Big(1 +
\frac{\alpha}{\pi}\,f_A(\bar{E}_e)\Big)\,\zeta(\bar{E}_e),
\end{eqnarray}
where $\bar{E}_e = E_0 - E_{\bar{\nu}}$ and $\bar{k}_e =
\sqrt{\bar{E}^2_e - m^2_e}$ are the energy and momentum of the
electron, $E_0 = (m^2_n - m^2_p + m^2_e)/2m_n= 1.2927\,{\rm MeV}$ is
the end--point energy of the electron energy spectrum of the neutron
$\beta^-$--decay and $F(\bar{E}_e, Z = 1)$ is the Fermi function,
describing the electron--proton Coulomb final-state interaction
\cite{Ivanov2013}. To leading order in the large $M$ expansion the
function $\zeta(\bar{E}_e)$ is equal to unity, $\zeta(\bar{E}_e) =
1$. The deviations from unity are defined to order $1/M$ and caused by
the contributions of the ``weak magnetism'' and the proton recoil. 

Thus, the analysis, carried out above, confirms a universality of the
radiative corrections to the cross sections for a class of neutrino
(antineutrino) reactions, induced by charged currents and calculated
to order $\alpha/\pi \sim 10^{-3}$. A nice review of the radiative
corrections in precision electroweak physics has been recently written
by Sirlin and Ferroglia \cite{Sirlin2012}.

We would like to emphasise that the function $\zeta(\bar{E}_e)$ cannot
be obtained from the function $\zeta(E_e)$, calculated in
\cite{Ivanov2013}, by the replacement $E_e \to \bar{E}_e$ that is
valid for the transition probability of the neutron $\beta^-$--decay
\cite{Sirlin2011}, calculated to leading order in the large $M$
expansion. A deviation of $\zeta(\bar{E}_e)$ from $\zeta(E_e)$ is
attributable to different phase--space factors
$\Phi_{\beta^-_c}(\vec{\bar{k}}_e, \vec{k}_{\bar{\nu}})$
\cite{Ivanov2013}, where $\vec{k}_{\bar{\nu}}$ is a 3--momentum of the
antineutrino. For the antineutrino energy spectrum of the neutron
$\beta^-$--decay the phase--space factor
$\Phi_{\beta^-_c}(\vec{\bar{k}}_e, \vec{k}\,)$ is defined by
\begin{eqnarray}\label{label16}
\hspace{-0.3in}&&\Phi_{\beta^-_c}(\vec{\bar{k}}_e, \vec{k}\,) =
\int^{\infty}_{m_e}\delta(f(E_e))\,\frac{m_n}{E_p}\,\frac{k_e
  E_e}{\bar{k}_e \bar{E}_e}\,dE_e = \frac{1}{\displaystyle 1 -
  \frac{E_{\bar{\nu}}}{m_n}\Big(1 -
  \frac{1}{\beta}\,\cos\theta_{e\bar{\nu}}\Big)}\,\frac{k_e
  E_e}{\bar{k}_e \bar{E}_e}\Big|_{f(E_e) = 0},
\end{eqnarray}
where $f(E_e) = m_n - E_p - E_{\bar{\nu}} - E_e$, $E_p =
\sqrt{(\vec{k}_e + \vec{k}_{\bar{\nu}})^2 + m^2_p}$, $k_e =
\sqrt{E^2_e - m^2_e}$ and the electron energy $E_e$ is equal to
\begin{eqnarray}\label{label17}
\hspace{-0.3in}E_e = \frac{E_0 - E_{\bar{\nu}}}{\displaystyle 1 -
  \frac{E_{\bar{\nu}}}{m_n}\Big(1 -
  \beta\,\cos\theta_{e\bar{\nu}}\Big)}.
\end{eqnarray}
The exact expression for the phase--space factor
$\Phi_{\beta^-_c}(\vec{\bar{k}}_e, \vec{k}_{\bar{\nu}})$ is
\begin{eqnarray}\label{label18}
\hspace{-0.3in}&&\Phi_{\beta^-_c}(\vec{\bar{k}}_e, \vec{k}_{\bar{\nu}}) =
\frac{\displaystyle \sqrt{1 + 2\,\frac{1 -
      \bar{\beta}^2}{\bar{\beta}^2}\,\frac{E_{\bar{\nu}}}{m_n}\,\Big(1 -
    \beta\,\cos\theta_{e\bar{\nu}}\Big)\Big(1 -
    \frac{1}{2}\,\frac{E_{\bar{\nu}}}{m_n}\Big(1 -
    \beta\,\cos\theta_{e\bar{\nu}}\Big)\Big)}}{\displaystyle \Big(1 -
  \frac{E_{\bar{\nu}}}{m_n}\Big(1 -
  \frac{1}{\beta}\,\cos\theta_{e\bar{\nu}}\Big)\Big)\,\Big(1 -
  \frac{E_{\bar{\nu}}}{m_n}\Big(1 -
  \beta\,\cos\theta_{e\bar{\nu}}\Big)\Big)^2}.
\end{eqnarray}
A detailed analysis of the antineutrino--energy and angular
$(E_{\bar{\nu}},\cos\theta_{e\bar{\nu}})$ distribution and the
antineutrino--energy spectrum we are planning to perform in our
forthcoming publication. Here we give the antineutrino--energy
spectrum in the antineutrino-- energy region $1 \gg
E_{\bar{\nu}}/\bar{\beta}^2M$ or $E_{\bar{\nu}} \ll (E_0 - m_e)(1 -
m_e/2M)$. In this energy region the phase--space factor
$\Phi_{\beta^-_c}(\vec{\bar{k}}_e, \vec{k}_{\bar{\nu}})$ takes the
form
\begin{eqnarray}\label{label19}
\hspace{-0.3in}&&\Phi_{\beta^-_c}(\vec{\bar{k}}_e, \vec{k}_{\bar{\nu}}) = 1 +
\frac{E_{\bar{\nu}}}{M}\,\Big(\frac{1 + 2
  \bar{\beta}^2}{\bar{\beta}^2} - \frac{2 +
  \bar{\beta}^2}{\bar{\beta}}\,\cos\theta_{e\bar{\nu}}\Big) = 1 +
\frac{E_{\bar{\nu}}}{M}\,\Big(\frac{1 + 2
  \bar{\beta}^2}{\bar{\beta}^2} - \frac{2 +
  \bar{\beta}^2}{\bar{\beta}^2}\,\frac{\vec{\bar{k}}_e\cdot \vec{k}_{\bar{\nu}}}{\bar{E}_e E_{\bar{\nu}}}\,\Big).
\end{eqnarray}
Using the results, obtained in \cite{Ivanov2013} (see Eq.(A-17) of
Ref.\cite{Ivanov2013}) and the expansions
\begin{eqnarray}\label{label20}
\hspace{-0.3in}E_e &=& \bar{E}_e\,\Big(1 +
\frac{E_{\bar{\nu}}}{M}\,\Big(1 -
\bar{\beta}\,\cos\theta_{e\bar{\nu}}\Big)\Big),\nonumber\\ k_e &=&
\bar{k}_e\,\Big(1 +
\frac{1}{\bar{\beta}^2}\,\frac{E_{\bar{\nu}}}{M}\,\Big(1 -
\bar{\beta}\,\cos\theta_{e\bar{\nu}}\Big)\Big),\nonumber\\ \beta &=&
\bar{\beta}\,\Big(1 + \frac{1 -
  \bar{\beta}^2}{\bar{\beta}^2}\,\frac{E_{\bar{\nu}}}{M}\,\Big(1 -
\bar{\beta}\,\cos\theta_{e\bar{\nu}}\Big)\Big)
\end{eqnarray}
for the function $\zeta(\bar{E}_e)$ we obtain the following expression
\begin{eqnarray}\label{label21}
\hspace{-0.3in}\zeta(\bar{E}_e) &=& 1 + \frac{1}{1 +
  3\lambda^2}\,\frac{1}{M}\,\Big[\Big((1 +
  3\lambda^2)\,\frac{E_{\bar{\nu}}}{\bar{\beta}^2} + \Big(7\lambda^2 +
    4(\kappa + 1)\,\lambda + 1\Big)\,E_{\bar{\nu}} -
    2\,\lambda\,\Big(\lambda + \kappa + 1\Big)\,E_0 \nonumber\\
\hspace{-0.3in}&-& \Big(\lambda^2 -
    2(\kappa + 1) + 1\Big)
    \,\frac{m^2_e}{\bar{E}_e}\,\Big].
\end{eqnarray}
Analysing the lifetime of the neutron by integrating the
antineutrino--energy spectrum one can show that in the vicinity of the
end--point energy $(E_0 - m_e)$ of the antineutrino--energy spectrum
the contribution of the term, proportional to $1/\bar{\beta}^2$,
behaves as $\alpha\,((E_0 - m_e)/M)\,{\ell n}\bar{\beta}$.

Taking into account in the function $\zeta(\bar{E}_e)$ the terms,
non--singular in the limit $\bar{\beta} \to 0$, we get the lifetime of
the neutron equal to $\tau_n = 880.6(1.1)\,{\rm s}$. The lifetime of
the neutron, calculated from the antineutrino--energy spectrum with
the function $\zeta(\bar{E}_e)$, given by Eq.(\ref{label21}), by the
integration over the antineutrino--energy spectrum in the limits $0
\le E_{\bar{\nu}} \le (E_0 - m_e)(1 - m_e/2M)$ is equal to $\tau_n =
879.9(1.1)\,{\rm s}$. Thus, the contribution of the term, proportional
to $1/\bar{\beta}^2$, is of order $0.7\,{\rm s}$ or $0.08\,\%$, which
is smaller compared with the theoretical uncertainty of the lifetime
of the neutron $\Delta_n = \pm 1.1\,{\rm s}$ or $\pm 0.13\,\%$. Both
values of the lifetime of the neutron $\tau_n = 880.6(1.1)\,{\rm s}$
and $\tau_n = 879.9(1.1)\,{\rm s}$, calculated from the
antineutrino--energy spectrum, agree well with the lifetime of the
neutron $\tau_n = 879.6(1.1)\,{\rm s}$, calculated in
\cite{Ivanov2013} by integrating over the electron--energy spectrum,
and the world average value $\tau_n = 880.1(1.1)\,{\rm s}$
\cite{PDG12}.

\section{Conclusion and discussion}
\label{sec:conclusion}

We have analysed a deficit of reactor antineutrinos at distances
smaller than $100\,{\rm m}$. We have carried out this analysis having
investigated the yield of antineutrinos in terms of the yield of
positrons from the inverse $\beta$--decay $\bar{\nu}_e + p \to n +
e^+$. The positrons are produced by reactor antineutrinos in the
antineutrino energy region $2\,{\rm MeV} \le E_{\bar{\nu}} \le 8\,{\rm
  MeV}$. The cross section for the inverse $\beta$--decay we have
calculated by taking into account i) the contributions of the ``weak
magnetism'' and neutron recoil to next--to--leading order in the large
baryon mass expansion and ii) the radiative corrections to order
$\alpha/\pi$, caused by one--virtual photon exchanges and the
radiative inverse $\beta$--decay, calculated to leading order in the
large baryon mass expansion. The radiative corrections defined by the
electroweak boson exchanges and the QCD corrections are taken in terms
of the parameter $\Delta_R = 0.02381$ \cite{RC1,Abele1} (see also
\cite{Ivanov2013}). We have shown that the calculation of the cross
section for the inverse $\beta$--decay for the axial coupling constant
$\lambda = - 1.2750$ and the lifetime of the neutron $\tau_n =
879.6(1.1)\,{\rm s}$ \cite{Ivanov2013} increases the yield of
positrons of about $0.734\,\%$ with respect to the yield of
positrons, calculated for the axial coupling constant $\lambda = -
1.2694$ and lifetime of the neutron $\tau_n = 885.7\,{\rm s}$
\cite{RNA1}. As a result, a deficit of reactor antineutrinos at
distances smaller than $100\,{\rm m}$ is of about $\Delta
Y_{\bar{\nu}} = 6,434\,\%$. Such an increase of a deficit of reactor
antineutrinos, which for the first time has been pointed out in
\cite{Serebrov2012}, makes meaningful a search for sterile
antineutrinos at distances $(6 - 13)\,{\rm m}$ from reactors, proposed
by Serebrov {\it et al.}  \cite{Serebrov2012}.

The dependence of the angular distribution for the inverse
$\beta$--decay on $\cos\theta_{e\bar{\nu}}$ allows to define the
asymmetry $B_{\exp}(E_{\bar{\nu}})$ of the position yields in the
$\theta_{e\bar{\nu}} = 0$ and $\theta_{e\bar{\nu}} = \pi$
directions. In the non--relativistic approximation or to leading order
in the large baryon mass expansion the asymmetry
$B_{\exp}(E_{\bar{\nu}})$ is proportional to the correlation
coefficient $a_0$, describing in the neutron $\beta^-$--decay and in
the non--relativistic approximation correlations between the
3--momenta of the electron and antineutrino \cite{Abele1} (see also
\cite{Ivanov2013}). The experimental analysis of the asymmetry
$B_{\exp}(E_{\bar{\nu}})$ may be treated as an alternative method for
a measurement of the axial coupling constant $\lambda$ and a
determination of the correlation coefficient $a_0$ (see \cite{Konrad}
and \cite{Ivanov2013}).  We have shown that the absolute values of the
asymmetry $B_{\exp}(E_{\bar{\nu}})$ are maximal in the antineutrino
energy region $2\,{\rm MeV} \le E_{\bar{\nu}} \le 4\,{\rm MeV}$, where
the antineutrino--energy spectrum is maximal. Since at experiment
positrons are emitted forward and backward into a solid angle $\Delta
\Omega_{12} = 2\pi(\cos\theta_1 - \cos\theta_2)$, we may define the
asymmetry $B_{\exp}(E_{\bar{\nu}},\theta_2, \theta_1)$ as follows
\begin{eqnarray}\label{label22}
\hspace{-0.3in}B_{\exp}(E_{\bar{\nu}},\theta_2, \theta_1) =
\frac{1}{2} \,\frac{B(E_{\bar{\nu}})\bar{\beta}\,(cos\theta_1 +
  cos\theta_2)}{\displaystyle A(E_{\bar{\nu}}) +
  \frac{1}{3}\,C(E_{\bar{\nu}})\,\bar{\beta}^2\,(\cos^2\theta_1 + \cos\theta_1
  \cos\theta_2 + \cos^2\theta_2)}\,\Big(1 +
\frac{\alpha}{\pi}\,(f_B(\bar{E}) - f_A(\bar{E}))\Big).
\end{eqnarray}
For $\theta_2 = \theta_1 = 0$ that corresponds an emission of
positrons into a zero solid angle forward and backward we arrive at
Eq.(\ref{label7}). A definition of the asymmetry
$B_{\exp}(E_{\bar{\nu}}; \theta_2, \theta_1)$ is similar to a
definition of the asymmetry $A_{\exp}(E_e)$ in the neutron
$\beta^-$--decay \cite{Ivanov2013}, describing correlations between
the neutron spin and the electron 3--momentum.

We have calculated the average value of $\langle
\cos\theta_{e\bar{\nu}} \rangle$ as a function of the antineutrino
energy $E_{\bar{\nu}}$. In comparison with the results, obtained by Vogel
and Beacom \cite{IBD3}, we have improved the average value $\langle
\cos\theta_{e\bar{\nu}}\rangle$ by taking into account the
contributions of the radiative corrections.

In our approach the radiative corrections are defined by the functions
$(\alpha/\pi)\,f_A(\bar{E})$ and $(\alpha/\pi)\,f_B(\bar{E})$. They
are calculated to leading order in the large baryon mass expansion.
The function $(\alpha/\pi)\,f_A(\bar{E})$ and
$(\alpha/\pi)\,f_B(\bar{E})$ define the radiative corrections to the
correlation coefficient $A(E_{\bar{\nu}})$ or to the cross section for
the inverse $\beta$--decay and the correlation coefficient
$B(E_{\bar{\nu}})$ or to the asymmetry $B_{\exp}(E_{\bar{\nu}})$,
respectively. They are in analytical agreement with the results,
obtained by Vogel \cite{IBD1}, Fayans \cite{IBD3}, Fukugita and Kubota
\cite{IBD4} and Raha, Myhrer and Kubodera \cite{IBD5}.  We would like
to note that the authors of Ref.\cite{IBD5} calculated the
phase--space factor Eq.(\ref{labelA.14}) by using a non--relativistic
approximation for a total energy of the neutron. This leads to an
increase of the cross section for the inverse $\beta$--decay by a
factor $(1 + \Delta/M)$ in comparison to our result and to the result,
obtained by Vogel and Beacom \cite{IBD3}. 

In Appendix D we have calculated the angular and photon--energy
distribution of the radiative inverse $\beta$--decay by taking into
account the contribution of the proton--photon interaction to leading
order in the large proton mass expansion. Such a contribution is
responsible for a gauge invariance of the amplitude of the radiative
inverse $\beta$--decay and a gauge invariant calculation of the
angular and photon--energy distribution of the radiative inverse
$\beta$--decay. The performed analysis confirms our results, obtained
in Appendix A by using the Coulomb gauge for a description of physical
degrees of freedom of an emitted real photon.

Finally we have confirmed a universality of the radiative corrections
to order $\alpha/\pi \sim 10^{-3}$, described by the function
$f_A(\bar{E})$ for the cross section for the inverse
$\beta$--decay. Such a universality of the radiative corrections to
order $\alpha/\pi$ for the neutrino (antineutrino) reactions, induced
by weak charged currents, has been proved by Kurylov, Ramsey-Musolf
and Vogel by example of the neutrino (antineutrino) disintegration of
the deuteron with the electron (positron) in the final state
\cite{RamseyM2003,RamseyM2002}. Following
\cite{RamseyM2003,RamseyM2002} and using the results, obtained by
Sirlin \cite{Sirlin2011}, one may extend such a universality of the
radiative corrections, given by the function $f_A(\bar{E})$, on the
neutrino (antineutrino) energy spectra. We have calculated the
antineutrino--energy spectrum of the neutron $\beta^-$--decay by
taking into account the radiative corrections to order $\alpha/\pi
\sim 10^{-3}$ and the contributions of the ``weak magnetism'' and the
proton recoil to next--to--leading order in the large $M$
expansion. We have found that the large $M$ expansion is not well
defined for the antineutrino--energy spectrum of the neutron
$\beta^-$--decay. Restricting the antineutrino energies from above by
the inequality $1\gg E_{\bar{\nu}}/\bar{\beta}^2 M$ we have calculated
the contributions of the ``weak magnetism'' and the proton recoil.  We
have shown that the non--singular terms of the antineutrino--energy
spectrum give the lifetime of the neutron equal to $\tau_n =
880.6(1.1)\,{\rm s}$. The contributions of the non--singular terms
together with the contribution of the term proportional to
$1/\bar{\beta}^2$, integrated over the antineutrino--energies in the
limits $0 \le E_{\bar{\nu}} \le (E_0 - m_e)(1 - m_e/2M)$, change the
lifetime of the neutron as $\tau_n = 879.9(1.1)\,{\rm s}$. Both values
of the lifetime of the neutron $\tau_n = 880.6(1.1)\,{\rm s}$ and
$\tau_n = 879.9(1.1)\,{\rm s}$ agree well with the lifetime of the
neutron $\tau_n = 879.6(1.1)\,{\rm s}$, calculated in
\cite{Ivanov2013} by integrating the electron--energy spectrum of the
neutron $\beta^-$--decay and the world average value $\tau_n =
880.1(1.1)\,{\rm s}$ \cite{PDG12}. A more detailed analysis of the
antineutrino--energy and angular distribution $(E_{\bar{\nu}},
\cos\theta_{e\bar{\nu}})$ of the neutron $\beta^-$--decay and the
lifetime of the neutron, obtained from the antineutrino--energy
spectrum, we are planning to perform in our forthcoming publication.

An alternative analysis of a total cross section for the inverse
$\beta$ decay has been proposed by Strumia and Vissani
\cite{Vissani2003}. The authors have calculated a cross section for
the inverse $\beta$ decay by using i) the radiative corrections,
calculated by Kurylov, Ramsey-Musolf and Vogel \cite{RamseyM2002}, ii)
vector, axial-vector and weak magnetism form-factors, taken in the
dipole approximation, and iii) the contribution of the pseudoscalar
form-factor, calculated within current algebra technique with PCAC
(Partial Conservation of Axial Current) hypothesis in the
one--pion--exchange approximation \cite{Nagels1979}. Because of these
form-factors the obtained expression of the cross section for the
inverse $\beta$ decay can be applied to an analysis of
antineutrino--proton inelastic scattering in the energy region, going
beyond the energy region of reactor antineutrinos. In the
non--relativistic approximation for baryons and to order 1/M a total
cross section for the inverse $\beta$ decay, calculated in
\cite{Vissani2003}, should be reduced to ours and as well as to the
expressions, obtained in \cite{IBD1}--\cite{IBD5}.

\section{Acknowledgements}

We are grateful to Hartmut Abele, Torleif Ericson, Manfried Faber and
Gertrud Konrad for fruitful discussions. We thank Ramsey-Musolf for
calling our attention to the papers \cite{RamseyM2003,RamseyM2002} and
fruitful discussions. This work was supported by the Austrian ``Fonds
zur F\"orderung der Wissenschaftlichen Forschung'' (FWF) under the
contracts I689-N16, I534-N20 PERC and I862-N20 and by the Russian
Foundation for Basic Research under the contract No. 11-02-91000
-ANF$_-$a.

\section{Appendix A: Amplitude and cross section of inverse 
$\beta$--decay} \renewcommand{\theequation}{A-\arabic{equation}}
\setcounter{equation}{0}

\subsection*{1. Next--to--leading order $1/M$ corrections, caused by 
weak magnetism and neutron recoil}

For the calculation of the cross section for the inverse
$\beta$--decay we use the following Hamiltonian of weak
lepton--nucleon interactions \cite{Ivanov2013}
\begin{eqnarray}\label{labelA.1}
\hspace{-0.3in}{\cal H}_W(x) =
\frac{G_F}{\sqrt{2}}\,V_{ud}\,\Big\{[\bar{\psi}_n(x)\gamma_{\mu}(1 +
  \lambda \gamma^5)\psi_p(x)] + \frac{\kappa}{2 M}
\partial^{\nu}[\bar{\psi}_n(x)\sigma_{\mu\nu}\psi_p(x)]\Big\}
        [\bar{\psi}_{\nu}(x)\gamma^{\mu}(1 - \gamma^5)\psi_e(x)]
\end{eqnarray}
invariant under time reversal, where $\psi_p(x)$, $\psi_n(x)$,
$\psi_e(x)$ and $\psi_{\nu}(x)$ are the field operators of the proton,
neutron, electron (positron) and neutrino (antineutrino),
respectively, $\gamma^{\mu}$, $\gamma^5$ and $\sigma^{\mu\nu} =
\frac{i}{2}(\gamma^{\mu}\gamma^{\nu} - \gamma^{\nu}\gamma^{\mu})$ are
the Dirac matrices \cite{IZ80} and $\kappa = \kappa_p - \kappa_n =
3.7058$ is the isovector anomalous magnetic moment of the nucleon,
defined by the anomalous magnetic moments of the proton $\kappa_p =
1.7928$ and the neutron $\kappa_n = - 1.9130$ and measured in nuclear
magneton \cite{PDG12}. For numerical calculations we use $\lambda = -
1.2750(9)$ \cite{Abele1}, $G_F = 1.1664\times 10^{-11}\,{\rm
  MeV}^{-2}$ and $|V_{ud}| = 0.97427(15)$ \cite{PDG12}. The value of
the CKM matrix element $|V_{ud}| = 0.97427(15)$ agrees well with
$|V_{ud}| = 0.97425(22)$, measured from the superallowed $0^+ \to 0^+$
nuclear $\beta^-$--decays \cite{Vud}. The inverse $\beta$--decay
possesses a threshold. In the laboratory frame or in the rest frame of
the proton the antineutrino threshold energy is equal to
$(E_{\bar{\nu}})_{\rm thr} = ((m_n + m_e)^2 - m^2_p)/2 m_p =
1.806067(30)\,{\rm MeV}$, calculated for $m_n = 939.565379(21)\,{\rm
  MeV}$, $m_p = 938.272046(21)\,{\rm MeV}$ and $m_e = 0.510999\,{\rm
  MeV}$ \cite{PDG12}.

Following \cite{Ivanov2013} we define the amplitude of the inverse
$\beta$-decay $\bar{\nu}_e + p \to n + e^+$ as
\begin{eqnarray}\label{labelA.2} 
M(\bar{\nu}_e p \to n e^+) =
 2 m_p\frac{G_F}{\sqrt{2}}\,V_{ud}\,{\cal M}_{\beta},
\end{eqnarray}
where the amplitude ${\cal M}_{\beta}$ is calculated in the rest
  frame of the proton and the non--relativistic approximation for the
  neutron, taking into account the contributions of the weak magnetism
  and the neutron recoil. It reads \cite{Ivanov2013a}
\begin{eqnarray}\label{labelA.3}
\hspace{-0.3in}&&{\cal M}_{\beta} = \Big(1 +
\frac{\Delta}{2M}\Big)\Big\{
     [\varphi^{\dagger}_n\varphi_p][\bar{v}_{\bar{\nu}}\gamma^0(1 -
       \gamma^5)v] - \lambda\,
     [\varphi^{\dagger}_n\vec{\sigma}\,\varphi_p]\cdot
     [\bar{v}_{\bar{\nu}}\vec{\gamma}\,(1 - \gamma^5)v] +
     \frac{\lambda}{2 M}\,[\varphi^{\dagger}_n(\vec{\sigma}\cdot
       \vec{k}_n)\varphi_p]\nonumber\\
 \hspace{-0.3in}&&\times\,[\bar{v}_{\bar{\nu}}\gamma^0(1 - \gamma^5)v]
 - i\,\frac{\kappa + 1}{2 M}\,[\varphi^{\dagger}_n(\vec{\sigma}\times
   \vec{k}_n)\varphi_p]\cdot [\bar{v}_{\bar{\nu}}\vec{\gamma}\,(1 -
   \gamma^5)v] - \frac{\vec{k}_n}{2 M}\cdot
 [\varphi^{\dagger}_n\varphi_p][\bar{v}_{\bar{\nu}} \vec{\gamma}\,(1 -
   \gamma^5)v]\Big\},
\end{eqnarray}
where $\varphi_p$ and $\varphi_n$ are Pauli spinor functions of the
proton and neutron, $v$ and $v_{\bar{\nu}}$ are Dirac bispinor
functions of the positron and antineutrino, respectively, $M = (m_n +
m_p)/2$ is the average nucleon mass \cite{Ivanov2013} and $\Delta =
m_n - m_p$. The factor $(1 + \Delta/2M)$ comes from the normalisation
factor $\sqrt{2m_p(E_n + m_n)}$ of the Dirac bispinor wave functions
of the proton and neutron, divided by $2 m_p$. To order $1/M$ this
gives
\begin{eqnarray}\label{labelA.4} 
\frac{\sqrt{2m_p(E_n + m_n)}}{2m_p} = 1 + \frac{\Delta}{2M}.
\end{eqnarray}
Using the Dirac equations for free antineutrino and positron
$\bar{v}_{\bar{\nu}}(\vec{k}_{\bar{\nu}}\cdot \vec{\gamma}\,) =
E_{\bar{\nu}}\bar{v}_{\bar{\nu}}\gamma^0$ and $(\vec{k}\cdot
\vec{\gamma}\,)v = (E \gamma^0 + m_e)v$, respectively, we transcribe
the last term in Eq.(\ref{labelA.3}) into the form
\begin{eqnarray}\label{labelA.5}
- \frac{\vec{k}_n}{2M}[\varphi^{\dagger}_n\varphi_p]\cdot
[\bar{v}_{\bar{\nu}}\vec{\gamma}\,(1 - \gamma^5)v] = - \frac{\Delta}{2
  M}[\varphi^{\dagger}_n\varphi_p] [\bar{v}_{\bar{\nu}}\gamma^0\,(1 -
  \gamma^5)v] +
\frac{m_e}{2M}\,[\varphi^{\dagger}_n\varphi_p][\bar{v}_{\bar{\nu}}(1 +
  \gamma^5)v].
\end{eqnarray}
Substituting Eq.(\ref{labelA.5}) into Eq.(\ref{labelA.3}) we obtain
\begin{eqnarray}\label{labelA.6}
\hspace{-0.3in}&&{\cal M}_{\beta} =
       [\varphi^{\dagger}_n\varphi_p][\bar{v}_{\bar{\nu}}\gamma^0(1 -
         \gamma^5)v] - \tilde{\lambda}\,
       [\varphi^{\dagger}_n\vec{\sigma}\,\varphi_p]\cdot
       [\bar{v}_{\bar{\nu}}\vec{\gamma}\,(1 - \gamma^5)v] +
       \frac{\tilde{\lambda}}{2 M}\,[\varphi^{\dagger}_n(\vec{\sigma}\cdot
         \vec{k}_n)\varphi_p]\nonumber\\
 \hspace{-0.3in}&&\times\,[\bar{v}_{\bar{\nu}}\gamma^0(1 -
   \gamma^5)v] - i\,\frac{\kappa + 1}{2
   M}\,[\varphi^{\dagger}_n(\vec{\sigma}\times
   \vec{k}_n)\varphi_p]\cdot [\bar{v}_{\bar{\nu}}\vec{\gamma}\,(1 -
   \gamma^5)v] + \frac{m_e}{2
   M}\,[\varphi^{\dagger}_n\varphi_p][\bar{v}_{\bar{\nu}} (1 +
   \gamma^5)v],
\end{eqnarray}
where we have denoted $\tilde{\lambda} = \lambda\,(1 +
\Delta/2M)$. The hermitian conjugate amplitude takes the form
\begin{eqnarray}\label{labelA.7}
\hspace{-0.3in}&&{\cal M}^{\dagger}_{\beta} =
       [\varphi^{\dagger}_p\varphi_n] [\bar{v}\gamma^0(1 -
         \gamma^5)v_{\bar{\nu}}] - \tilde{\lambda}\,
       [\varphi^{\dagger}_p\vec{\sigma}\,\varphi_n]\cdot
       [\bar{v}\vec{\gamma}\,(1 - \gamma^5)v_{\bar{\nu}}] +
       \frac{\tilde{\lambda}}{2
         M}\,[\varphi^{\dagger}_p(\vec{\sigma}\cdot
         \vec{k}_n)\varphi_n]\nonumber\\
 \hspace{-0.3in}&&\times\,[\bar{v}\gamma^0(1 - \gamma^5)
   v_{\bar{\nu}}] + i\,\frac{\kappa + 1}{2
   M}\,[\varphi^{\dagger}_p(\vec{\sigma}\times
   \vec{k}_n)\varphi_n]\cdot [\bar{v}\vec{\gamma}\,(1 -
   \gamma^5) v_{\bar{\nu}}] + \frac{m_e}{2
   M}[\varphi^{\dagger}_p\varphi_n][\bar{v} (1 - \gamma^5)
   v_{\bar{\nu}}].
\end{eqnarray}
In the rest frame of the proton the cross section for the inverse
$\beta$--decay is defined by
\begin{eqnarray}\label{labelA.8}
\hspace{-0.3in} \sigma(E_{\bar{\nu}}) = G^2_F|V_{ud}|^2\,\frac{m_p}{2
  E_{\bar{\nu}}}\int \frac{1}{2}\sum_{\rm pol}|{\cal M}_{\beta}|^2
(2\pi)^4\,\delta^{(4)}(k + k_n - k_p - k_{\bar{\nu}})\,\frac{d^3
  k}{(2\pi)^3 2 E}\frac{d^3 k_n}{(2\pi)^3 2 E_n},
\end{eqnarray}
where we have summed over the polarisations of the neutron and
positron and averaged over the polarisations of the proton. The
4--momenta of the neutron, positron, proton and antineutrino are
defined by $k_n = (E_n, \vec{k}_n)$, $k = (E, \vec{k})$, $k_p=
(m_p,\vec{0}\,)$ and $k_{\bar{\nu}} = (E_{\bar{\nu}},
\vec{k}_{\bar{\nu}})$, respectively. Having integrated over the
3--momentum of the neutron and the positron energy we arrive at the
expression
\begin{eqnarray}\label{labelA.9}
\hspace{-0.3in} \sigma(E_{\bar{\nu}}) =
G^2_F|V_{ud}|^2\int^{(\cos\theta_{e\bar{\nu}})_{\rm max}}_{(\cos\theta_{e\bar{\nu}})_{\rm min}}
\frac{1}{2}\sum_{\rm pol}|M(\bar{\nu}_e p \to n e^+)|^2
\Phi(E_{\bar{\nu}}, k, \cos\theta_{e\bar{\nu}})\,
d\cos\theta_{e\bar{\nu}},
\end{eqnarray}
where $\cos\theta_{e\bar{\nu}}= \vec{k}_{\bar{\nu}}\cdot
\vec{k}/E_{\bar{\nu}}k$ and $\Phi(E_{\bar{\nu}}, k,
\cos\theta_{e\bar{\nu}})$ is the phase--space factor equal to
\begin{eqnarray}\label{labelA.10}
\hspace{-0.3in}&&\Phi(E_{\bar{\nu}}, k, \cos\theta_{e\bar{\nu}}) =
\frac{1}{16\pi}\,\frac{m_p}{E_{\bar{\nu}}}\int \delta(E + \sqrt{m^2_n
  + (\vec{k}_{\bar{\nu}} - \vec{k})^2} - m_p -
E_{\bar{\nu}})\frac{k}{E_n}\,dE =
\frac{1}{16\pi}\,\frac{m_p}{E_{\bar{\nu}}}\,\frac{k}{\displaystyle E_n
  + E - E_{\bar{\nu}}\frac{E}{k}\,\cos\theta_{e\bar{\nu}}} =\nonumber\\
\hspace{-0.3in}&&=
\frac{1}{16\pi}\,\frac{m_p}{E_{\bar{\nu}}}\,\frac{k}{\displaystyle m_p
  + E_{\bar{\nu}} - E_{\bar{\nu}}\frac{E}{k}\cos\theta_{e\bar{\nu}}} =
\frac{1}{16\pi}\,\frac{k}{E_{\bar{\nu}}}\frac{1}{\displaystyle 1 +
  \frac{E_{\bar{\nu}}}{m_p}\Big(1 -
  \frac{1}{\beta}\cos\theta_{e\bar{\nu}}\Big)} =
\frac{1}{16\pi}\,\frac{k}{E_{\bar{\nu}}}\frac{1}{\displaystyle 1 +
  \frac{E_{\bar{\nu}}}{m_p}\Big(1 -
  \frac{1}{\beta}\cos\theta_{e\bar{\nu}}\Big)},
\end{eqnarray}
where $\beta = k/E$ is the positron velocity. For the derivation of
the phase--factor $\Phi(E_{\bar{\nu}}, k, \cos\theta_{e\bar{\nu}})$ we
have used energy $E_n + E = m_p + E_{\bar{\nu}}$ and 3--momentum
$\vec{k}_{\bar{\nu}} = \vec{k}_n + \vec{k}$ conservation. The positron energy is given by
\begin{eqnarray}\label{labelA.11}
\hspace{-0.3in}E = \frac{\displaystyle E_{\bar{\nu}}- \frac{m^2_n -
    m^2_p - m^2_e}{2 m_p}}{\displaystyle 1
  + \frac{E_{\bar{\nu}}}{m_p}\,(1 - \beta \cos\theta_{e\bar{\nu}})}.
\end{eqnarray}
A behaviour of the phase--space factor $\Phi(E_{\bar{\nu}}, k,
\cos\theta_{e\bar{\nu}})$ as a function of the antineutrino energy
$E_{\bar{\nu}}$ may affect the upper limit
$(\cos\theta_{e\bar{\nu}})_{\rm max}$ and may lead to a deviation from
$(\cos\theta_{e\bar{\nu}})_{\rm max} = + 1$. Indeed, the phase--factor
becomes negative for
\begin{eqnarray}\label{labelA.12}
\hspace{-0.3in}\cos\theta_{e\bar{\nu}} > \beta \Big(1 +
\frac{m_p}{E_{\bar{\nu}}}\Big).
\end{eqnarray}
Thus, if $\beta \ge 1/(1 + m_p/E_{\bar{\nu}})$ the integration over
$\cos\theta_{e\bar{\nu}}$ may be carried out in the limits $- 1 \le
\cos\theta_{e\bar{\nu}} \le + 1$. In turn, for $\beta < 1/(1 +
m_p/E_{\bar{\nu}})$ the region of the integration over
$\cos\theta_{e\bar{\nu}}$ is restricted from above as $- 1 \le
\cos\theta_{e\bar{\nu}} \le (\cos\theta_{e\bar{\nu}})_{\rm
  max}$. However, one may show that the antineutrino energy region,
obeying the constraint $\beta < 1/(1 + m_p/E_{\bar{\nu}})$, is located
very close to threshold and does not affect on the value of the cross
section. Below we integrate over $\cos\theta_{e\bar{\nu}}$ in the
limits $- 1 \le \cos\theta_{e\bar{\nu}} \le + 1$. As a result, the
cross section Eq.(\ref{labelA.9}) is defined by
\begin{eqnarray}\label{labelA.13}
\hspace{-0.3in} \sigma(E_{\bar{\nu}}) = G^2_F|V_{ud}|^2\int^{+1}_{-1}
\frac{1}{2}\sum_{\rm pol}|M(\bar{\nu}_e p \to n e^+)|^2
\Phi(E_{\bar{\nu}}, k, \cos\theta_{e\bar{\nu}})\,
d\cos\theta_{e\bar{\nu}}.
\end{eqnarray}
Now we may proceed to the analysis of the cross section for the
inverse $\beta$--decay to next--to--leading order in the large $M$
expansion keeping the terms of order $1/M$.

To order $1/M$ of the large $M$ expansion the phase--space factor
$\Phi(E_{\bar{\nu}}, k, \cos\theta_{e\bar{\nu}})$ is
\begin{eqnarray}\label{labelA.14}
\hspace{-0.3in}\Phi(E_{\bar{\nu}}, k, \cos\theta_{e\bar{\nu}}) =
\frac{1}{16\pi}\,\frac{k}{E_{\bar{\nu}}}\,\Big[1 -
\frac{E_{\bar{\nu}}}{M}\Big(1 -
\frac{1}{\beta}\,\cos\theta_{e\bar{\nu}}\Big)\Big].
\end{eqnarray}
The phase--space factor, taken in the form of the expansion
Eq.(\ref{labelA.14}), is always positive for all antineutrino energies
$E_{\bar{\nu}} \ge (E_{\bar{\nu}})_{\rm thr}$.  The positron energy
$E$, calculated to next--to--leading order in the large $M$ expansion,
is equal to
\begin{eqnarray}\label{labelA.15}
\hspace{-0.3in}E = \bar{E}\Big[1 - \frac{
  E_{\bar{\nu}}}{M}\,\Big(1 -
 \bar{\beta}\,\cos\theta_{e\bar{\nu}} +
  \frac{\Delta^2 - m^2_e}{2\bar{E} E_{\bar{\nu}}}\Big)\Big],
\end{eqnarray}
where we have denoted $\bar{E} = E_{\bar{\nu}} - \Delta$, $\bar{\beta}
= \bar{k}/\bar{E}$ and $\bar{k} = \sqrt{\bar{E}^2 - m^2_e}$.  For
numerical calculations we use $\Delta = m_n - m_p = 1.293332\,{\rm
  MeV}$ \cite{PDG12}. For the absolute value of the momentum $k$ and
velocity $\beta$ of the positron we obtain the following expressions
\begin{eqnarray}\label{labelA.16}
\hspace{-0.3in}k = \bar{k}\Big[1 -
  \frac{E_{\bar{\nu}}}{M}\,\frac{1}{\bar{\beta}^2}\, \Big(1 -
  \bar{\beta}\,\cos\theta_{e\bar{\nu}} + \frac{\Delta^2 - m^2_e}{2
    \bar{E} E_{\bar{\nu}}}\Big)\Big].
\end{eqnarray}
and 
\begin{eqnarray}\label{labelA.17}
\hspace{-0.3in}\beta = \bar{\beta}\Big[1 -
  \frac{E_{\bar{\nu}}}{M}\,\frac{1 -
    \bar{\beta}^2}{\bar{\beta}^2}\,\Big(1 -
  \bar{\beta}\,\cos\theta_{e\bar{\nu}} + \frac{\Delta^2 - m^2_e}{2
    \bar{E} E_{\bar{\nu}}}\Big)\Big].
\end{eqnarray}
The squared absolute value of the amplitude ${\cal M}_{\beta}$, summed
over the polarisations of interacting particles, is equal to
\begin{eqnarray}\label{labelA.18}
\hspace{-0.3in}\sum_{\rm pol}|{\cal M}_{\beta}|^2 &=& 16\,\Big\{(1 +
3\tilde{\lambda}^2)\,E E_{\bar{\nu}} + (1 -
\tilde{\lambda}^2)\,\vec{k}\cdot \vec{k}_{\bar{\nu}} +
\frac{1}{M}\,\Big[- m^2_e E_{\bar{\nu}} - \lambda^2 (E
  E^2_{\bar{\nu}} - E_{\bar{\nu}} k^2) - \lambda^2 ( E_{\bar{\nu}} -
  E)\,\vec{k}\cdot \vec{k}_{\bar{\nu}}\nonumber\\
\hspace{-0.3in}&+& 2\,(\kappa + 1)\,\lambda\,(E E^2_{\bar{\nu}} +
E_{\bar{\nu}} k^2) - 2\,(\kappa + 1)\,\lambda\,( E_{\bar{\nu}} +
E)\,\vec{k}\cdot \vec{k}_{\bar{\nu}}\Big]\Big\}.
\end{eqnarray}
For the calculation of the cross section we transcribe the r.h.s. of
Eq.(\ref{labelA.15}) into the form
\begin{eqnarray}\label{labelA.19}
\hspace{-0.3in}\sum_{\rm pol}|{\cal M}_{\beta}|^2 &=& 16\,(1 +
3 \lambda^2)\,E E_{\bar{\nu}}\,\Big(\bar{A}(E_{\bar{\nu}}) +
\bar{B}(E_{\bar{\nu}})\,\frac{\vec{k}\cdot \vec{k}_{\bar{\nu}}}{E E_{\bar{\nu}}}\Big) = \nonumber\\&=& 16\,(1 +
3 \lambda^2)\,E E_{\bar{\nu}}\,\Big(\bar{A}(E_{\bar{\nu}}) +
\bar{B}(E_{\bar{\nu}})\,\beta\,\cos\theta_{e\bar{\nu}}\Big),
\end{eqnarray}
where we have denoted
\begin{eqnarray}\label{labelA.20}
\hspace{-0.3in}\bar{A}(E_{\bar{\nu}}) &=& 1 + \frac{1}{M}\,\frac{1}{1 + 3
  \lambda^2}\,\Big(3\lambda^2 \Delta - (\lambda^2 - 2(\kappa +
  1)\,\lambda)\,E_{\bar{\nu}} + (\lambda^2 + 2(\kappa +
  1)\,\lambda)\,E - (\lambda^2 + 2 (\kappa + 1)\,\lambda +
  1)\,\frac{m^2_e}{E}\Big),\nonumber\\
\hspace{-0.3in}\bar{B}(E_{\bar{\nu}}) &=& \frac{1 - \lambda^2}{1 + 3
  \lambda^2} +
\frac{1}{M}\,\frac{1}{1 + 3 \lambda^2}\,\Big( - \lambda^2 \Delta -
(\lambda^2 + 2(\kappa + 1)\,\lambda)\,E_{\bar{\nu}} + (\lambda^2 -
2(\kappa + 1)\,\lambda)\,E\Big) = \nonumber\\
\hspace{-0.3in}&=& a_0 + \frac{1}{M}\,\frac{1}{1 + 3 \lambda^2}\,\Big(
- \lambda^2 \Delta - (\lambda^2 + 2(\kappa +
1)\,\lambda)\,E_{\bar{\nu}} + (\lambda^2 - 2(\kappa +
1)\,\lambda)\,E\Big).
\end{eqnarray}
In terms of the antineutrino energy $E_{\bar{\nu}}$ the correlation
coefficients $\bar{A}(E_{\bar{\nu}})$ and $\bar{B}(E_{\bar{\nu}})$
read
\begin{eqnarray}\label{labelA.21}
\hspace{-0.3in}\bar{A}(E_{\bar{\nu}}) &=& 1 + \frac{1}{M}\,\frac{1}{1
  + 3 \lambda^2}\,\Big(2 \,(\lambda^2 - (\kappa + 1)\,\lambda)\,\Delta
+ 4 (\kappa + 1)\,\lambda\,E_{\bar{\nu}} - (\lambda^2 + 2 (\kappa +
1)\,\lambda + 1)\,\frac{m^2_e}{\bar{E}}\Big),\nonumber\\
\hspace{-0.3in}\bar{B}(E_{\bar{\nu}}) &=& a_0 + \frac{1}{M}\,\frac{1}{1 + 3
  \lambda^2}\,\Big(- 2\,(\lambda^2 - 2(\kappa + 1)\,\lambda)\,\Delta - 4
(\kappa + 1)\,\lambda\,E_{\bar{\nu}}\Big),
\end{eqnarray}
where $\bar{E} = E_{\bar{\nu}} - \Delta$. The angular distribution for
the inverse $\beta$--decay takes the form
\begin{eqnarray}\label{labelA.22}
\hspace{-0.3in}\frac{d\sigma(E_{\bar{\nu}},\cos\theta_{e\bar{\nu}})}{
  d\cos\theta_{e\bar{\nu}}} = (1 + 3
\lambda^2)\,\frac{G^2_F|V_{ud}|^2}{2\pi}\,\Big(\bar{A}(E_{\bar{\nu}})
+ \bar{B}(E_{\bar{\nu}})\,\beta\,\cos\theta_{e\bar{\nu}}\Big)\,k
E\,\Big[1 - \frac{E_{\bar{\nu}}}{M}\Big(1 -
\frac{1}{\beta}\,\cos\theta_{e\bar{\nu}}\Big)\Big].
\end{eqnarray}
Using the expansions Eq.(\ref{labelA.15}) - Eq.(\ref{labelA.17}) we
may transcribe the angular distribution Eq.(\ref{labelA.24}) into the
form
\begin{eqnarray}\label{labelA.23}
\hspace{-0.3in}\frac{d\sigma(E_{\bar{\nu}},\cos\theta_{e\bar{\nu}})}{
  d\cos\theta_{e\bar{\nu}}} = (1 + 3
\lambda^2)\,\frac{G^2_F|V_{ud}|^2}{2\pi}\,\Big(A(E_{\bar{\nu}}) +
B(E_{\bar{\nu}})\,\bar{\beta}\,\cos\theta_{e\bar{\nu}} +
C(E_{\bar{\nu}})\,\bar{\beta}^2\,\cos^2\theta_{e\bar{\nu}}\Big)\,\bar{k}
\bar{E},
\end{eqnarray}
where we have denoted
\begin{eqnarray}\label{labelA.24}
\hspace{-0.3in}&&A(E_{\bar{\nu}}) = \bar{A}(E_{\bar{\nu}}) - \frac{1 + 2
  \bar{\beta}^2}{\bar{\beta}^2}\,\frac{E_{\bar{\nu}}}{M}\,\Big(1 +
\frac{1 + \bar{\beta}^2}{1 + 2\bar{\beta}^2}\,\frac{\Delta^2 -
  m^2_e}{2 \bar{E} E_{\bar{\nu}}}\Big),\nonumber\\
\hspace{-0.3in}&&B(E_{\bar{\nu}}) = \bar{B}(E_{\bar{\nu}}) + \frac{2 +
  \bar{\beta}^2}{\bar{\beta}^2}\,\frac{E_{\bar{\nu}}}{M}\,\Big[1 -
  a_0\,\Big(1 + \frac{2}{2 + \bar{\beta}^2}\,\frac{\Delta^2 - m^2_e}{2
    \bar{E} E_{\bar{\nu}}}\Big)\Big],\nonumber\\
\hspace{-0.3in}&&C(E_{\bar{\nu}}) =
a_0\,\frac{3}{\bar{\beta}^2}\,\frac{E_{\bar{\nu}}}{M}.
\end{eqnarray}
For the derivation of these
expansions we have used the expansions of the following products
\begin{eqnarray}\label{labelA.25}
\hspace{-0.3in}&&k E =  \bar{k}\bar{E}\Big[1 - \frac{1
    + \bar{\beta}^2}{\bar{\beta}^2}\,\frac{E_{\bar{\nu}}}{M}\Big(1 +
  \frac{\Delta^2 - m^2_e}{2 \bar{E} E_{\bar{\nu}}}\Big) + \frac{1 +
    \bar{\beta}^2}{\bar{\beta}}\,\frac{E_{\bar{\nu}}}{M}\,
  \cos\theta_{e\bar{\nu}}\Big],\nonumber\\
\hspace{-0.3in}&&k E \beta =\bar{k}\bar{E}\bar{\beta}\Big[1 - \frac{2
  }{\bar{\beta}^2}\,\frac{E_{\bar{\nu}}}{M}\Big(1 + \frac{\Delta^2 -
    m^2_e}{2 \bar{E} E_{\bar{\nu}}}\Big) +
  \frac{2}{\bar{\beta}}\,\frac{E_{\bar{\nu}}}{M}\,
  \cos\theta_{e\bar{\nu}}\Big],\nonumber\\
\hspace{-0.3in}&&k E \Big[1 - \frac{E_{\bar{\nu}}}{M}\Big(1 -
  \frac{1}{\beta}\,\cos\theta_{e\bar{\nu}}\Big)\Big] =
\bar{k}\bar{E}\Big[1 - \frac{1 +
    2\bar{\beta}^2}{\bar{\beta}^2}\,\frac{E_{\bar{\nu}}}{M}\Big(1 +
  \frac{1 + \bar{\beta}^2}{1 + 2 \bar{\beta}^2}\,\frac{\Delta^2 -
    m^2_e}{2 \bar{E} E_{\bar{\nu}}}\Big) + \frac{2 +
    \bar{\beta}^2}{\bar{\beta}}\,\frac{E_{\bar{\nu}}}{M}\,
  \cos\theta_{e\bar{\nu}}\Big],\nonumber\\
\hspace{-0.3in}&&k E \beta \Big[1 - \frac{E_{\bar{\nu}}}{M}\Big(1 -
  \frac{1}{\beta}\,\cos\theta_{e\bar{\nu}}\Big)\Big]
=\bar{k}\bar{E}\bar{\beta}\Big[1 - \frac{2 + \bar{\beta}^2
  }{\bar{\beta}^2}\,\frac{E_{\bar{\nu}}}{M}\Big(1 + \frac{2}{2 +
    \bar{\beta}^2}\frac{\Delta^2 - m^2_e}{2 \bar{E}
    E_{\bar{\nu}}}\Big) +
  \frac{3}{\bar{\beta}}\,\frac{E_{\bar{\nu}}}{M}\,
  \cos\theta_{e\bar{\nu}}\Big].
\end{eqnarray}
The correlation coefficients $A(E_{\bar{\nu}})$, $B(E_{\bar{\nu}})$
and $C(E_{\bar{\nu}})$ are calculated in agreement with the results,
obtained by Vogel and Beacom \cite{IBD3}. Now let us take into account
the contributions of the radiative corrections, which we calculate to
leading order in the large $M$ expansion.

\subsection*{2. Radiative corrections, caused by one--virtual photon exchanges}

For this aim of the calculation of the contributions of one--virtual
photon exchanges we use the results, obtained in \cite{Ivanov2013} for
the neutron $\beta^-$--decay. The amplitude of the inverse
$\beta$--decay with one--virtual photon exchanges we write as follows
\begin{eqnarray}\label{labelA.26}
M^{(\gamma)}(\bar{\nu}_e p \to n e^+) =
2m_p\,\frac{G_F}{\sqrt{2}}V_{ud}\Big( {\cal M}^{(\gamma)}_{pp} + {\cal
  M}^{(\gamma)}_{e^+e^+} + {\cal M}^{(\gamma)}_{pe^+}\Big),
\end{eqnarray}
where the amplitudes ${\cal M}^{(\gamma)}_{pp}$ and ${\cal
  M}^{(\gamma)}_{e^+e^+}$ are related to the contributions of the
proton and positron self--energy corrections, and the amplitude ${\cal
  M}^{(\gamma)}_{pe^+}$ is induced by the proton--positron vertex
correction. They are defined by
\begin{eqnarray}\label{labelA.27}
\hspace{-0.3in}&& 2 m_p {\cal M}^{(\gamma)}_{pp} =  e^2
\int\frac{d^4q}{(2\pi)^4i}\,D_{\alpha\beta}(q)\,
\Big[\bar{u}_n\gamma^{\mu}(1 + \lambda \gamma^5)\,
  \frac{1}{m_p - \hat{k}_p - i0}\,\gamma^{\alpha}\,\frac{1}{m_p -
    \hat{k}_p - \hat{q} - i0}\,\gamma^{\beta}\, u_p\Big]
    [\bar{v}_{\bar{\nu}}\gamma_{\mu}(1 - \gamma^5)v]\nonumber\\
\hspace{-0.3in}&& + \Big[\bar{u}_n\gamma^{\mu}(1 + \lambda
  \gamma^5)\,\frac{1}{m_p - \hat{k}_p - i0}\,\Big( - \delta m_p +
  \frac{Z^{(p)}_2 - 1}{2}(m_p - \hat{k}_p)\Big)\,u_p\Big]
       [\bar{v}_{\bar{\nu}}\gamma_{\mu} (1 - \gamma^5)
         v],\nonumber\\
\hspace{-0.3in}&&2 m_p {\cal M}^{(\gamma)}_{e^+e^+} =  e^2\, [\bar{u}_n
  \gamma^{\mu}(1 + \lambda
  \gamma^5)u_p]\int\frac{d^4q}{(2\pi)^4i}\,D_{\alpha\beta}(q)\,
\Big[\bar{v}_{\bar{\nu}}\gamma_{\mu}(1 - \gamma^5) \frac{1}{m_e +
    \hat{k} - i0} \gamma^{\alpha}\frac{1}{m_e + \hat{k} + \hat{q}
    -i0}\gamma^ {\beta} v\Big]\nonumber\\
\hspace{-0.3in}&& + [\bar{u}_n \gamma^{\mu}(1 + \lambda
  \gamma^5) u_p]\Big[\bar{v}_{\bar{\nu}}\gamma_{\mu}(1 -
  \gamma^5)\,\frac{1}{m_e + \hat{k} - i0}\Big( - \delta m_e +
  \frac{Z^{(e)}_2 - 1}{2}(m_e + \hat{k})\Big)
  v\Big],\nonumber\\
\hspace{-0.3in}&&2 m_p {\cal M}^{(\gamma)}_{pe^+} = -\,
e^2\int\frac{d^4q}{(2\pi)^4i}\,D_{\alpha\beta}(q)\,\Big[\bar{u}_n
  \,\gamma^{\mu}(1 + \lambda \gamma^5)\,\frac{1}{m_p - \hat{k}_p -
    \hat{q} - i0} \gamma^{\alpha}
  u_p\Big]\Big[\bar{v}_{\bar{\nu}}\gamma_{\mu}(1 -
  \gamma^5)\frac{1}{m_e + \hat{k} + \hat{q} -i0}\gamma^{\beta}
  v\Big],\nonumber\\
\hspace{-0.3in}&&
\end{eqnarray}
where $(\delta m_p, Z^{(p)}_2)$ and $(\delta m_e, Z^{(e)}_2)$ are
renormalisation constants of masses and wave functions of the proton
and positron, respectively, and $D_{\alpha\beta}(q)$ is the photon
propagator
\begin{eqnarray}\label{labelA.28}
D_{\alpha\beta}(q) = \frac{1}{q^2 + i0}\,\Big(g_{\alpha\beta} -
\xi\,\frac{q_{\alpha} q_{\beta}}{q^2}\Big),
\end{eqnarray}
and $\xi$ is a gauge parameter. Following \cite{Ivanov2013} we replace
the amplitudes ${\cal M}^{(\gamma)}_{pp}$, ${\cal
  M}^{(\gamma)}_{ee}$ and ${\cal M}^{(\gamma)}_{pe}$ by the expressions
\begin{eqnarray}\label{labelA.29}
\hspace{-0.3in}&& 2 m_p {\cal M}^{(\gamma)}_{pp} \to 2 m_p {\cal
  M}^{(\rm SC)}_{pp} = 
- \frac{\alpha}{8\pi}[\bar{u}_nW^{\mu}u_p][\bar{v}_{\bar{\nu}}O_{\mu}v]\int
\frac{d^4q}{\pi^2 i}\, D_{\alpha\beta}(q)\,\frac{(2k_p +
  q)^{\alpha}(2k_p + q)^{\beta}}{(q^2 + 2 k_p\cdot q + i0)^2},
\nonumber\\
\hspace{-0.3in}&&2 m_p {\cal M}^{(\gamma)}_{e^+e^+} \to 2 m_p {\cal
  M}^{(\rm SC)}_{e^+e^+} = \frac{\alpha}{8\pi m_e}\,[\bar{u}_n W^{\mu}
  u_p]\int\frac{d^4q}{\pi^2 i}\,D_{\alpha\beta}(q)\,
\frac{[\bar{v}_{\bar{\nu}}O_{\mu}(2 k^{\alpha} +
    \gamma^{\alpha}\hat{q})\hat{k}(2k^{\beta} +
    \hat{q}\gamma^{\beta}) v]}{(q^2 + 2 k \cdot q + i
  0)^2},\nonumber\\
\hspace{-0.3in}&&2 m_p {\cal M}^{(\gamma)}_{pe^+} \to 2 m_p {\cal
  M}^{(\rm SC)}_{pe^+} = \frac{\alpha}{4\pi}\int\frac{d^4q}{\pi^2
  i}\,D_{\alpha\beta}(q)\, \frac{[\bar{u}_nW^{\mu}(2 k^{\alpha}_p +
    q^{\alpha})u_p]}{q^2 + 2 k_p\cdot q + i
  0}\,\frac{[\bar{v}_{\bar{\nu}}O_{\mu}(2 k^{\beta} +
    \hat{q}\gamma^{\beta}) v]}{q^2 + 2 k\cdot q + i 0},
\end{eqnarray}
where the abbreviation (SC) means ``Sirlin's Correction'' with
$W^{\mu} = \gamma^{\mu}(1 + \lambda \gamma^5)$ and $O_{\mu} =
\gamma_{\mu}(1 - \gamma^5)$ \cite{Ivanov2013}. The sum of the
amplitudes
\begin{eqnarray}\label{labelA.30}
{\cal M}^{(\rm SC)}_{\rm RC} = {\cal M}^{(\rm SC)}_{pp} + {\cal
  M}^{(\rm SC)}_{e^+e^+} + {\cal M}^{(\rm SC)}_{pe^+}
\end{eqnarray}
is invariant under a gauge transformation $D_{\alpha\beta}(q) \to
D_{\alpha\beta}(q) + c(q^2) q_{\alpha} q_{\beta}$, where $c(q^2)$ is
an arbitrary function. As has been shown in \cite{Ivanov2013} the
deviations of the amplitudes ${\cal M}^{(\gamma)}_{pp}$, ${\cal
  M}^{(\gamma)}_{e^+e^+}$ and ${\cal M}^{(\gamma)}_{pe^+}$ from the
amplitudes ${\cal M}^{(\rm SC)}_{pp}$, ${\cal M}^{(\rm RC)}_{e^+e^+}$
and ${\cal M}^{(\rm SC)}_{pe^+}$ do not depend on the positron energy
and can be absorbed by the renormalisation constants of the Fermi
coupling constant $G_F$ and the axial coupling constant $\lambda$
\cite{Ivanov2013}. Following \cite{Ivanov2013} and using for the
regularisation of the ultra--violet and infrared divergent
contributions the Pauli-Villars regularisation and the finite--photon
mass (FPM) regularisation, respectively, we obtain
\begin{eqnarray*}
\hspace{-0.3in}&& 2 m_p {\cal M}^{(\rm SC)}_{pp} =
       [\bar{u}_nW^{\mu}u_p][\bar{v}_{\bar{\nu}}O_{\mu}v]\,\frac{\alpha}{2\pi}\,
       \Big[- \frac{1}{2}\,{\ell n}\Big(\frac{\Lambda}{m_p}\Big) -
         {\ell n}\Big(\frac{\mu}{m_p}\Big) -
         \frac{3}{4}\Big]. \nonumber\\
\hspace{-0.3in}&&2 m_p {\cal
  M}^{(\rm SC)}_{e^+e^+} =  [\bar{u}_nW^{\mu}u_p][\bar{v}_{\bar{\nu}}O_{\mu}v]\,
\frac{\alpha}{2\pi}\,\Big[- \frac{1}{2}\,{\ell
           n}\Big(\frac{\Lambda}{m_e}\Big) - {\ell
           n}\Big(\frac{\mu}{m_e}\Big) - \frac{9}{8}\Big],
\end{eqnarray*}
\begin{eqnarray}\label{labelA.31}
\hspace{-0.3in}&&2 m_p {\cal M}^{(\rm SC)}_{pe^+} = [\bar{u}_n W^{\mu}
  u_p] \frac{\alpha}{2\pi}\,\Big\{[\bar{v}_{\bar{\nu}} O_{\mu}
  v]\,\Big[{\ell n}\Big(\frac{\Lambda}{m_e}\Big) + \frac{1}{2}+ {\ell
    n}\Big(\frac{\mu}{m_e}\Big)\,\frac{1}{~\beta}\,{\ell n}\Big(\frac{1
    + \beta}{1 - \beta}\Big) - \frac{1}{~4\beta}\,{\ell
    n}^2\Big(\frac{1 + \beta}{1 - \beta}\Big)\nonumber\\
\hspace{-0.3in}&& + \frac{1}{~\beta}\,L\Big(\frac{2\beta}{1 +
  \beta}\Big) + \frac{1}{~2\beta}\,{\ell n}\Big(\frac{1 + \beta}{1 -
  \beta}\Big)\Big] + [\bar{v}_{\bar{\nu}} O_{\mu} \gamma^0
  v]\,\frac{\sqrt{1 - \beta^2}}{2\beta}\,{\ell n}\Big(\frac{1 +
  \beta}{1 - \beta}\Big)\Big\},
\end{eqnarray}
where $\Lambda$ and $\mu$ are the ultra--violet cut--off and the
finite--photon mass, respectively, and $L(z)$ is the Spence function,
defined by \cite{HMF72}--\cite{PolyLog3} 
\begin{eqnarray}\label{labelA.40}
L(z) = \int^z_0\frac{dt}{t}\,{\ell n}|1 - t|.
\end{eqnarray}
Due to gauge invariance of the sum of the
amplitudes ${\cal M}^{(\rm SC)}_{pp}$, ${\cal M}^{(\rm SC)}_{e^+e^+}$ and
${\cal M}^{(\rm SC)}_{p^+e}$ we have used the Feynman gauge $\xi = 0$
\cite{Ivanov2013}.

Summing up the amplitudes ${\cal M}^{(\rm SC)}_{pp}$, ${\cal M}^{(\rm
  SC)}_{e^+e^+}$ and ${\cal M}^{(\rm SC)}_{pe^+}$ we obtain the
radiative corrections to the amplitude of the inverse $\beta$--decay,
caused by one--virtual photon exchanges. We get
\begin{eqnarray}\label{labelA.32}
\hspace{-0.3in}&&2m_p {\cal M}^{(\rm SC)}_{\rm RC} =
\frac{\alpha}{2\pi}\,[\bar{u}_nW^{\mu}u_p]\Big\{[\bar{v}_{\bar{\nu}}O_{\mu}v]
\Big[\frac{3}{2}\,{\ell n}\Big(\frac{m_p}{m_e}\Big) - \frac{11}{8} +
  {\ell n}\Big(\frac{\mu}{m_e}\Big)\,\Big[\frac{1}{\beta}\,{\ell
      n}\Big(\frac{1 + \beta}{1 - \beta}\Big) - 2\Big]\nonumber\\
\hspace{-0.3in}&& +
\frac{1}{~\beta}\,L\Big(\frac{2\beta}{1 +
  \beta}\Big) - \frac{1}{~4\beta}\,{\ell n}^2\Big(\frac{1
  + \beta}{1 - \beta}\Big) +
\frac{1}{~2\beta}\,{\ell n}\Big(\frac{1 + \beta}{1 -
  \beta}\Big)\Big] + [\bar{v}_{\bar{\nu}} O_{\mu} \gamma^0
  v]\,\frac{\sqrt{1 - \beta^2}}{2\beta}\,{\ell
  n}\Big(\frac{1 + \beta}{1 - \beta}\Big)\Big\}.
\end{eqnarray}
Since the amplitude $2 m_p {\cal M}^{(\rm SC)}_{p e^+}$ defines the
energy dependence of the radiative corrections we give the calculation
of the amplitude $2 m_p {\cal M}^{(\rm SC)}_{p e^+}$ in more detail.
Using the Feynman gauge $\xi = 0$ we transcribe $2 m_p {\cal M}^{(\rm
  SC)}_{p e^+}$ into the form
\begin{eqnarray}\label{labelA.33}
\hspace{-0.3in}&& 2 m_p {\cal M}^{(\rm SC)}_{pe^+} =
\frac{\alpha}{4\pi}\,[\bar{u}_nW^{\mu} u_p]\Big\{
     [\bar{v}_{\bar{\nu}}O_{\mu} v]\Big[\int\frac{d^4q}{\pi^2
         i}\,\frac{1}{q^2 + i 0}\,\frac{1}{q^2 + 2 k_p \cdot q + i 0}
       + \int\frac{d^4q}{\pi^2 i}\,\frac{1}{q^2 + i 0}\,\frac{1}{q^2 +
         2 k \cdot q + i 0}\nonumber\\
\hspace{-0.3in}&& - \int\frac{d^4q}{\pi^2 i}\,\frac{1}{q^2 + 2 k_p
  \cdot q + i 0}\,\frac{1}{q^2 + 2 k \cdot q + i 0} +
4(k\cdot k_p)\int\frac{d^4q}{\pi^2 i}\,\frac{1}{q^2 + i
  0}\,\frac{1}{q^2 + 2 k_p \cdot q + i 0}\,\frac{1}{q^2 + 2 k
  \cdot q + i 0}\Big]\nonumber\\
\hspace{-0.3in}&& - 2i\int\frac{d^4q}{\pi^2 i}\,\frac{1}{q^2 + i
  0}\,\frac{1}{q^2 + 2 k_p \cdot q + i 0}\,\frac{1}{q^2 + 2 k \cdot
  q + i 0}[\bar{v}_{\bar{\nu}}O_{\mu}
  \sigma_{\alpha\beta}q^{\alpha}k^{\beta}_p v]\Big\}.
\end{eqnarray}
The contribution of the first three integrals is equal to (see
Appendix D and Eq.({\rm D-12}) of Ref. \cite{Ivanov2013})
\begin{eqnarray}\label{labelA.34}
\hspace{-0.3in}&& \int\frac{d^4q}{\pi^2
         i}\,\frac{1}{q^2 + i 0}\,\frac{1}{q^2 + 2 k_p \cdot q + i 0}
       + \int\frac{d^4q}{\pi^2 i}\,\frac{1}{q^2 + i 0}\,\frac{1}{q^2 +
         2 k \cdot q + i 0}\nonumber\\
\hspace{-0.3in}&& - \int\frac{d^4q}{\pi^2 i}\,\frac{1}{q^2 + 2 k_p
  \cdot q + i 0}\,\frac{1}{q^2 + 2 k \cdot q + i 0} = 2\,{\ell
  n}\Big(\frac{\Lambda}{m_e}\Big) + 1,
\end{eqnarray}
where $\Lambda$ is an ultra--violet cut--off. After the merging of the
denominators (see Eq.(C-3) - Eq.(C-5) of Ref.\cite{Ivanov2013}), the
shift of the virtual momentum and the Wick rotation we reduce the
fourth and fifth integrals in Eq.(\ref{labelA.33}) to the form
\cite{Ivanov2013}
\begin{eqnarray}\label{labelA.35}
\hspace{-0.3in}4(k \cdot k_p)\int\frac{d^4q}{\pi^2
  i}\,\frac{1}{q^2 + i 0}\,\frac{1}{q^2 + 2 k_p \cdot q + i
  0}\,\frac{1}{q^2 + 2 k \cdot q + i 0} = - 2 Em_p
\int^1_0\frac{dx}{p^2(x)}\,{\ell
  n}\Big[\frac{p^2(x)}{\mu^2}\Big]
\end{eqnarray}
and 
\begin{eqnarray}\label{labelA.36}
\hspace{-0.3in}- 2i\int\frac{d^4q}{\pi^2 i}\,\frac{1}{q^2 + i
  0}\,\frac{1}{q^2 + 2 k_p \cdot q + i 0}\,\frac{1}{q^2 + 2 k \cdot q
  + i 0}[\bar{v}_{\bar{\nu}}O_{\mu}
  \sigma_{\alpha\beta}q^{\alpha}k^{\beta}_p v] = -
2i[\bar{v}_{\bar{\nu}}O_{\mu} \sigma_{\alpha\beta}k^{\alpha}
  k^{\beta}_p v] \int^1_0\frac{dx\,x}{p^2(x)},
\end{eqnarray}
respectively, where $p(x) = kx + k_p(1 - x)$, $p^2(x) = m^2_e x^2 +
m^2_p (1 - x)^2 + 2 m_e m_p \gamma x(1 - x)$ and $\gamma = 1/\sqrt{1 -
  \beta^2}$.  For the calculation of the integral over $x$ in
Eq.(\ref{labelA.35}) we transform it as follows
\begin{eqnarray}\label{labelA.37}
\hspace{-0.3in}&&\int^1_0\frac{dx}{p^2(x)}\,{\ell
  n}\Big[\frac{p^2(x)}{\mu^2}\Big] = \frac{1}{m_e m_p
  c}\int^1_0\frac{dx}{(a - x)^2 - b^2}\,{\ell n}\Big[\frac{m_em_p
    c}{\mu^2}\Big((a - x)^2 - b^2\Big)\Big],
\end{eqnarray}
where we have denoted
\begin{eqnarray}\label{labelA.38}
\hspace{-0.3in} a = \frac{\rho - \gamma}{c}\;,\, b =
\frac{\sqrt{\gamma^2 - 1}}{c}\,,\,c = \frac{1}{\rho} + \rho - 2
\gamma
\end{eqnarray}
with $\rho = m_p/m_e$. Making a change of variables $a - x =
b\,\coth\varphi$ \cite{Ivanov2013} we obtain
\begin{eqnarray}\label{labelA.39}
\hspace{-0.3in}&&\int^1_0\frac{dx}{p^2(x)}\,{\ell
  n}\Big[\frac{p^2(x)}{\mu^2}\Big] = \frac{1}{m_e m_p b
  c}\int^{\varphi_2}_{\varphi_1}d\varphi\,{\ell n}\Big[\frac{m_em_p
    c}{\mu^2}\,\frac{b^2}{\sinh^2\varphi}\Big] =\nonumber\\
\hspace{-0.3in}&&= \frac{1}{m_e m_p b c}\Big\{{\ell
  n}\Big[\frac{4 m_em_p c}{\mu^2}\,b^2\Big]\,(\varphi_2 - \varphi_1) -
(\varphi^2_2 - \varphi^2_1) + L(e^{\, -2
\varphi_2}) - L(e^{\, -2 \varphi_1})\Big\},
\end{eqnarray}
where the last two terms are the Spence functions
\cite{HMF72}--\cite{PolyLog3}.  The limits of the integration
$\varphi_2$ and $\varphi_1$ are equal to
\begin{eqnarray}\label{labelA.41}
\hspace{-0.3in}&&\varphi_2 = \frac{1}{2}\,{\ell n}\Big(\frac{1 - a -
  b}{1 - a + b}\Big) = \frac{1}{2}\,{\ell n}\Bigg(\frac{\displaystyle
  \frac{1}{\rho} - \gamma - \sqrt{\gamma^2 - 1}}{\displaystyle
  \frac{1}{\rho} - \gamma + \sqrt{\gamma^2 - 1}}\Bigg),\nonumber\\
\hspace{-0.3in}&&\varphi_1 = \frac{1}{2}\,{\ell n}\Big(\frac{a + b}{a
- b}\Big) = \frac{1}{2}\,{\ell n}\Big(\frac{\rho - \gamma +
\sqrt{\gamma^2 - 1}}{\rho - \gamma - \sqrt{\gamma^2 - 1}}\Big).
\end{eqnarray}
Keeping the leading order contributions in the large $\rho$
expansion we arrive at the expression
\begin{eqnarray}\label{labelA.42}
\hspace{-0.3in}\int^1_0\frac{dx}{p^2(x)}\,{\ell
  n}\Big[\frac{p^2(x)}{\mu^2}\Big] &=& \frac{1}{m_e m_p \sqrt{\gamma^2
    - 1}}\Big\{{\ell n}\Big[\frac{4 m^2_e}{\mu^2}\,(\gamma^2 -
  1)\Big]\,\frac{1}{2}\,{\ell n}\Big(\frac{\gamma + \sqrt{\gamma^2 -
    1}}{\gamma - \sqrt{\gamma^2 - 1}}\Big) - \frac{1}{4}\,{\ell
  n}^2\Big(\frac{\gamma + \sqrt{\gamma^2 - 1}}{\gamma - \sqrt{\gamma^2
    - 1}}\Big)\nonumber\\
\hspace{-0.3in}&+& L\Big(\frac{\gamma - \sqrt{\gamma^2 - 1}}{\gamma +
  \sqrt{\gamma^2 - 1}}\Big) - L(1)\Big\}.
\end{eqnarray}
In terms of the positron velocity $\beta$ it reads
\begin{eqnarray}\label{labelA.43}
\hspace{-0.3in}&&\int^1_0\frac{dx}{p^2(x)}\,{\ell
  n}\Big[\frac{p^2(x)}{\mu^2}\Big] = \frac{1}{E m_p
  \beta}\Big\{ {\ell n}\Big[\frac{4
    m^2_e}{\mu^2}\,\frac{\beta^2}{1 -
    \beta^2}\Big]\,\frac{1}{2}\,{\ell n}\Big(\frac{1 +
  \beta}{1 - \beta}\Big) - \frac{1}{4}\,{\ell
  n}^2\Big(\frac{1 + \beta}{1 - \beta}\Big) +
L\Big(\frac{1 - \beta}{1 + \beta}\Big) -
L(1)\Big\},\nonumber\\
\hspace{-0.3in}&&
\end{eqnarray}
where $E = m_e/\sqrt{1 - \beta^2}$. Using the relation for 
the Spence functions \cite{HMF72}--\cite{PolyLog3}
\begin{eqnarray}\label{labelA.44}
\hspace{-0.3in}L\Big(\frac{1 - \beta}{1 + \beta}\Big) -
L(1) = - L\Big(\frac{2\beta}{1 + \beta}\Big) - {\ell
  n}\Big(\frac{2\beta}{1 + \beta}\Big) {\ell
  n}\Big(\frac{1 + \beta}{1 - \beta}\Big)
\end{eqnarray}
the r.h.s. of Eq.(\ref{labelA.43}) may be written in the following form
\begin{eqnarray}\label{labelA.45}
\hspace{-0.3in}&&\int^1_0\frac{dx}{p^2(x)}\,{\ell
  n}\Big[\frac{p^2(x)}{\mu^2}\Big] = \frac{1}{E m_p
  \beta}\Big\{ {\ell n}\Big[\frac{4
    m^2_e}{\mu^2}\,\frac{\beta^2}{1 -
    \beta^2}\Big]\,\frac{1}{2}\,{\ell n}\Big(\frac{1 +
  \beta}{1 - \beta}\Big) - \frac{1}{4}\,{\ell
  n}^2\Big(\frac{1 + \beta}{1 - \beta}\Big) -
L\Big(\frac{2\beta}{1 + \beta}\Big)\nonumber\\
\hspace{-0.3in}&& - {\ell n}\Big(\frac{2\beta}{1 +
  \beta}\Big) {\ell n}\Big(\frac{1 + \beta}{1 -
  \beta}\Big)\Big\} = \frac{1}{E m_p \beta}\Big\{ -
       {\ell n}\Big(\frac{\mu}{m_e}\Big)\,{\ell n}\Big(\frac{1 +
         \beta}{1 - \beta}\Big) + \frac{1}{4}\,{\ell
         n}^2\Big(\frac{1 + \beta}{1 - \beta}\Big) -
       L\Big(\frac{2\beta}{1 + \beta}\Big)\Big\}.\nonumber\\
\hspace{-0.3in}&&
\end{eqnarray}
This gives
\begin{eqnarray}\label{labelA.46}
\hspace{-0.3in}&& 4(k\cdot k_p)\int\frac{d^4q}{\pi^2
  i}\,\frac{1}{q^2 + i 0}\,\frac{1}{q^2 + 2 k_p \cdot q + i
  0}\,\frac{1}{q^2 + 2 k \cdot q + i 0} =\nonumber\\
\hspace{-0.3in}&&= 
 2\,{\ell n}\Big(\frac{\mu}{m_e}\Big)\,\frac{1}{\beta}\,{\ell
  n}\Big(\frac{1 + \beta}{1 - \beta}\Big) -
\frac{1}{~2\beta}\,{\ell n}^2\Big(\frac{1 + \beta}{1 -
  \beta}\Big) + \frac{2}{\beta}\,L\Big(\frac{2\beta}{1 +
  \beta}\Big).
\end{eqnarray}
The integral over $x$ in Eq.(\ref{labelA.36}) is equal to
\begin{eqnarray}\label{labelA.47}
\hspace{-0.3in}\int^1_0\frac{dx\,x}{p^2(x)} = \frac{1}{2 m_e m_p
  bc}\,\Big[(b + a)\,{\ell n}\Big(\frac{b + a - 1}{b + a}\Big) - (b -
  a)\,{\ell n}\Big(\frac{b - a + 1}{b - a}\Big)\Big].
\end{eqnarray}
Keeping the leading order contributions in the large $\rho$ expansion
we obtain
\begin{eqnarray}\label{labelA.48}
\hspace{-0.3in}\int^1_0\frac{dx\,x}{p^2(x)} = \frac{1}{E
  m_p}\,\frac{1}{2 \beta}\,{\ell n}\Big(\frac{1 + \beta}{1
  - \beta}\Big).
\end{eqnarray}
As a result, the integral in Eq.(\ref{labelA.36}) is
\begin{eqnarray}\label{labelA.49}
\hspace{-0.3in}&& - 2i\int\frac{d^4q}{\pi^2 i}\,\frac{1}{q^2 + i
  0}\,\frac{1}{q^2 + 2 k_p \cdot q + i 0}\,\frac{1}{q^2 + 2 k \cdot
  q + i 0}[\bar{v}_{\bar{\nu}}O_{\mu}
  \sigma_{\alpha\beta}q^{\alpha}k^{\beta}_p v] = -
2i[\bar{v}_{\bar{\nu}}O_{\mu} \sigma_{\alpha\beta}k^{\alpha} k^{\beta}_p v] \nonumber\\
\hspace{-0.3in}&&\times\, \frac{1}{E m_p}\, \frac{1}{2
  \beta}\,{\ell n}\Big(\frac{1 + \beta}{1 -
  \beta}\Big) =
       [\bar{v}_{\bar{\nu}}O_{\mu}v]\,\frac{1}{\beta}\,{\ell
         n}\Big(\frac{1 + \beta}{1 - \beta}\Big) +
       [\bar{v}_{\bar{\nu}}O_{\mu}\gamma^0 v]\,\frac{\sqrt{1 -
           \beta^2}}{\beta}\,{\ell n}\Big(\frac{1 +
         \beta}{1 - \beta}\Big),
\end{eqnarray}
where we have used the Dirac equation $(\vec{k}\cdot
\vec{\gamma}\,)\,v = (E \gamma^0 + m_e)\,v$ for the positron wave
function. Summing up the contributions, defined by
Eq.(\ref{labelA.35}), Eq.(\ref{labelA.46}) and Eq.(\ref{labelA.49}),
we arrive at the expression for the amplitude $2 m_p {\cal M}^{(\rm
  SC)}_{p e^+}$, given in Eq.(\ref{labelA.31}).

The angular distribution for the inverse $\beta$--decay, including the
contributions of  one--virtual photon exchanges, takes the form
\begin{eqnarray}\label{labelA.50}
\hspace{-0.3in}\frac{d\sigma(E_{\bar{\nu}},\cos\theta_{e\bar{\nu}})}{
  d\cos\theta_{e\bar{\nu}}} &=& (1 + 3
\lambda^2)\,\frac{G^2_F|V_{ud}|^2}{2\pi}\,\Big(A(E_{\bar{\nu}},\mu) +
B(E_{\bar{\nu}},\mu)\,\bar{\beta}\,\cos\theta_{e\bar{\nu}} +
C(E_{\bar{\nu}})\,\bar{\beta}^2\,\cos^2\theta_{e\bar{\nu}}\Big)\,
\bar{k} \bar{E},
\end{eqnarray}
where we have denoted
\begin{eqnarray}\label{labelA.51}
\hspace{-0.3in}A(E_{\bar{\nu}}, \mu) &=& \Big(1 +
\frac{\alpha}{\pi}\,f(\bar{E},\mu)\Big) + \frac{1}{M}\,\frac{1}{1 + 3
  \lambda^2}\,\Big(2 \,(\lambda^2 - (\kappa + 1)\,\lambda)\,\Delta
+ 4 (\kappa + 1)\,\lambda\,E_{\bar{\nu}} \nonumber\\
\hspace{-0.3in}&-& (\lambda^2 + 2 (\kappa + 1)\,\lambda +
  1)\,\frac{m^2_e}{E}\Big) - \frac{1 + 2
  \bar{\beta}^2}{\bar{\beta}^2}\,\frac{E_{\bar{\nu}}}{M}\,\Big(1 +
\frac{1 + \bar{\beta}^2}{1 + 2\bar{\beta}^2}\,\frac{\Delta^2 -
  m^2_e}{2 \bar{E} E_{\bar{\nu}}}\Big),\nonumber\\
\hspace{-0.3in}B(E_{\bar{\nu}}, \mu) &=& a_0\,\Big(1 +
\frac{\alpha}{\pi}\,g(\bar{E},\mu)\Big)+ \frac{1}{M}\,\frac{1}{1 + 3
  \lambda^2}\,\Big(- 2\,(\lambda^2 - 2(\kappa + 1)\,\lambda)\,\Delta -
4 (\kappa + 1)\,\lambda\,E_{\bar{\nu}}\Big)\nonumber\\
\hspace{-0.3in}&+& \frac{2 +
  \bar{\beta}^2}{\bar{\beta}^2}\,\frac{E_{\bar{\nu}}}{M}\,\Big[1 -
  a_0\,\Big(1 + \frac{2}{2 +\bar{\beta}^2}\,\frac{\Delta^2 - m^2_e}{2
    \bar{E} E_{\bar{\nu}}}\Big)\Big].
\end{eqnarray}
The functions $f(\bar{E},\mu)$ and  $g(\bar{E},\mu)$ are given by
\begin{eqnarray}\label{labelA.52}
\hspace{-0.3in}&&f(\bar{E},\mu) = \frac{3}{2}\,{\ell
  n}\Big(\frac{m_p}{m_e}\Big) - \frac{11}{8} + {\ell
  n}\Big(\frac{\mu}{m_e}\Big)\,\Big[\frac{1}{\bar{\beta}}\,{\ell
    n}\Big(\frac{1 + \bar{\beta}}{1 - \bar{\beta}}\Big) - 2\Big] +
\frac{1}{~\bar{\beta}}\,L\Big(\frac{2\bar{\beta}}{1 +
  \bar{\beta}}\Big) - \frac{1}{~4\bar{\beta}}\,{\ell n}^2\Big(\frac{1
  + \bar{\beta}}{1 - \bar{\beta}}\Big) + \frac{\bar{\beta}}{2}\,{\ell
  n}\Big(\frac{1 + \bar{\beta}}{1 - \bar{\beta}}\Big),\nonumber\\
\hspace{-0.3in}&&g(\bar{E},\mu) = \frac{3}{2}\,{\ell
  n}\Big(\frac{m_p}{m_e}\Big) - \frac{11}{8} + {\ell
  n}\Big(\frac{\mu}{m_e}\Big)\,\Big[\frac{1}{\bar{\beta}}\,{\ell
    n}\Big(\frac{1 + \bar{\beta}}{1 - \bar{\beta}}\Big) - 2\Big] +
\frac{1}{~\bar{\beta}}\,L\Big(\frac{2\bar{\beta}}{1 +
  \bar{\beta}}\Big) - \frac{1}{~4\bar{\beta}}\,{\ell n}^2\Big(\frac{1
  + \bar{\beta}}{1 - \bar{\beta}}\Big) + \frac{1}{2\bar{\beta}}\,{\ell
  n}\Big(\frac{1 + \bar{\beta}}{1 - \bar{\beta}}\Big).\nonumber\\
\hspace{-0.3in}&&
\end{eqnarray}
The functions $f(\bar{E},\mu)$ and $g(\bar{E},\mu)$ are different,
since the term, proportional to $[\bar{v}_{\bar{\nu}}O_{\mu}\gamma^0
  v]$, gives a contribution only to the correlation coefficient
$A(\bar{E}, \mu)$.  A dependence of the functions $f(\bar{E},\mu)$ and
$g(\bar{E},\mu)$ on the infrared regularisation scale $\mu$ should be
removed by taking into account the contribution of the radiative
inverse $\beta$--decay, i.e. the contribution of the reaction
$\bar{\nu}_e + p \to n + e^+ + \gamma$, where $\gamma$ is a photon on
mass--shell. 

\subsection*{3. Radiative corrections, caused by radiative 
inverse $\beta$--decay}

The amplitude of the radiative inverse $\beta$--decay is 
\begin{eqnarray}\label{labelA.53}
\hspace{-0.3in}&&M(\bar{\nu}_e p \to n e^+ \gamma) =
\varepsilon^{*\alpha} M_{\alpha}(\bar{\nu}_e p \to n e^+ \gamma) =
\nonumber\\
\hspace{-0.3in}&&= e\,\frac{G_F}{\sqrt{2}}\,V_{ud}\Big\{[\bar{u}_nW^{\mu}u_p]
\Big[\bar{v}_{\bar{\nu}}O_{\mu}\,\frac{1}{m_e + \hat{k} + \hat{q} -
    i0}\,\hat{\varepsilon}^*\,v\Big] -
\Big[\bar{u}_nW^{\mu}\,\frac{1}{m_p - \hat{k}_p + \hat{q} -
    i0}\,\hat{\varepsilon}^*\,u_p\Big][\bar{v}_{\bar{\nu}}O_{\mu}
  v]\Big\},
\end{eqnarray}
where $q = (\omega, \vec{q} = \omega\,\vec{n}\,)$ and $\varepsilon^*$
are the 4--momentum and 4--polarisation vector of a photon, obeying
the constraint $q\cdot \varepsilon^* = 0$. The amplitude
$M_{\alpha}(\bar{\nu}_e p \to n e^+ \gamma)$ is gauge invariant
$q^{\alpha}M_{\alpha}(\bar{\nu}_e p \to n e^+ \gamma) = 0$. For the
calculation of the cross section for the reaction $\bar{\nu}_e + p \to
n + e^+ + \gamma$ we use the Coulomb gauge $\varepsilon =
(0,\vec{\varepsilon}\,)$ and keep the leading order contributions in
the large baryon mass expansion \cite{Ivanov2013}. In such an
approximation we define the amplitude of the reaction $\bar{\nu}_e + p
\to n + e^+ + \gamma$ as follows \cite{Ivanov2013}
\begin{eqnarray}\label{labelA.54}
\hspace{-0.3in}M(\bar{\nu}_e p \to n e^+ \gamma) =
- \,e\,\frac{G_F}{\sqrt{2}}\,V_{ud}\,\frac{m_p}{\omega}\,\frac{{\cal
    M}_{\beta \gamma}}{E - \vec{k}\cdot \vec{n}}.
\end{eqnarray}
The amplitude ${\cal M}_{\beta \gamma}$ and its hermitian conjugate
${\cal M}^{\dagger}_{\beta \gamma}$ are equal to
\begin{eqnarray}\label{labelA.55}
\hspace{-0.3in}{\cal M}_{\beta \gamma} =
       [\varphi^{\dagger}_n\varphi_p][\bar{v}_{\bar{\nu}}\,\gamma^0(1 -
         \gamma^5)\, Q\,v] - \lambda\,[\varphi^{\dagger}_n
         \vec{\sigma}\,\varphi_p]\cdot
       [\bar{v}_{\bar{\nu}}\,\vec{\gamma}\,(1 - \gamma^5)\, Q\,v],
\end{eqnarray}
and 
\begin{eqnarray}\label{labelA.56}
\hspace{-0.3in}{\cal M}^{\dagger}_{\beta \gamma} =
       [\varphi^{\dagger}_p\varphi_n][\bar{v}\,\bar{Q}\,\gamma^0 (1 -
         \gamma^5) v_{\bar{\nu}}] - \lambda\,[\varphi^{\dagger}_p
         \vec{\sigma}\,\varphi_n]\cdot
       [\bar{v}\, \bar{Q}\,\vec{\gamma}\,(1 - \gamma^5) v_{\bar{\nu}}],
\end{eqnarray}
where $Q = 2 k\cdot \varepsilon^* + \hat{q}\hat{\varepsilon}^*$ and
$\bar{Q} = \gamma^0 Q^{\dagger}\gamma^0 = 2 k\cdot \varepsilon +
\hat{\varepsilon}\hat{q} $ \cite{Ivanov2013}.

The absolute squared value of the amplitude ${\cal M}_{\beta \gamma}$,
summed over the polarisations of the neutron and positron and averaged
over polarisations of the proton, is given by
\begin{eqnarray}\label{labelA.57}
\hspace{-0.3in}\frac{1}{2}\sum_{\rm pol}|{\cal M}_{\beta
  \gamma}|^2 &=& 2\,{\rm tr}\{\hat{k} \bar{Q} \gamma^0
\hat{k}_{\bar{\nu}}\gamma^0 Q(1 - \gamma^5)\} + 2\,\lambda^2\,{\rm
  tr}\{\hat{k} \bar{Q} \vec{\gamma}\cdot
\hat{k}_{\bar{\nu}}\vec{\gamma}\, Q (1 - \gamma^5)\}=\nonumber\\
\hspace{-0.3in}&=& 2 \,(1 + 3\lambda^2) \,E_{\bar{\nu}} \Big\{{\rm
  tr}\{\hat{k} \bar{Q} \gamma^0 Q (1 - \gamma^5)\} +
a_0\,\frac{\vec{k}_{\bar{\nu}}}{E_{\bar{\nu}}}\cdot {\rm tr}\{\hat{k}
\bar{Q} \,\vec{\gamma}\, Q (1 - \gamma^5)\}\Big\}.
\end{eqnarray}
The calculation of the traces we carry out by using the following formula
\begin{eqnarray}\label{labelA.58}
\hspace{-0.3in}\frac{1}{4}\,{\rm tr}\{\hat{k} Q \gamma^{\mu}
\bar{Q}(1 - \gamma^5)\} = 4 (k\cdot \varepsilon^*)(k\cdot
\varepsilon)(k + q)^{\mu} - 2\,(k \cdot q)[(k\cdot
  \varepsilon^*)\,\varepsilon^{\mu} + (k\cdot
  \varepsilon)\,\varepsilon^{*\mu}] - 2 (k\cdot q)( \varepsilon^*\cdot
  \varepsilon) q^{\mu} + \ldots,
\end{eqnarray}
where the ellipsis denotes the terms, which do not contribute to the
cross section for the radiative inverse $\beta$--decay with
unpolarised photons. One may obtain these terms using Eq.(B-9) in
Appendix B of Ref.\cite{Ivanov2013}. The angular and photon--energy
distribution of the radiative inverse $\beta$--decay is
\begin{eqnarray}\label{labelA.59}
\hspace{-0.3in}\frac{d^2
  \sigma^{(\gamma)}(E_{\bar{\nu}},\cos\theta_{e\bar{\nu}})}{ d\omega\,
  d\cos\theta_{e\bar{\nu}}} &=& (1 +
3\lambda^2)\,\frac{\alpha}{\pi}\,\frac{G^2_F|V_{ud}|^2}{2\pi}
\frac{1}{\omega}\,k\,E
\int\frac{d\Omega_{\vec{n}}}{4\pi}\,\Big\{\frac{\beta^2
  -(\vec{\beta}\cdot \vec{n}\,)^2}{(1 - \vec{\beta}\cdot
  \vec{n}\,)^2}\,\Big(1 + \frac{\omega}{E}\Big) + \frac{1}{1 -
  \vec{\beta}\cdot \vec{n}}\,\frac{\omega^2}{E^2}\nonumber\\
\hspace{-0.3in}&+&
a_0\,\frac{\vec{k}_{\bar{\nu}}}{E_{\bar{\nu}}}\cdot\Big[\frac{\beta^2
    - (\vec{\beta}\cdot \vec{n}\,)^2}{(1 - \vec{\beta}\cdot
    \vec{n}\,)^2}\,\Big(\vec{\beta} + \vec{n}\,\frac{\omega}{E}\Big) +
  \frac{\vec{\beta} - \vec{n}\,(\vec{\beta}\cdot \vec{n}\,)}{1 -
    \vec{\beta}\cdot \vec{n}}\,\frac{\omega}{E} + \frac{\vec{n}}{1 -
    \vec{\beta}\cdot \vec{n}}\,\frac{\omega^2}{E^2}\Big]\Big\},
\end{eqnarray}
where $E = \bar{E} - \omega$ and $\beta = k/E = \sqrt{1 -
  m^2_e/E^2}$. Up to a common factor the expression in curl brackets
coincides with Eq.(29) of Ref.\cite{IBD4} and Eq.(9) of
Ref.\cite{IBD5}. The integrals over directions of a photon momentum
are equal to
\begin{eqnarray}\label{labelA.60}
\hspace{-0.3in}&& \int\frac{d\Omega_{\vec{n}}}{4\pi}\,\frac{\beta^2
  -(\vec{\beta}\cdot \vec{n}\,)^2}{(1 - \vec{\beta}\cdot \vec{n}\,)^2}
= 2 \Big[\frac{1}{2\beta}\,{\ell n}\Big(\frac{1 + \beta}{1 - \beta}\Big) -
1\Big],\nonumber\\
\hspace{-0.3in}&&\int\frac{d\Omega_{\vec{n}}}{4\pi}\,\frac{1}{1 -
  \vec{\beta}\cdot \vec{n}} = 1 + \Big[\frac{1}{2 \beta}\,{\ell n}\Big(\frac{1
  + \beta}{1 - \beta}\Big) - 1\Big],\nonumber\\
\hspace{-0.3in}&&\int\frac{d\Omega_{\vec{n}}}{4\pi}\,\frac{\vec{n}\,(\beta^2
  -(\vec{\beta}\cdot \vec{n}\,)^2)}{(1 - \vec{\beta}\cdot
  \vec{n}\,)^2} = \vec{\beta}\,\Big\{ -1 - \Big(1 -
\frac{3}{\beta^2}\Big)\,\Big[\frac{1}{2 \beta}\,{\ell n}\Big(\frac{1 +
    \beta}{1 - \beta}\Big) - 1\Big]\Big\},\nonumber\\
\hspace{-0.3in}&&\int\frac{d\Omega_{\vec{n}}}{4\pi}\,
\frac{\vec{\beta} - \vec{n}\,(\vec{\beta}\cdot \vec{n}\,)}{1 -
  \vec{\beta}\cdot \vec{n}} =  \vec{\beta}\,\Big\{ 1 + \Big(1 -
\frac{1}{\beta^2}\Big)\,\Big[\frac{1}{2 \beta}\,{\ell n}\Big(\frac{1 +
    \beta}{1 - \beta}\Big) - 1\Big]\Big\},\nonumber\\
\hspace{-0.3in}&&\int\frac{d\Omega_{\vec{n}}}{4\pi}\,\frac{\vec{n}}{1
  - \vec{\beta}\cdot \vec{n}} =
\frac{\vec{\beta}}{\beta^2}\,\Big[\frac{1}{2 \beta}\,{\ell
    n}\Big(\frac{1 + \beta}{1 - \beta}\Big) - 1\Big].
\end{eqnarray}
Substituting Eq.(\ref{labelA.60}) into Eq.(\ref{labelA.59}) the
angular and photon--energy distribution of the radiative inverse
$\beta$--decay is
\begin{eqnarray}\label{labelA.61}
\hspace{-0.3in}\frac{d
  \sigma^{(\gamma)}(E_{\bar{\nu}},\cos\theta_{e\bar{\nu}})}{ d\omega
  d\cos\theta_{e\bar{\nu}}} &=& (1 +
3\lambda^2)\,\frac{\alpha}{\pi}\,\frac{G^2_F|V_{ud}|^2}{2\pi}
\frac{1}{\omega}\,k\,E\,\Big\{\Big[\Big(1 + \frac{\omega}{E} +
  \frac{1}{2}\,\frac{\omega^2}{E^2}\Big)\Big[\frac{1}{\beta}\,{\ell
      n}\Big(\frac{1 + \beta}{1 - \beta}\Big) - 2\Big] +
  \frac{\omega^2}{E^2}\Big]\nonumber\\
\hspace{-0.3in}&+& a_0\,\Big(1 + \frac{1}{\beta^2} \frac{\omega}{E} +
\frac{1}{2
  \beta^2}\,\frac{\omega^2}{E^2}\Big)\Big[\frac{1}{\beta}\,{\ell
    n}\Big(\frac{1 + \beta}{1 - \beta}\Big) -
  2\Big]\,\beta\,\cos\theta_{e\bar{\nu}}\Big\}.
\end{eqnarray}
Apart from kinematic factors, the r.h.s. of Eq.(\ref{labelA.61})
reproduces the photon--electron energy spectrum of the radiative
$\beta^-$--decay of the neutron (see Eq.(B-11) of
Ref.\cite{Ivanov2013}). Having integrated over the photon energy we
obtain the following angular distribution
\begin{eqnarray}\label{labelA.62}
\hspace{-0.3in}\frac{d
  \sigma^{(\gamma)}(E_{\bar{\nu}},\cos\theta_{e\bar{\nu}})}{
  d\cos\theta_{e\bar{\nu}}} &=& (1 +
3\lambda^2)\,\frac{\alpha}{\pi}\,\frac{G^2_F|V_{ud}|^2}{2\pi}
\int^{\bar{E}- m_e}_0\frac{d\omega}{\omega}\,k\,E\,\Big\{\Big[\Big(1 +
  \frac{\omega}{E} +
  \frac{1}{2}\,\frac{\omega^2}{E^2}\Big)\Big[\frac{1}{\beta}\,{\ell
      n}\Big(\frac{1 + \beta}{1 - \beta}\Big) - 2\Big] +
  \frac{\omega^2}{E^2}\Big]\nonumber\\
\hspace{-0.3in}&+& a_0\,\Big(1 + \frac{1}{\beta^2} \frac{\omega}{E} +
\frac{1}{2
  \beta^2}\,\frac{\omega^2}{E^2}\Big)\Big[\frac{1}{\beta}\,{\ell
    n}\Big(\frac{1 + \beta}{1 - \beta}\Big) -
  2\Big]\,\beta\,\cos\theta_{e\bar{\nu}}\Big\}.
\end{eqnarray}
The integral over the photon energy is infrared divergent. For the
regularisation of the infrared divergences we propose to rewrite the
r.h.s. of Eq.(\ref{labelA.62}) as follows
\begin{eqnarray}\label{labelA.63}
\hspace{-0.3in}\frac{d
  \sigma^{(\gamma)}(E_{\bar{\nu}},\cos\theta_{e\bar{\nu}})}{
  d\cos\theta_{e\bar{\nu}}} &=& (1 +
3\lambda^2)\,\frac{\alpha}{\pi}\,\frac{G^2_F|V_{ud}|^2}{2\pi}\,\bar{k}\,
\bar{E}\, \Big\{\int^{\bar{E}-
  m_e}_0\frac{d\omega}{\omega}\,\Big[\frac{1}{\bar{\beta}}\,{\ell
    n}\Big(\frac{1 + \bar{\beta}}{1 - \bar{\beta}}\Big) - 2\Big] +
    f^{\gamma}_A(\bar{E})\nonumber\\
 \hspace{-0.3in}&+& a_0\Big(\int^{\bar{E}- m_e}_0
 \frac{d\omega}{\omega}\,\Big[\frac{1}{\bar{\beta}}\,{\ell
     n}\Big(\frac{1 + \bar{\beta}}{1 - \bar{\beta}}\Big) - 2\Big] +
 f^{\gamma}_B(\bar{E})\Big)\,\bar{\beta}\,\cos\theta_{e\bar{\nu}}\Big\}.
\end{eqnarray}
The functions $f^{\gamma}_A(\bar{E})$ and $f^{\gamma}_B(\bar{E})$ are
defined by
\begin{eqnarray}\label{labelA.64}
\hspace{-0.3in}f^{\gamma}_A(\bar{E}) &=&
\frac{1}{\bar{k}\bar{E}}\int^{\bar{E}}_{m_e}\frac{d E}{\bar{E} -
  E}\Big\{k E\,\Big[\frac{1}{\beta}\,{\ell n}\Big(\frac{1 + \beta}{1 -
    \beta}\Big) - 2\Big] - \bar{k}\bar{E}\,
\Big[\frac{1}{\bar{\beta}}\,{\ell n}\Big(\frac{1 + \bar{\beta}}{1 -
    \bar{\beta}}\Big) - 2\Big]\Big\}\nonumber\\
\hspace{-0.3in}&+&\frac{1}{\bar{k}\bar{E}}\int^{\bar{E}}_{m_e}d E\,k
\Big(\frac{\bar{E} +  E}{2 E}\,\Big[\frac{1}{\beta}\,{\ell
    n}\Big(\frac{1 + \beta}{1 - \beta}\Big) - 2\Big] + \frac{\bar{E} -
  E}{E}\Big)
\end{eqnarray}
and 
\begin{eqnarray}\label{labelA.65}
\hspace{-0.3in}f^{\gamma}_B(\bar{E}) &=&
\frac{1}{\bar{k}\bar{E}\bar{\beta}}\int^{\bar{E}}_{m_e}\frac{d
  E}{\bar{E} - E}\Big\{k E \beta\,\Big[\frac{1}{\beta}\,{\ell
    n}\Big(\frac{1 + \beta}{1 - \beta}\Big) - 2\Big] -
\bar{k}\bar{E}\bar{\beta}\, \Big[\frac{1}{\bar{\beta}}\,{\ell
    n}\Big(\frac{1 + \bar{\beta}}{1 - \bar{\beta}}\Big) -
  2\Big]\Big\}\nonumber\\
\hspace{-0.3in}&+&\frac{1}{\bar{k}\bar{E}\bar{\beta}}\int^{\bar{E}}_{m_e}
d E\, \frac{1}{2}\,(\bar{E} +  E)\Big[\frac{1}{\beta}\,{\ell
    n}\Big(\frac{1 + \beta}{1 - \beta}\Big) - 2\Big],
\end{eqnarray}
where $k = \sqrt{E^2 - m^2_e}$, $\beta = k/E$, $\bar{k} =
\sqrt{\bar{E}^2 - m^2_e}$, $\bar{\beta} = \bar{k}/\bar{E}$ and
$\bar{E} = E_{\bar{\nu}} - \Delta$. In Appendices B and C we give
analytical expressions for functions $f^{\gamma}_A(\bar{E})$ and
$f^{\gamma}_B(\bar{E})$, respectively.

The regularisation of the infrared divergent contribution runs as
follows (see Eq.(B-18) - Eq.(B-26) in Appendix B of
Ref.\cite{Ivanov2013})
\begin{eqnarray}\label{labelA.66}
\hspace{-0.3in}&& \int^{\bar{E}-
  m_e}_0\frac{d\omega}{\omega}\,\Big[\frac{1}{\bar{\beta}}\,{\ell
    n}\Big(\frac{1 + \bar{\beta}}{1 - \bar{\beta}}\Big) - 2\Big] \to
\int^{\bar{E}- m_e}_0\frac{dq
  q^2}{\omega^3}\int\frac{d\Omega_{\vec{v}}}{4\pi}\,\frac{\beta^2
  -(\vec{\beta}\cdot \vec{v}\,)^2}{(1 - \vec{\beta}\cdot \vec{v}\,)^2}
= \int^{v_{\rm max}}_0 \frac{dv v^2}{1 -
  v^2}\frac{\bar{\beta}^2}{2}\int^{+1}_{-1}\frac{dx(1 - v^2x^2)}{(1 -
  \bar{\beta}v x)^2} =\nonumber\\
\hspace{-0.3in}&&= {\ell n}\Big(\frac{2(\bar{E} -
  m_e)}{\mu}\Big)\Big[\frac{1}{\bar{\beta}}{\ell n}\Big(\frac{1 +
    \bar{\beta}}{1 - \bar{\beta}}\Big) - 2\Big] + 1 +
\frac{1}{2\bar{\beta}}\,{\ell n}\Big(\frac{1 + \bar{\beta}}{1 -
  \bar{\beta}}\Big) - \frac{1}{4\bar{\beta}}\,{\ell n}^2\Big(\frac{1 +
  \bar{\beta}}{1 - \bar{\beta}}\Big) +
\frac{1}{\bar{\beta}}\,L\Big(\frac{2 \bar{\beta}}{1 + \bar{\beta}}
\Big),
\end{eqnarray}
where $q = |\vec{q}\,|$, $\omega = \sqrt{q^2 + \mu^2}$ and $\vec{v} =
\vec{q}/\omega$. As a result, we define the angular distribution for
the radiative inverse $\beta$--decay as follows \cite{Ivanov2013}
\begin{eqnarray}\label{labelA.67}
\hspace{-0.3in}\frac{d
  \sigma^{(\gamma)}(E_{\bar{\nu}},\cos\theta_{e\bar{\nu}})}{
  d\cos\theta_{e\bar{\nu}}} = (1 +
3\lambda^2)\,\frac{\alpha}{\pi}\,\frac{G^2_F|V_{ud}|^2}{2\pi}\,\bar{k}\,
\bar{E}\, \Big(g^{(1)}_{\beta\gamma}(\bar{E},\mu) +
a_0\,g^{(2)}_{\beta\gamma}(\bar{E},\mu)
\,\bar{\beta}\,\cos\theta_{e\bar{\nu}}\Big),
\end{eqnarray}
where the functions $g^{(1)}_{\beta\gamma}(\bar{E},\mu)$ and
$g^{(2)}_{\beta\gamma}(\bar{E},\mu)$ are equal to
\begin{eqnarray}\label{labelA.68}
\hspace{-0.3in}&&g^{(1)}_{\beta\gamma}(\bar{E},\mu) = {\ell
  n}\Big(\frac{2(\bar{E} -
  m_e)}{\mu}\Big)\Big[\frac{1}{\bar{\beta}}{\ell n}\Big(\frac{1 +
    \bar{\beta}}{1 - \bar{\beta}}\Big) - 2\Big] + 1 +
\frac{1}{2\bar{\beta}}\,{\ell n}\Big(\frac{1 + \bar{\beta}}{1 -
  \bar{\beta}}\Big) - \frac{1}{4\bar{\beta}}\,{\ell n}^2\Big(\frac{1 +
  \bar{\beta}}{1 - \bar{\beta}}\Big) +
\frac{1}{\bar{\beta}}\,L\Big(\frac{2 \bar{\beta}}{1 + \bar{\beta}}
\Big) + f^{\gamma}_A(\bar{E})\nonumber\\
\hspace{-0.3in}&&
\end{eqnarray}
and 
\begin{eqnarray}\label{labelA.69}
\hspace{-0.3in}&&g^{(2)}_{\beta\gamma}(\bar{E},\mu) = {\ell
  n}\Big(\frac{2(\bar{E} -
  m_e)}{\mu}\Big)\Big[\frac{1}{\bar{\beta}}{\ell n}\Big(\frac{1 +
    \bar{\beta}}{1 - \bar{\beta}}\Big) - 2\Big] + 1 +
\frac{1}{2\bar{\beta}}\,{\ell n}\Big(\frac{1 + \bar{\beta}}{1 -
  \bar{\beta}}\Big) - \frac{1}{4\bar{\beta}}\,{\ell n}^2\Big(\frac{1 +
  \bar{\beta}}{1 - \bar{\beta}}\Big) +
\frac{1}{\bar{\beta}}\,L\Big(\frac{2 \bar{\beta}}{1 + \bar{\beta}}
\Big) + f^{\gamma}_B(\bar{E}).\nonumber\\
\hspace{-0.3in}&&
\end{eqnarray}
Summing up the contributions of one--virtual photon exchanges and the
radiative inverse $\beta$--decay we obtain the following angular
distribution of the inverse $\beta$--decay
\begin{eqnarray}\label{labelA.70}
\hspace{-0.3in}\frac{d\sigma(E_{\bar{\nu}},\cos\theta_{e\bar{\nu}})}{
  d\cos\theta_{e\bar{\nu}}} &=& (1 + 3
\lambda^2)\,\frac{G^2_F|V_{ud}|^2}{2\pi}\,(1 +
\Delta_R)\,\Big[A(E_{\bar{\nu}})\,\Big(1 +
\frac{\alpha}{\pi}\,f_A(\bar{E})\Big) + B(E_{\bar{\nu}})\,\Big(1 +
\frac{\alpha}{\pi}\,f_B(\bar{E})\Big)\,\bar{\beta}\,\cos\theta_{e\bar{\nu}}
\nonumber\\
\hspace{-0.3in}&+&
C(E_{\bar{\nu}})\,\bar{\beta}^2\,\cos^2\theta_{e\bar{\nu}}\Big]\,
\bar{k} \bar{E},
\end{eqnarray}
where the functions $f_A(\bar{E})$ and $f_B(\bar{E})$, describing the
radiative corrections, are equal to
\begin{eqnarray}\label{labelA.71}
\hspace{-0.3in}f_A(\bar{E}) &=& \frac{3}{2}\,{\ell
  n}\Big(\frac{m_p}{m_e}\Big) - \frac{3}{8} + {\ell
  n}\Big(\frac{2(\bar{E} -
  m_e)}{m_e}\Big)\Big[\frac{1}{\bar{\beta}}{\ell n}\Big(\frac{1 +
    \bar{\beta}}{1 - \bar{\beta}}\Big) - 2\Big] -
\frac{1}{2\bar{\beta}}\,{\ell n}^2\Big(\frac{1 + \bar{\beta}}{1 -
  \bar{\beta}}\Big) + \frac{2}{\bar{\beta}}\,L\Big(\frac{2
  \bar{\beta}}{1 + \bar{\beta}} \Big) \nonumber\\
\hspace{-0.3in}&+& \frac{1 +
  \bar{\beta}^2}{2\bar{\beta}}\,{\ell n}\Big(\frac{1 + \bar{\beta}}{1
  - \bar{\beta}}\Big) + f^{\gamma}_A(\bar{E})
\end{eqnarray}
and
\begin{eqnarray}\label{labelA.72}
\hspace{-0.3in}f_B(\bar{E}) &=& \frac{3}{2}\,{\ell
  n}\Big(\frac{m_p}{m_e}\Big) - \frac{3}{8} + {\ell
  n}\Big(\frac{2(\bar{E} -
  m_e)}{m_e}\Big)\Big[\frac{1}{\bar{\beta}}{\ell n}\Big(\frac{1 +
    \bar{\beta}}{1 - \bar{\beta}}\Big) - 2\Big] -
\frac{1}{2\bar{\beta}}\,{\ell n}^2\Big(\frac{1 + \bar{\beta}}{1 -
  \bar{\beta}}\Big) + \frac{2}{\bar{\beta}}\,L\Big(\frac{2
  \bar{\beta}}{1 + \bar{\beta}} \Big)\nonumber\\
\hspace{-0.3in}&+& \frac{1}{\bar{\beta}}\,{\ell
  n}\Big(\frac{1 + \bar{\beta}}{1 - \bar{\beta}}\Big) +
f^{\gamma}_B(\bar{E}).
\end{eqnarray}
The contribution $\Delta_R = (\alpha/\pi)\,C_{WZ} = 0.0238$ with
$C_{WZ} = 10.249$ is defined by electroweak boson exchanges and QCD
corrections. Such a contribution has been calculated in
\cite{RC16,RC1,RC17}. The numerical value $\Delta_R = 0.0238$,
calculated in Appendix D of Ref.\cite{Ivanov2013}, agrees also well
with that $\Delta_R = 0.024$, used in \cite{RNA1}.

The cross section for the inverse $\beta$--decay is defined by the
integral
\begin{eqnarray}\label{labelA.73}
\hspace{-0.3in}\sigma(E_{\bar{\nu}}) = \int^{+1)_{\rm max}}_{-1}\frac{d
  \sigma(E_{\bar{\nu}},\cos\theta_{e\bar{\nu}})}{
  d\cos\theta_{e\bar{\nu}}}\, d\cos\theta_{e\bar{\nu}}.
\end{eqnarray}
Having integrated over $\cos\theta_{e\bar{\nu}}$ we obtain
\begin{eqnarray}\label{labelA.74}
\hspace{-0.3in}\sigma(E_{\bar{\nu}}) = (1 + 3
\lambda^2)\,\frac{G^2_F|V_{ud}|^2}{\pi}\,(1 +
\Delta_R)\,\bar{k}\bar{E}\,\Big(A(E_{\bar{\nu}}) +
  \frac{1}{3}\,C(E_{\bar{\nu}})\,\bar{\beta}^2\Big)\,\Big(1 +
\frac{\alpha}{\pi}\,f_A(\bar{E})\Big).
\end{eqnarray}
Using Eq.(\ref{labelA.21}) and Eq.(\ref{labelA.24}) the cross section
for the inverse $\beta$--decay we transcribe into the form
\begin{eqnarray}\label{labelA.75}
\hspace{-0.3in}\sigma(E_{\bar{\nu}}) &=& (1 + 3
\lambda^2)\,\frac{G^2_F|V_{ud}|^2}{\pi}\,(1 +
\Delta_R)\,\bar{k}\bar{E}\,\Big[1 + \frac{1}{M}\,\frac{1}{1 +
    3\lambda^2}\Big(2(\lambda^2 - (\kappa + 1)\lambda)\Delta +
  4(\kappa + 1)\lambda E_{\bar{\nu}}\nonumber\\ \hspace{-0.3in}&-&
  (\lambda^2 + 2(\kappa + 1)\lambda + 1)\,\frac{m^2_e}{\bar{E}}\Big) -
  \frac{1 + 2
    \bar{\beta}^2}{\bar{\beta}^2}\,\frac{E_{\bar{\nu}}}{M}\,\Big(1 +
  \frac{1 + \bar{\beta}^2}{1 + 2\bar{\beta}^2}\,\frac{\Delta^2 -
    m^2_e}{2 \bar{E} E_{\bar{\nu}}}\Big) +
  a_0\,\frac{E_{\bar{\nu}}}{M}\Big]\,\Big(1 +
\frac{\alpha}{\pi}\,f_A(\bar{E})\Big),
\end{eqnarray}
where $\bar{E} = E_{\bar{\nu}} - \Delta$ and $\bar{k} =
\sqrt{(E_{\bar{\nu}} - \Delta)^2 - m^2_e}$.  

The cross section Eq.(\ref{labelA.75}) is in analytical agreement with
the cross sections, calculated by Vogel and Beacom \cite{IBD3} (see
Eq.(12) and Eq.(13) of Ref.\cite{IBD3}) and by Raha, Myhrer and
Kubodera \cite{IBD5} (see Eq.(20) of Ref.\cite{IBD5}) and Erratum to
Ref.\cite{IBD5}), with the replacement $f_A(\bar{E}) \to f_V(\bar{E})$
and $f_A(\bar{E}) \to f_R(\bar{E})$, respectively. The functions
$f_V(\bar{E})$ and $\textstyle f_R(\bar{E}) = \frac{1}{2}\,\delta_{\rm
  out}(\bar{E})$ describe the radiative corrections, calculated by
Vogel \cite{IBD1} and by Raha, Myhrer and Kubodera \cite{IBD5}. Using
the results, obtained in \cite{IBD1} and \cite{IBD2}, one may show
that the radiative corrections to the correlation coefficient
$A(E_{\bar{\nu}})$, calculated by Vogel \cite{IBD1} (see Eq.(19) of
Ref.\cite{IBD1}), Fayans \cite{IBD2} (see Eq.(25) and Eq.($\Pi$.14) of
Ref.\cite{IBD2}) and Raha, Myhrer and Kubodera \cite{IBD5} (see
Eq.(12) of Ref.\cite{IBD5}), are defined by the function $f(\bar{E}) =
f_V(\bar{E})= f_F(\bar{E}) = f_R(\bar{E})$ equal to
\begin{eqnarray}\label{labelA.76}
\hspace{-0.3in}f(\bar{E}) &=&\frac{3}{2}\,{\ell
  n}\Big(\frac{m_p}{m_e}\Big) + \frac{23}{8} + 2\,{\ell
  n}\Big(\frac{2\bar{\beta}}{1 +
  \bar{\beta}}\Big)\Big[\frac{1}{\bar{\beta}}\,{\ell n}\Big(\frac{1 +
    \bar{\beta}}{1 - \bar{\beta}}\Big) - 2\Big] +
\frac{4}{\bar{\beta}}\,L\Big(\frac{2\bar{\beta}}{1 + \bar{\beta}}\Big)
+ \frac{3}{8}\Big(\bar{\beta}^2
+\frac{7}{3}\Big)\,\frac{1}{\bar{\beta}}\,{\ell n}\Big(\frac{1 +
  \bar{\beta}}{1 - \bar{\beta}}\Big)\nonumber\\
\hspace{-0.3in}&-& 2\,{\ell n}\Big(\frac{1 +
  \bar{\beta}}{1 - \bar{\beta}}\Big).
\end{eqnarray}
In Appendix B (see Eq.(\ref{labelB.20})) we show that the function
$f_A(\bar{E})$ agrees fully with the function $f(\bar{E})$.

\section{Appendix B: Analytical expressions for functions
$f^{(\gamma)}_A(\bar{E})$ and $f_A(\bar{E})$ }
\renewcommand{\theequation}{B-\arabic{equation}}
\setcounter{equation}{0}

In this Appendix we propose a detailed calculation of the integrals,
defining the function $f^{(\gamma)}_A(\bar{E})$, and give the
analytical expression for the function $f_A(\bar{E})$.  Making a
change of variables $E = m_e\cosh\varphi$ and $\bar{E} = m_e
\cosh\bar{\varphi}$, we get
\begin{eqnarray}\label{labelB.1}
\hspace{-0.3in}&&\frac{1}{\bar{k}\bar{E}}\int^{\bar{E}}_{m_e}\frac{d
  E}{\bar{E} - E}\Big\{k E\,\Big[\frac{1}{\beta}\,{\ell n}\Big(\frac{1
    + \beta}{1 - \beta}\Big) - 2\Big] - \bar{k}\bar{E}\,
\Big[\frac{1}{\bar{\beta}}\,{\ell n}\Big(\frac{1 + \bar{\beta}}{1 -
    \bar{\beta}}\Big) - 2\Big]\Big\} =\nonumber\\
\hspace{-0.3in}&&=
\frac{2}{\sinh\bar{\varphi}\cosh\bar{\varphi}}\int^{\bar{\varphi}}_0
\frac{d\varphi\,\sinh\varphi}{\cosh\bar{\varphi} -
  \cosh\varphi}\Big[\sinh\varphi \cosh\varphi\,
  \Big(\frac{\varphi}{\tanh\varphi} - 1\Big) - \sinh\bar{\varphi}
  \cosh\bar{\varphi}\, \Big(\frac{\bar{\varphi}}{\tanh\bar{\varphi}} -
  1\Big)\Big]\nonumber\\
\hspace{-0.3in}&&=
\frac{2}{\sinh\bar{\varphi}\cosh\bar{\varphi}}\int^{\bar{\varphi}}_0
\frac{d\varphi\,\sinh\varphi}{\cosh\bar{\varphi} -
  \cosh\varphi}\Big[\varphi\, \cosh^2\varphi - \sinh\varphi
  \cosh\varphi - \bar{\varphi}\,\cosh^2\bar{\varphi} +
  \sinh\bar{\varphi} \cosh\bar{\varphi}\Big] =\nonumber\\
\hspace{-0.3in}&&=
\frac{2}{\sinh\bar{\varphi}\cosh\bar{\varphi}}\int^{\bar{\varphi}}_0
\frac{d\varphi\,\sinh\varphi}{\cosh\bar{\varphi} -
  \cosh\varphi}\Big[\varphi\, (\cosh^2\varphi - \cosh^2\bar{\varphi})
  + (\varphi - \bar{\varphi})\,\cosh^2\bar{\varphi} - (\sinh\varphi
  \cosh\varphi - \sinh\bar{\varphi}
  \cosh\bar{\varphi})\Big]=\nonumber\\
\hspace{-0.3in}&& = I_1 + I_2 + I_3.
\end{eqnarray}
The calculation of the integral $I_1$ runs as follows
\begin{eqnarray}\label{labelB.2}
\hspace{-0.3in}&&I_1 =
\frac{2}{\sinh\bar{\varphi}\cosh\bar{\varphi}}\int^{\bar{\varphi}}_0
\frac{d\varphi\,\sinh\varphi}{\cosh\bar{\varphi} -
  \cosh\varphi}\,\varphi\, (\cosh^2\varphi - \cosh^2\bar{\varphi}) = -
\frac{2}{\sinh\bar{\varphi}\cosh\bar{\varphi}}\int^{\bar{\varphi}}_0
d\varphi\,\varphi\,\sinh\varphi\, (\cosh\varphi + \cosh\bar{\varphi})
=\nonumber\\
\hspace{-0.3in}&&= - 3\,\frac{\bar{\varphi}}{\tanh\bar{\varphi}} +
\frac{1}{2}\,\frac{\bar{\varphi}}{\sinh\bar{\varphi}
  \cosh\bar{\varphi}} + \frac{5}{2} = -\frac{5 + \bar{\beta}^2}{4
  \bar{\beta}}\,{\ell n}\Big(\frac{1 + \bar{\beta}}{1 -
  \bar{\beta}}\Big) + \frac{5}{2},
\end{eqnarray}
The integral $I_1$ is equal to
\begin{eqnarray}\label{labelB.3}
\hspace{-0.3in}I_1 = - \frac{5 + \bar{\beta}^2}{4 \bar{\beta}}\,{\ell
  n}\Big(\frac{1 + \bar{\beta}}{1 - \bar{\beta}}\Big) + \frac{5}{2}.
\end{eqnarray}
For the integral $I_2$ we propose the following calculation
\begin{eqnarray}\label{labelB.4}
\hspace{-0.3in}&&
I_2 = \frac{2}{\sinh\bar{\varphi}\cosh\bar{\varphi}}\int^{\bar{\varphi}}_0
\frac{d\varphi\,\sinh\varphi}{\cosh\bar{\varphi} -
  \cosh\varphi}\,(\varphi - \bar{\varphi})\,\cosh^2\bar{\varphi}
=\frac{2}{\tanh\bar{\varphi}}\int^{\bar{\varphi}}_0
\frac{d\varphi\,\sinh\varphi}{\cosh\bar{\varphi} -
  \cosh\varphi}(\varphi - \bar{\varphi}) = \nonumber\\
\hspace{-0.3in}&&=\frac{2}{\tanh\bar{\varphi}}\Big[ -
  \bar{\varphi}\,{\ell n}(\cosh\bar{\varphi} - 1) +
  \int^{\bar{\varphi}}_0d\varphi\,{\ell n}(\cosh\bar{\varphi} -
  \cosh\varphi)\Big].
\end{eqnarray}
The remained integral in Eq.(\ref{labelB.4}) we calculate by changing
of variable $u = e^{\varphi}$ and $\bar{u} = e^{\bar{\varphi}}$. This
gives
\begin{eqnarray}\label{labelB.5}
\hspace{-0.3in}&& \int^{\bar{\varphi}}_0d\varphi\,{\ell
  n}(\cosh\bar{\varphi} - \cosh\varphi) = - {\ell n}2\,{\ell n}\bar{u}
+ \int^{\bar{u}}_1\frac{du}{u}\,{\ell n}(\bar{u} - u) +
\int^{\bar{u}}_1\frac{du}{u}\,{\ell n}\Big(1 - \frac{1}{\bar{u}u}\Big)
=\nonumber\\
\hspace{-0.3in}&& = - {\ell n}2\,{\ell n}\bar{u} + {\ell n}^2\bar{u} +
\int^1_{1/\bar{u}^2}\frac{dx}{x}\,{\ell n}(1 - x) = - {\ell n}2\,{\ell
  n}\bar{u} + {\ell n}^2\bar{u} + L(1) - L(1/\bar{u}^2) = - {\ell
  n}2\,\bar{\varphi} + \bar{\varphi}^2 + L(1) - L(e^{-2\bar{\varphi}})
=\nonumber\\
\hspace{-0.3in}&& = - {\ell n}2\,\frac{1}{2}\,{\ell n}\Big(\frac{1 +
  \bar{\beta}}{1 - \bar{\beta}}\Big) + \frac{1}{4}\,{\ell
  n}^2\Big(\frac{1 + \bar{\beta}}{1 - \bar{\beta}}\Big) + L(1) -
L\Big(\frac{1 - \bar{\beta}}{1 + \bar{\beta}}\Big) = - {\ell
  n}2\,\frac{1}{2}\,{\ell n}\Big(\frac{1 + \bar{\beta}}{1 -
  \bar{\beta}}\Big) + \frac{1}{4}\,{\ell n}^2\Big(\frac{1 +
  \bar{\beta}}{1 - \bar{\beta}}\Big) + L\Big(\frac{2 \bar{\beta}}{1 +
  \bar{\beta}}\Big)\nonumber\\
\hspace{-0.3in}&& + {\ell n}\Big(\frac{2 \bar{\beta}}{1 +
  \bar{\beta}}\Big)\,{\ell n}\Big(\frac{1 + \bar{\beta}}{1 -
  \bar{\beta}}\Big),
\end{eqnarray}
where we have used Eq.(\ref{labelA.44}). Thus for the integral $I_2$
we obtain the expression
\begin{eqnarray}\label{labelB.6}
\hspace{-0.3in}&&
I_2 = \frac{2}{\sinh\bar{\varphi}\cosh\bar{\varphi}}\int^{\bar{\varphi}}_0
\frac{d\varphi\,\sinh\varphi}{\cosh\bar{\varphi} -
  \cosh\varphi}\,(\varphi - \bar{\varphi})\,\cosh^2\bar{\varphi}
=\frac{2}{\tanh\bar{\varphi}}\int^{\bar{\varphi}}_0
\frac{d\varphi\,\sinh\varphi}{\cosh\bar{\varphi} -
  \cosh\varphi}(\varphi - \bar{\varphi}) = \nonumber\\
\hspace{-0.3in}&&= \frac{2}{\tanh\bar{\varphi}}\Big[ -
  \bar{\varphi}\,{\ell n}(\cosh\bar{\varphi} - 1) - {\ell
    n}2\,\bar{\varphi} + \bar{\varphi}^2 + L(1) -
  L(e^{-2\bar{\varphi}}) \Big] = \frac{2}{\bar{\beta}}\,\Big[ -
  \frac{1}{2}\,{\ell n}\Big(\frac{1 + \bar{\beta}}{1 -
    \bar{\beta}}\Big)\,{\ell n}\Big(\frac{1 - \sqrt{1 -
        \bar{\beta}^2}}{\sqrt{1 - \bar{\beta}^2}}\Big) \nonumber\\
\hspace{-0.3in}&& - \frac{{\ell n}2}{2}\,{\ell n}\Big(\frac{1 +
  \bar{\beta}}{1 - \bar{\beta}}\Big) + \frac{1}{4}\,{\ell
  n}^2\Big(\frac{1 + \bar{\beta}}{1 - \bar{\beta}}\Big) +
L\Big(\frac{2 \bar{\beta}}{1 + \bar{\beta}}\Big) + {\ell
  n}\Big(\frac{2 \bar{\beta}}{1 + \bar{\beta}}\Big)\,{\ell
  n}\Big(\frac{1 + \bar{\beta}}{1 - \bar{\beta}}\Big)\Big]
\end{eqnarray}
where we have used the relation
\begin{eqnarray}\label{labelB.7}
\hspace{-0.3in}{\ell n}\Big(\frac{1 - \sqrt{1 -
    \bar{\beta}^2}}{\sqrt{1 - \bar{\beta}^2}}\Big) = -
\frac{1}{2}\,{\ell n}\Big(\frac{1 + \sqrt{1 - \bar{\beta}^2}}{1 -
  \sqrt{1 - \bar{\beta}^2}}\Big) + {\ell n}\Big(\frac{\bar{\beta}}{1 +
  \bar{\beta}}\Big) + \frac{1}{2}\,{\ell n}\Big(\frac{1 +
  \bar{\beta}}{1 - \bar{\beta}}\Big).
\end{eqnarray}
The integral $I_2$ is equal to
\begin{eqnarray}\label{labelB.8}
\hspace{-0.3in}I_2 = \frac{1}{2\bar{\beta}}\,{\ell n}\Big(\frac{1 +
  \bar{\beta}}{1 - \bar{\beta}}\Big)\,{\ell n}\Big(\frac{1 + \sqrt{1 -
    \bar{\beta}^2}}{1 - \sqrt{1 - \bar{\beta}^2}}\Big) + \frac{2}{\bar{\beta}}\, L\Big(\frac{2
  \bar{\beta}}{1 + \bar{\beta}}\Big) + \frac{1}{\bar{\beta}}\,{\ell
  n}\Big(\frac{2 \bar{\beta}}{1 + \bar{\beta}}\Big)\,{\ell
  n}\Big(\frac{1 + \bar{\beta}}{1 - \bar{\beta}}\Big).
\end{eqnarray}
We calculate the integral $I_3$ as follows
\begin{eqnarray*}
\hspace{-0.3in}&&I_3 = -
\frac{2}{\sinh\bar{\varphi}\cosh\bar{\varphi}}\int^{\bar{\varphi}}_0
\frac{d\varphi\,\sinh\varphi}{\cosh\bar{\varphi} -
  \cosh\varphi}\,(\sinh\varphi\, \cosh\varphi - \sinh\bar{\varphi}\,
\cosh\bar{\varphi}) = -
\frac{2}{\sinh\bar{\varphi}\cosh\bar{\varphi}}\int^{\bar{\varphi}}_0
\frac{d\varphi\,\sinh\varphi}{\cosh\bar{\varphi} -
  \cosh\varphi}\nonumber\\
\hspace{-0.3in}&&\times\,\Big[\sinh\varphi\, (\cosh\varphi -
  \cosh\bar{\varphi}) + \cosh\bar{\varphi}\,(\sinh\varphi -
  \sinh\bar{\varphi})\Big] = -
\frac{2}{\sinh\bar{\varphi}\cosh\bar{\varphi}} \Big[-
\int^{\bar{\varphi}}_0 d\varphi\,\sinh^2\varphi + \cosh\bar{\varphi}
\end{eqnarray*}
\begin{eqnarray}\label{labelB.9}
\hspace{-0.3in}&&\times \int^{\bar{\varphi}}_0
\frac{d\varphi\,\sinh\varphi}{\cosh\bar{\varphi} -
  \cosh\varphi}(\sinh\varphi - \sinh\bar{\varphi})\Big] = -
\frac{2}{\sinh\bar{\varphi}\cosh\bar{\varphi}} \Big\{-\frac{1}{2}
\int^{\bar{\varphi}}_0 d\varphi\,(\cosh 2\varphi - 1) +
\cosh\bar{\varphi}\Big[- \sinh\bar{\varphi}\,{\ell
    n}(\cosh\bar{\varphi} - 1)\nonumber\\
\hspace{-0.3in}&& + \int^{\bar{\varphi}}_0
d\varphi\,\cosh\varphi\,{\ell n}(\cosh\bar{\varphi} -
\cosh\varphi)\Big]\Big\} = -
\frac{2}{\sinh\bar{\varphi}\cosh\bar{\varphi}}
\Big[\frac{1}{2}\,(\bar{\varphi} - \sinh\bar{\varphi}\cosh\bar{\varphi})
  - \sinh\bar{\varphi}\,\cosh\bar{\varphi}\,{\ell
    n}(\cosh\bar{\varphi} - 1)\nonumber\\
\hspace{-0.3in}&& + \cosh\bar{\varphi}\int^{\bar{\varphi}}_0
d\varphi\,\cosh\varphi\,{\ell n}(\cosh\bar{\varphi} -
\cosh\varphi)\Big].
\end{eqnarray}
For the calculation of the last integral in Eq.(\ref{labelB.9}) we
make a change of variables $u = e^{\varphi}$ and $\bar{u} =
e^{\,\bar{\varphi}}$. This gives
\begin{eqnarray}\label{labelB.10}
\hspace{-0.3in}&&\int^{\bar{\varphi}}_0 d\varphi\,\cosh\varphi\,{\ell
  n}(\cosh\bar{\varphi} - \cosh\varphi) = - \frac{1}{2}\,{\ell n}2
\int^{\bar{u}}_1du\,\Big(1 + \frac{1}{u^2}\Big) + \frac{1}{2}
\int^{\bar{u}}_1du\,\Big(1 + \frac{1}{u^2}\Big)\,{\ell n}\Big
    [(\bar{u} - u)\Big(1 - \frac{1}{\bar{u}u}\Big)\Big] =\nonumber\\
\hspace{-0.3in}&&= - {\ell n}2\,\sinh\bar{\varphi} + \frac{1}{2}
\int^{\bar{u}}_1du\,{\ell n} (\bar{u} - u) + \frac{1}{2}
  \int^{\bar{u}}_1\frac{du}{u^2}\,{\ell n}(\bar{u} - u) + \frac{1}{2}
\int^{\bar{u}}_1du\,{\ell n}\Big(1 -
  \frac{1}{\bar{u}u}\Big)  + \frac{1}{2}
  \int^{\bar{u}}_1\frac{du}{u^2}\,{\ell n}\Big(1 -
  \frac{1}{\bar{u}u}\Big).
\end{eqnarray}
The integrals over $u$ are equal to
\begin{eqnarray*}
\hspace{-0.3in}&&\frac{1}{2} \int^{\bar{u}}_1du\,{\ell n} (\bar{u} -
u) = \frac{1}{2}\,(\bar{u} - 1)\,{\ell n}\bar{u} +
\frac{1}{2}\,(\bar{u} - 1)\,{\ell n} \Big(1 - \frac{1}{\bar{u}}\Big) -
  \frac{1}{2}\,(\bar{u} - 1),\nonumber\\
\hspace{-0.3in}&&\frac{1}{2} \int^{\bar{u}}_1\frac{du}{u^2}\,{\ell
  n}(\bar{u} - u) = \frac{1}{2}\,\Big(1 -
\frac{1}{\bar{u}}\Big)\,{\ell n}\bar{u}+ \frac{1}{2}\,\Big(1 -
\frac{1}{\bar{u}}\Big)\,{\ell n} \Big(1 - \frac{1}{\bar{u}}\Big) -
\frac{1}{2\bar{u}}\,{\ell n}\bar{u},\nonumber\\
\end{eqnarray*}
\begin{eqnarray}\label{labelB.11}
\hspace{-0.3in}&&\frac{1}{2} \int^{\bar{u}}_1du\,{\ell n}\Big(1 -
\frac{1}{\bar{u}u}\Big) = \frac{1}{2}\,\Big(\bar{u} -
\frac{1}{\bar{u}}\Big)\,{\ell n}\Big(\bar{u} - \frac{1}{\bar{u}}\Big)
- \frac{1}{2}\,\Big(1 - \frac{1}{\bar{u}}\Big)\,{\ell n}\Big(1 -
\frac{1}{\bar{u}}\Big) - \frac{1}{2}\,\bar{u}\,{\ell
  n}\bar{u},\nonumber\\
\hspace{-0.3in}&&\frac{1}{2} \int^{\bar{u}}_1\frac{du}{u^2}\,{\ell
  n}\Big(1 - \frac{1}{\bar{u}u}\Big) = \frac{1}{2}\,\Big(\bar{u} -
\frac{1}{\bar{u}}\Big)\,{\ell n}\Big(\bar{u} - \frac{1}{\bar{u}}\Big)
- \frac{1}{2}\,(\bar{u} - 1)\,{\ell n}\Big(1 - \frac{1}{\bar{u}}\Big)
- \frac{1}{2}\,\Big(\bar{u} - \frac{1}{\bar{u}}\Big)\,{\ell n}\bar{u}-
\frac{1}{2}\,\Big(1 - \frac{1}{\bar{u}}\Big).
\end{eqnarray}
The sum of the integrals over $u$ is 
\begin{eqnarray}\label{labelB.12}
\hspace{-0.3in}&&\frac{1}{2} \int^{\bar{u}}_1du\,{\ell n} (\bar{u} -
u) + \frac{1}{2} \int^{\bar{u}}_1\frac{du}{u^2}\,{\ell n}(\bar{u} - u)
+ \frac{1}{2} \int^{\bar{u}}_1du\,{\ell n}\Big(1 -
\frac{1}{\bar{u}u}\Big) + \frac{1}{2}
\int^{\bar{u}}_1\frac{du}{u^2}\,{\ell n}\Big(1 -
\frac{1}{\bar{u}u}\Big) =\nonumber\\
\hspace{-0.3in}&&= \Big(\bar{u} - \frac{1}{\bar{u}}\Big)\,{\ell
  n}\Big(\bar{u} - \frac{1}{\bar{u}}\Big) - \frac{1}{2}\Big(\bar{u} +
\frac{1}{\bar{u}}\Big)\,{\ell n}\bar{u} - \frac{1}{2}\Big(\bar{u} -
\frac{1}{\bar{u}}\Big) = (2\,{\ell n}2 - 1)\,\sinh\bar{\varphi} +
2\,\sinh\bar{\varphi}\,{\ell n}\sinh\bar{\varphi} -
\bar{\varphi}\,\cosh\bar{\varphi}.
\end{eqnarray}
The integral Eq.(\ref{labelB.10}) is 
\begin{eqnarray}\label{labelB.13}
\hspace{-0.3in}\int^{\bar{\varphi}}_0 d\varphi\,\cosh\varphi\,{\ell
  n}(\cosh\bar{\varphi} - \cosh\varphi) = ({\ell n}2 - 1)\,\sinh\bar{\varphi} +
2\,\sinh\bar{\varphi}\,{\ell n}\sinh\bar{\varphi} -
\bar{\varphi}\,\cosh\bar{\varphi}.
\end{eqnarray}
In terms of $\bar{\varphi}$ and $\bar{\beta}$ the integral $I_3$ reads
\begin{eqnarray}\label{labelB.14}
\hspace{-0.3in}&&I_3 = 3 - 2\,{\ell n}2 +
\bar{\varphi}\,\Big(\frac{1}{\tanh\bar{\varphi}} +
\tanh\bar{\varphi}\Big) + 2\,{\ell n}(\cosh\bar{\varphi} - 1) -
4\,{\ell n}(\sinh\bar{\varphi}) = \nonumber\\
\hspace{-0.3in}&& = 3 - 2\,{\ell n}2 +
\frac{1}{2}\Big(\frac{1}{\bar{\beta}} + \bar{\beta}\Big)\,{\ell
  n}\Big(\frac{1 + \bar{\beta}}{1 - \bar{\beta}}\Big) + 2\,{\ell
  n}\Big(\frac{1 - \sqrt{1 - \bar{\beta}^2}}{\sqrt{1 -
    \bar{\beta}^2}}\Big) - 4\,{\ell n}\Big(\frac{\bar{\beta}}{\sqrt{1
    - \bar{\beta}^2}}\Big) =\nonumber\\
\hspace{-0.3in}&& = 3 + \Big(-1 + \frac{1}{2\bar{\beta}} +
\frac{\bar{\beta}}{2}\Big)\,{\ell n}\Big(\frac{1 + \bar{\beta}}{1 -
  \bar{\beta}}\Big) - 2\,{\ell n}\Big(\frac{2\bar{\beta}}{1 +
  \bar{\beta}}\Big) - {\ell n}\Big(\frac{1 + \sqrt{1 -
    \bar{\beta}^2}}{1 - \sqrt{1 - \bar{\beta}^2}}\Big).
\end{eqnarray}
For the integral $I_3$ we obtain the expression
\begin{eqnarray}\label{labelB.15}
\hspace{-0.3in}I_3 = 3 + \Big(-1 + \frac{1}{2\bar{\beta}} +
\frac{\bar{\beta}}{2}\Big)\,{\ell n}\Big(\frac{1 + \bar{\beta}}{1 -
  \bar{\beta}}\Big) - 2\,{\ell n}\Big(\frac{2\bar{\beta}}{1 +
  \bar{\beta}}\Big) - {\ell n}\Big(\frac{1 + \sqrt{1 -
    \bar{\beta}^2}}{1 - \sqrt{1 - \bar{\beta}^2}}\Big).
\end{eqnarray}
Substituting Eq.(\ref{labelB.3}), Eq.(\ref{labelB.8}) and
Eq.(\ref{labelB.15}) into Eq.(\ref{labelB.1}) we obtain
\begin{eqnarray}\label{labelB.16}
\hspace{-0.3in}&&\frac{1}{\bar{k}\bar{E}}\int^{\bar{E}}_{m_e}\frac{d
  E}{\bar{E} - E}\Big\{k E\,\Big[\frac{1}{\beta}\,{\ell n}\Big(\frac{1
    + \beta}{1 - \beta}\Big) - 2\Big] - \bar{k}\bar{E}\,
\Big[\frac{1}{\bar{\beta}}\,{\ell n}\Big(\frac{1 + \bar{\beta}}{1 -
    \bar{\beta}}\Big) - 2\Big]\Big\} = \frac{11}{2} + \Big(- 1 -
\frac{3}{4\bar{\beta}} + \frac{\bar{\beta}}{4}\Big)\,{\ell
  n}\Big(\frac{1 + \bar{\beta}}{1 - \bar{\beta}}\Big)\nonumber\\
\hspace{-0.3in}&&+ \frac{2}{\bar{\beta}}\,L\Big(\frac{2\bar{\beta}}{1
  + \bar{\beta}}\Big) + {\ell n}\Big(\frac{2 \bar{\beta}}{1 +
  \bar{\beta}}\Big)\,\Big[\frac{1}{\bar{\beta}}\,{\ell n}\Big(\frac{1
    + \bar{\beta}}{1 - \bar{\beta}}\Big) - 2\Big] + \frac{1}{2}\,{\ell
  n}\Big(\frac{1 + \sqrt{1 - \bar{\beta}^2}}{1 - \sqrt{1 -
    \bar{\beta}^2}}\Big)\,\Big[\frac{1}{\bar{\beta}}\,{\ell
      n}\Big(\frac{1 + \bar{\beta}}{1 - \bar{\beta}}\Big) - 2\Big].
\end{eqnarray}
Now we calculate the integral
\begin{eqnarray}\label{labelB.17}
\hspace{-0.3in}&&\frac{1}{\bar{k}\bar{E}}\int^{\bar{E}}_{m_e}d E\,k
\Big(\frac{\bar{E} + E}{2 E}\,\Big[\frac{1}{\beta}\,{\ell
    n}\Big(\frac{1 + \beta}{1 - \beta}\Big) - 2\Big] + \frac{\bar{E} -
  E}{E}\Big) =
\frac{1}{\sinh\bar{\varphi}\cosh\bar{\varphi}}\int^{\bar{\varphi}_0}
d\varphi\,\sinh^2\varphi\,\Big[\frac{\cosh\bar{\varphi} +
    \cosh\varphi}{\cosh\varphi}\,\Big(\frac{\varphi}{\tanh\varphi} -
  1\Big)\nonumber\\
\hspace{-0.3in}&& + \frac{\cosh\bar{\varphi} -
  \cosh\varphi}{\cosh\varphi}\Big] =
\frac{1}{\sinh\bar{\varphi}\cosh\bar{\varphi}}\int^{\bar{\varphi}}
d\varphi\,\Big[\varphi\,\sinh\varphi\,(\cosh\bar{\varphi} +
  \cosh\varphi) - 2\,\sinh^2\varphi\Big] =
\frac{1}{\sinh\bar{\varphi}\cosh\bar{\varphi}}\int^{\bar{\varphi}}
d\varphi\,\Big[\cosh\bar{\varphi}\nonumber\\
\hspace{-0.3in}&&\times\,\varphi\,\sinh\varphi +
\frac{1}{2}\,\varphi\,\sinh 2\varphi - \cosh 2\varphi + 1\Big] =
\frac{1}{\sinh\bar{\varphi}\cosh\bar{\varphi}}\Big[\bar{\varphi}\,
  \Big(\frac{3}{4}+ \frac{3}{2}\,\cosh^2\bar{\varphi}\Big) -
  \frac{9}{4}\,\sinh\bar{\varphi}\,\cosh\bar{\varphi}\Big]
=\nonumber\\
\hspace{-0.3in}&&= \bar{\varphi}\,\Big(
\frac{9}{4}\,\frac{1}{\tanh\bar{\varphi}} -
\frac{3}{4}\,\tanh\bar{\varphi}\Big) - \frac{9}{4} =
\frac{1}{2}\,\Big( \frac{9}{4}\,\frac{1}{\bar{\beta}} -
\frac{3}{4}\,\bar{\beta}\Big)\,{\ell n}\Big(\frac{1 + \bar{\beta}}{1 -
  \bar{\beta}}\Big) - \frac{9}{4}.
\end{eqnarray}
Thus, the integral under consideration is equal to
\begin{eqnarray}\label{labelB.18}
\hspace{-0.3in}\frac{1}{\bar{k}\bar{E}}\int^{\bar{E}}_{m_e}d E\,k
\Big(\frac{\bar{E} + E}{2 E}\,\Big[\frac{1}{\beta}\,{\ell
    n}\Big(\frac{1 + \beta}{1 - \beta}\Big) - 2\Big] + \frac{\bar{E} -
  E}{E}\Big) = \frac{1}{2}\,\Big( \frac{9}{4}\,\frac{1}{\bar{\beta}} -
\frac{3}{4}\,\bar{\beta}\Big)\,{\ell n}\Big(\frac{1 + \bar{\beta}}{1 -
  \bar{\beta}}\Big) - \frac{9}{4}.
\end{eqnarray}
As a result, the function $f^{(\gamma)}_A(\bar{E})$ is given by
\begin{eqnarray*}
\hspace{-0.3in}f^{(\gamma)}_A(\bar{E}) &=& \frac{13}{4} + \Big(- 1 +
\frac{3}{8\bar{\beta}} - \frac{\bar{\beta}}{8}\,\Big)\,{\ell
  n}\Big(\frac{1 + \bar{\beta}}{1 - \bar{\beta}}\Big) +
\frac{2}{\bar{\beta}}\,L\Big(\frac{2\bar{\beta}}{1 + \bar{\beta}}\Big)
+ {\ell n}\Big(\frac{2 \bar{\beta}}{1 +
  \bar{\beta}}\Big)\,\Big[\frac{1}{\bar{\beta}}\,{\ell n}\Big(\frac{1
    + \bar{\beta}}{1 - \bar{\beta}}\Big) - 2\Big]\nonumber\\
\end{eqnarray*}
\begin{eqnarray}\label{labelB.19}
 \hspace{-0.3in}&+& \frac{1}{2}\,{\ell
  n}\Big(\frac{1 + \sqrt{1 - \bar{\beta}^2}}{1 - \sqrt{1 -
    \bar{\beta}^2}}\Big)\,\Big[\frac{1}{\bar{\beta}}\,{\ell
      n}\Big(\frac{1 + \bar{\beta}}{1 - \bar{\beta}}\Big) - 2\Big].
\end{eqnarray}
Summing up Eq.(\ref{labelA.71}) and Eq.(\ref{labelB.19}) for the
function $f_A(\bar{E})$ we obtain the following analytical expression
\begin{eqnarray}\label{labelB.20}
\hspace{-0.3in}f_A(\bar{E}) &=& \frac{3}{2}\,{\ell
  n}\Big(\frac{m_p}{m_e}\Big) + \frac{23}{8} + 2\,{\ell n}\Big(\frac{2
  \bar{\beta}}{1 + \bar{\beta}}\Big)\,
\Big[\frac{1}{\bar{\beta}}\,{\ell n}\Big(\frac{1 + \bar{\beta}}{1 -
    \bar{\beta}}\Big) - 2\Big] +
\frac{4}{\bar{\beta}}\,L\Big(\frac{2\bar{\beta}}{1 + \bar{\beta}}\Big)
+ \frac{3}{8}\,\Big(\bar{\beta}^2 +
\frac{7}{3}\Big)\,\frac{1}{\bar{\beta}}\,{\ell n}\Big(\frac{1 +
  \bar{\beta}}{1 - \bar{\beta}}\Big)\nonumber\\
 \hspace{-0.3in}&-& 2\,{\ell n}\Big(\frac{1 + \bar{\beta}}{1 -
   \bar{\beta}}\Big).
\end{eqnarray}
Thus, we have shown that the function $f_A(\bar{E})$ coincides with
the function $f(\bar{E})$, calculated by Vogel \cite{IBD1}, Fayans
\cite{IBD2} and Raha, Myhrer and Kudobera \cite{IBD5} (see
Eq.(\ref{labelA.76})).

\section{Appendix C: Analytical expression for functions
$f^{(\gamma)}_B(\bar{E})$ and $f_B(\bar{E})$ }
\renewcommand{\theequation}{C-\arabic{equation}}
\setcounter{equation}{0}

In this Appendix we give a detailed calculation of the integrals,
defining the function $f^{(\gamma)}_B(\bar{E})$. Making a change of
variables $E = m_e\cosh\varphi$ and $\bar{E} = m_e \cosh\bar{\varphi}$
for the first integral in Eq.(\ref{labelA.65}) we obtain
\begin{eqnarray}\label{labelC.1}
\hspace{-0.3in}&&\frac{1}{\bar{k}\bar{E}\bar{\beta}}\int^{\bar{E}}_{m_e}\frac{d
  E}{\bar{E} - E}\Big\{k E\beta\,\Big[\frac{1}{\beta}\,{\ell
    n}\Big(\frac{1 + \beta}{1 - \beta}\Big) - 2\Big] -
\bar{k}\bar{E}\bar{\beta}\, \Big[\frac{1}{\bar{\beta}}\,{\ell
    n}\Big(\frac{1 + \bar{\beta}}{1 - \bar{\beta}}\Big) - 2\Big]\Big\}
= \frac{2}{\sinh^2\bar{\varphi}}\int^{\bar{\varphi}}_0
\frac{d\varphi\,\sinh\varphi}{\cosh\bar{\varphi} -
  \cosh\varphi}\,\Big[\sinh^2\varphi\nonumber\\
\hspace{-0.3in}&&\times \, \Big(\frac{\varphi}{\tanh\varphi} - 1\Big)
- \sinh^2\bar{\varphi} \,
\Big(\frac{\bar{\varphi}}{\tanh\bar{\varphi}} - 1\Big)\Big] =
\frac{2}{\sinh^2\bar{\varphi}}\int^{\bar{\varphi}}_0
\frac{d\varphi\,\sinh\varphi}{\cosh\bar{\varphi} - \cosh\varphi}\Big[-
  \varphi\,\sinh\varphi\,(\cosh\bar{\varphi} - \cosh\varphi) +
  \cosh\bar{\varphi}\nonumber\\
\hspace{-0.3in}&&\times\,(\varphi - \bar{\varphi})\,\sinh\varphi +
\bar{\varphi}\,\cosh\bar{\varphi}\,(\sinh\varphi - \sinh\bar{\varphi})
+ (\cosh\bar{\varphi} - \cos\varphi)\, (\cosh\bar{\varphi} +
\cosh\varphi)\Big] = I_1 + I_2 + I_3 + I_4.
\end{eqnarray}
The calculation of the integral $I_1$:
\begin{eqnarray}\label{labelC.2}
\hspace{-0.3in}&&I_1 = -
\frac{2}{\sinh^2\bar{\varphi}}\int^{\bar{\varphi}}_0
d\varphi\,\varphi\,\sinh^2\varphi = -
\frac{2}{\sinh^2\bar{\varphi}}\Big(\bar{\varphi}\,\frac{1}{4}\,\sinh
2\bar{\varphi} - \frac{1}{8}\,\cosh 2\bar{\varphi} + \frac{1}{8} -
\frac{1}{4}\,\bar{\varphi}^2\Big) =\nonumber\\
\hspace{-0.3in}&&= \frac{1}{2} -
\frac{\bar{\varphi}}{\tanh\bar{\varphi}} +
\frac{1}{2}\,\frac{\bar{\varphi}^2}{\sinh^2\bar{\varphi}} =
\frac{1}{2} - \frac{1}{2\bar{\beta}}\,{\ell n}\Big(\frac{1 +
    \bar{\beta}}{1 - \bar{\beta}}\Big) + \frac{1 -
    \bar{\beta}^2}{8\bar{\beta}^2}\,{\ell n}^2\Big(\frac{1 +
    \bar{\beta}}{1 - \bar{\beta}}\Big).
\end{eqnarray}
The integral $I_1$ is equal to
\begin{eqnarray}\label{labelC.3}
\hspace{-0.3in}&&I_1 = \frac{1}{2} - \frac{1}{2\bar{\beta}}\,{\ell
  n}\Big(\frac{1 + \bar{\beta}}{1 - \bar{\beta}}\Big) + \frac{1 -
  \bar{\beta}^2}{8\bar{\beta}^2}\,{\ell n}^2\Big(\frac{1 +
  \bar{\beta}}{1 - \bar{\beta}}\Big).
\end{eqnarray}
The calculation of the integral $I_2$:
\begin{eqnarray}\label{labelC.4}
\hspace{-0.3in}&&I_2 =
\frac{2}{\sinh\bar{\varphi}\tanh\bar{\varphi}}\int^{\bar{\varphi}}_0
\frac{d\varphi\,(\varphi - \bar{\varphi})}{\cosh\bar{\varphi} -
  \cosh\varphi}\,\sinh^2\varphi =
\frac{2}{\sinh\bar{\varphi}\tanh\bar{\varphi}}\Big[\int^{\bar{\varphi}}_0
  \frac{d\varphi\,(\varphi - \bar{\varphi})}{\cosh\bar{\varphi} -
    \cosh\varphi}\,(\cosh^2\varphi - \cosh^2\bar{\varphi}) + \sinh^2\bar{\varphi}\nonumber\\
\hspace{-0.3in}&&\times \int^{\bar{\varphi}}_0
\frac{d\varphi\,(\varphi - \bar{\varphi})}{\cosh\bar{\varphi} -
  \cosh\varphi}\Big] = \frac{2}{\sinh\bar{\varphi}\tanh\bar{\varphi}}
\Big[- \int^{\bar{\varphi}}_0 d\varphi\,(\varphi -
  \bar{\varphi})(\cosh\bar{\varphi} + \cosh\varphi) +
  \sinh^2\bar{\varphi}\int^{\bar{\varphi}}_0 \frac{d\varphi\,(\varphi
    - \bar{\varphi})}{\cosh\bar{\varphi} - \cosh\varphi}\Big]
=\nonumber\\
\hspace{-0.3in}&& =
\frac{2}{\sinh\bar{\varphi}\tanh\bar{\varphi}}\,\Big[- 1 +
  \cosh\bar{\varphi} +
  \frac{1}{2}\,\bar{\varphi}^2\,\cosh\bar{\varphi} +
  \sinh^2\bar{\varphi}\int^{\bar{\varphi}}_0 \frac{d\varphi\,(\varphi
    - \bar{\varphi})}{\cosh\bar{\varphi} - \cosh\varphi}\Big].
\end{eqnarray}
The calculation of the integral in Eq.(\ref{labelC.4}) we carry out by
changing variables $u = e^{\varphi}$ and $\bar{u} = e^{\bar{\varphi}}$. We get
\begin{eqnarray}\label{labelC.5}
\hspace{-0.3in}&&\int^{\bar{\varphi}}_0 \frac{d\varphi\,(\varphi -
  \bar{\varphi})}{\cosh\bar{\varphi} - \cosh\varphi} =
2\int^{\bar{u}}_1\frac{\displaystyle du\,{\ell
    n}\Big(\frac{u}{\bar{u}}\Big)}{\displaystyle (\bar{u} - u)\Big(u -
  \frac{1}{\bar{u}}\Big)} =
\frac{1}{\sinh\bar{\varphi}}\int^{\bar{u}}_1\frac{\displaystyle
  du\,{\ell n}\Big(\frac{u}{\bar{u}}\Big)}{\displaystyle \bar{u} - u}
- \frac{1}{\sinh\bar{\varphi}}\int^{\bar{u}}_1\frac{\displaystyle
  du\,{\ell n}\Big(\frac{\bar{u}}{u}\Big)}{\displaystyle u\Big(1 -
  \frac{1}{\bar{u}u}\Big)} =\nonumber\\
\hspace{-0.3in}&&=
\frac{1}{\sinh\bar{\varphi}}\int^{1}_{1/\bar{u}}\frac{ dt\,{\ell
    n}t}{1 - t} +
\frac{2\bar{\varphi}}{\sinh\bar{\varphi}}\int^{1/\bar{u}^2}_{1/\bar{u}}\frac{
  dt }{t(1 - t)} +
\frac{1}{\sinh\bar{\varphi}}\int^{1/\bar{u}^2}_{1/\bar{u}}\frac{
  dt\,{\ell n}t }{t(1 - t)} = \frac{1}{\sinh\bar{\varphi}}\,L(1 -
e^{\,-\bar{\varphi}}) - \frac{2\bar{\varphi}^2}{\sinh\bar{\varphi}}\nonumber\\
\hspace{-0.3in}&& - \frac{2\bar{\varphi}}{\sinh\bar{\varphi}}\,{\ell
  n}(1 + e^{\, - \bar{\varphi}}) +
\frac{3}{2}\,\frac{\bar{\varphi}^2}{\sinh\bar{\varphi}} +
\frac{1}{\sinh\bar{\varphi}}\,L(1 - e^{\,-\bar{\varphi}}) -
\frac{1}{\sinh\bar{\varphi}}\,L(1 - e^{\,- 2\bar{\varphi}}) = -
\frac{1}{2}\,\frac{\bar{\varphi}^2}{\sinh\bar{\varphi}} -
\frac{2\bar{\varphi}}{\sinh\bar{\varphi}}\,{\ell n}(1 + e^{\, -
  \bar{\varphi }})\nonumber\\
\hspace{-0.3in}&&+ \frac{2}{\sinh\bar{\varphi}}\,L(1 - e^{\,-\bar{\varphi}}) - \frac{1}{\sinh\bar{\varphi}}\,L(1 - e^{\,-
  2\bar{\varphi}}).
\end{eqnarray}
In terms $\bar{\varphi}$ the integral $I_2$ is 
\begin{eqnarray}\label{labelC.6}
\hspace{-0.3in}&&I_2 =
\frac{2}{\sinh\bar{\varphi}\tanh\bar{\varphi}}\,\Big[- 1 +
  \cosh\bar{\varphi} - 2\,\bar{\varphi}\,\sinh\bar{\varphi}\,{\ell
    n}(1 + e^{\,- \bar{\varphi}}) +
  \frac{1}{2}\,\bar{\varphi}^2\,(\cosh\bar{\varphi} -
  \sinh\bar{\varphi}) + 2\,\sinh\bar{\varphi}\,L(1 - e^{\,-\bar{\varphi}})\nonumber\\
\hspace{-0.3in}&& - \sinh\bar{\varphi}\,L(1 -
e^{\,-2\bar{\varphi}})\Big] = -
\frac{2}{\sinh\bar{\varphi}\tanh\bar{\varphi}} +
\frac{2}{\tanh^2\bar{\varphi}} - \frac{4
  \bar{\varphi}}{\tanh\bar{\varphi}}\,{\ell n}(1 + e^{\,-
  \bar{\varphi}}) +
\bar{\varphi}^2\,\Big(\frac{1}{\tanh^2\bar{\varphi}} -
\frac{1}{\tanh\bar{\varphi}}\Big)\nonumber\\
\hspace{-0.3in}&& + \frac{4}{\tanh\bar{\varphi}}\,L(1 -
e^{\,-\bar{\varphi}})- \frac{2}{\tanh\bar{\varphi}}\,L(1 -
e^{\,-2\bar{\varphi}}).
\end{eqnarray}
The integral $I_2$ is equal to
\begin{eqnarray}\label{labelC.7}
\hspace{-0.3in}&&I_2 = 2\,\frac{1 - \sqrt{1 -
    \bar{\beta}^2}}{\bar{\beta}^2} - \frac{2}{\bar{\beta}}\,{\ell
  n}\Big(\frac{1 + \bar{\beta}}{1 - \bar{\beta}}\Big)\,\,{\ell
  n}\Big(1 + \sqrt{\frac{1 - \bar{\beta}}{1 + \bar{\beta}}}\;\Big) +
\frac{1 - \bar{\beta}}{4 \bar{\beta}^2}\,\,{\ell n}^2\Big(\frac{1 +
  \bar{\beta}}{1 - \bar{\beta}}\Big) + \frac{4}{\bar{\beta}}\,L\Big(1
- \sqrt{\frac{1 - \bar{\beta}}{1 + \bar{\beta}}}\;\Big)\nonumber\\
\hspace{-0.3in}&& - \frac{2}{\bar{\beta}}\,L\Big(\frac{2
  \bar{\beta}}{1 + \bar{\beta}}\Big).
\end{eqnarray}
The calculation of the integral $I_3$:
\begin{eqnarray}\label{labelC.8}
\hspace{-0.3in}&&I_3 =
\frac{2\,\bar{\varphi}}{\sinh\bar{\varphi}\tanh\bar{\varphi}}
\int^{\bar{\varphi}}_0
\frac{d\varphi\,\sinh\varphi}{\cosh\bar{\varphi} -
  \cosh\varphi}\,(\sinh\varphi - \sinh\bar{\varphi}) =
\frac{2\,\bar{\varphi}}{\sinh\bar{\varphi}\tanh\bar{\varphi}}\Big[ -
  \sinh\bar{\varphi}\,{\ell n}(\cosh\bar{\varphi} - 1)\nonumber\\
\hspace{-0.3in}&& + \int^{\bar{\varphi}}_0
d\varphi\,\cosh\varphi\,{\ell n}(\cosh\bar{\varphi} -
\cosh\varphi)\Big] =
\frac{2\,\bar{\varphi}}{\sinh\bar{\varphi}\tanh\bar{\varphi}}\Big[ -
  \sinh\bar{\varphi}\,{\ell n}(\cosh\bar{\varphi} - 1)+ ({\ell n}2 -
  1)\,\sinh\bar{\varphi}\nonumber\\
\hspace{-0.3in}&& + 2\,\sinh\bar{\varphi}\,{\ell n}\sinh\bar{\varphi}
- \bar{\varphi}\,\cosh\bar{\varphi}\Big],
\end{eqnarray}
where we have used Eq.(\ref{labelB.13}). In terms of $\bar{\varphi}$
the integral $I_3$ is equal to
\begin{eqnarray}\label{labelC.9}
\hspace{-0.3in}&&I_3 = -
\frac{2\,\bar{\varphi}}{\tanh\bar{\varphi}}\,{\ell
  n}(\cosh\bar{\varphi} - 1) + 2\,({\ell n}2 -
1)\,\frac{\bar{\varphi}}{\tanh\bar{\varphi}} +
\frac{4\,\bar{\varphi}}{\tanh\bar{\varphi}}\,{\ell
  n}(\sinh\bar{\varphi}) -
\frac{2\,\bar{\varphi}^2}{\tanh^2\bar{\varphi}}.
\end{eqnarray}
For the integral $I_3$ we obtain the following expression
\begin{eqnarray*}
\hspace{-0.3in}I_3 &=& -\frac{1}{\bar{\beta}}\,{\ell n}\Big(\frac{1 +
  \bar{\beta}}{1 - \bar{\beta}}\Big)\,{\ell n}\Big(\frac{1 - \sqrt{1 -
    \bar{\beta}^2}}{\sqrt{1 - \bar{\beta}^2}}\Big) + ({\ell n}2 -
1)\,\frac{1}{\bar{\beta}}\,{\ell n}\Big(\frac{1 + \bar{\beta}}{1 -
  \bar{\beta}}\Big) + \frac{2}{\bar{\beta}}\,{\ell n}\Big(\frac{1 +
  \bar{\beta}}{1 - \bar{\beta}}\Big)\,{\ell
  n}\Big(\frac{\bar{\beta}}{\sqrt{1 - \bar{\beta}^2}}\Big)
\end{eqnarray*}
\begin{eqnarray}\label{labelC.10}
\hspace{-0.3in}&&- \frac{1}{2\bar{\beta}^2}\,{\ell n}^2\Big(\frac{1 +
  \bar{\beta}}{1 - \bar{\beta}}\Big) = \frac{1}{2\bar{\beta}}\,{\ell
  n}\Big(\frac{1 + \bar{\beta}}{1 - \bar{\beta}}\Big)\,{\ell
  n}\Big(\frac{1 + \sqrt{1 - \bar{\beta}^2}}{1 - \sqrt{1 -
    \bar{\beta}^2}}\Big) + \frac{1}{\bar{\beta}}\,{\ell n}\Big(\frac{2
  \bar{\beta}}{1 + \bar{\beta}}\Big)\,{\ell n}\Big(\frac{1 +
  \bar{\beta}}{1 - \bar{\beta}}\Big) - \frac{1 -
  \bar{\beta}}{2\bar{\beta}^2}\,{\ell n}^2\Big(\frac{1 +
  \bar{\beta}}{1 - \bar{\beta}}\Big) \nonumber\\
\hspace{-0.3in}&&- \frac{1}{\bar{\beta}}\,{\ell n}\Big(\frac{1 +
  \bar{\beta}}{1 - \bar{\beta}}\Big).
\end{eqnarray}
The calculation of the integral $I_4$ is as follows:
\begin{eqnarray}\label{labelC.11}
\hspace{-0.3in}&&I_4 =
\frac{2}{\sinh^2\bar{\varphi}}\int^{\bar{\varphi}}_0
d\varphi\,\sinh\varphi\, (\cosh\bar{\varphi} + \cosh\varphi) =
\frac{2}{\sinh^2\bar{\varphi}}\Big[1 - \cosh\bar{\varphi} +
  \frac{3}{2}\,\sinh^2\bar{\varphi}\Big] = 3 - 2\,\frac{\sqrt{1 -
  \bar{\beta}^2}}{\bar{\beta}^2}\Big(1 - \sqrt{1 -
  \bar{\beta}^2}\,\Big).\nonumber\\
\hspace{-0.3in}&&
\end{eqnarray}
The integral $I_4$ is equal to
\begin{eqnarray}\label{labelC.12}
\hspace{-0.3in}I_4 =  3 - 2\,\frac{\sqrt{1 -
  \bar{\beta}^2}}{\bar{\beta}^2}\Big(1 - \sqrt{1 -
  \bar{\beta}^2}\,\Big).
\end{eqnarray}
Summing up the contributions we obtain 
\begin{eqnarray}\label{labelC.13}
\hspace{-0.3in}&&\frac{1}{\bar{k}\bar{E}\bar{\beta}}\int^{\bar{E}}_{m_e}\frac{d
  E}{\bar{E} - E}\Big\{k E\beta\,\Big[\frac{1}{\beta}\,{\ell
    n}\Big(\frac{1 + \beta}{1 - \beta}\Big) - 2\Big] -
\bar{k}\bar{E}\bar{\beta}\, \Big[\frac{1}{\bar{\beta}}\,{\ell
    n}\Big(\frac{1 + \bar{\beta}}{1 - \bar{\beta}}\Big) - 2\Big]\Big\}
= \frac{7}{2} + \frac{2}{\bar{\beta}^2}\,(1 -
\sqrt{1 - \bar{\beta}^2}\,)^2 \nonumber\\
\hspace{-0.3in}&& - \frac{2}{\bar{\beta}}\,{\ell n}\Big(\frac{1 +
  \bar{\beta}}{1 - \bar{\beta}}\Big)\,\,{\ell n}\Big(1 + \sqrt{\frac{1
    - \bar{\beta}}{1 + \bar{\beta}}}\;\Big) + \frac{1}{2
  \bar{\beta}}\,{\ell n}\Big(\frac{1 + \bar{\beta}}{1 -
  \bar{\beta}}\Big)\,{\ell n}\Big(\frac{1 + \sqrt{1 -
    \bar{\beta}^2}}{1 - \sqrt{1 - \bar{\beta}^2} }\Big)
+ \frac{1}{\bar{\beta}}\,{\ell n}\Big(\frac{2\bar{\beta}}{1 +
  \bar{\beta}}\Big)\,{\ell n}\Big(\frac{1 + \bar{\beta}}{1 -
  \bar{\beta}}\Big) \nonumber\\
\hspace{-0.3in}&& - \frac{3}{2 \bar{\beta}}\,{\ell n}\Big(\frac{1 +
  \bar{\beta}}{1 - \bar{\beta}}\Big) - \frac{(1 -
  \bar{\beta})^2}{8 \bar{\beta}^2}\,{\ell n}^2\Big(\frac{1 +
  \bar{\beta}}{1 - \bar{\beta}}\Big) + \frac{4}{\bar{\beta}}\,L\Big(1
- \sqrt{\frac{1 - \bar{\beta}}{1 + \bar{\beta}}}\;\Big) -
\frac{2}{\bar{\beta}}\,L\Big(\frac{2\bar{\beta}}{1 +
  \bar{\beta}}\Big).
\end{eqnarray}
The calculation of the second integral Eq.(\ref{labelA.65})is as follows:
\begin{eqnarray}\label{labelC.14}
\hspace{-0.3in}&&\frac{1}{\bar{k}\bar{E}\bar{\beta}}\int^{\bar{E}}_{m_e}
d E\, \frac{1}{2}\,(\bar{E} + E)\Big[\frac{1}{\beta}\,{\ell
    n}\Big(\frac{1 + \beta}{1 - \beta}\Big) - 2\Big] =
\frac{1}{\sinh^2\bar{\varphi}}\int^{\bar{\varphi}}_0 d\varphi\,
\Big[\cosh\bar{\varphi}\,\varphi\,\cosh\varphi +
  \varphi\,\cosh^2\varphi - \cosh\bar{\varphi}\,\sinh\varphi
  \nonumber\\
\hspace{-0.3in}&&- \sinh\varphi\,\cosh\varphi\Big] =
  \frac{1}{\sinh^2\bar{\varphi}}\Big[- 2 + 2\,\cosh\bar{\varphi} +
    \frac{1}{4}\,\bar{\varphi}^2 - \frac{11}{4}\,\sinh^2\bar{\varphi}
    + \frac{3}{2}\,\bar{\varphi}\, \sinh\bar{\varphi}\,
    \cosh\bar{\varphi}\Big] = - \frac{11}{4} +
  2\,\frac{\sqrt{1 - \bar{\beta}^2}}{\bar{\beta}^2}\nonumber\\ 
\hspace{-0.3in}&&\times\,\Big(1 - \sqrt{1 - \bar{\beta}^2}\Big) + 
\frac{1 - \bar{\beta}^2}{16\bar{\beta}^2}\,{\ell n}^2\Big(\frac{1 +
  \bar{\beta}}{1 - \bar{\beta}}\Big) + \frac{3}{4\bar{\beta}}\,{\ell
  n}\Big(\frac{1 + \bar{\beta}}{1 - \bar{\beta}}\Big).
\end{eqnarray}
Summing up the contributions we obtain the function $f^{(\gamma)}_B(\bar{E})$
\begin{eqnarray}\label{labelC.15}
\hspace{-0.3in}f^{(\gamma)}_B(\bar{E}) &=& \frac{3}{4} +
\frac{2}{\bar{\beta}^2}\,(1 - \sqrt{1 - \bar{\beta}^2}) -
\frac{2}{\bar{\beta}}\,{\ell n}\Big(\frac{1 + \bar{\beta}}{1 -
  \bar{\beta}}\Big)\,{\ell n}\Big(1 + \sqrt{\frac{1 - \bar{\beta}}{1 +
    \bar{\beta}}}\;\Big) + \frac{1}{2 \bar{\beta}}\,{\ell n}\Big(\frac{1 +
  \bar{\beta}}{1 - \bar{\beta}}\Big)\,{\ell n}\Big(\frac{1 + \sqrt{1 -
    \bar{\beta}^2}}{1 - \sqrt{1 - \bar{\beta}^2} }\Big)\nonumber\\
\hspace{-0.3in}&+& \frac{1}{\bar{\beta}}\,{\ell
  n}\Big(\frac{2\bar{\beta}}{1 + \bar{\beta}}\Big)\,{\ell
  n}\Big(\frac{1 + \bar{\beta}}{1 - \bar{\beta}}\Big) -
\frac{3}{4\bar{\beta}}\,{\ell n}\Big(\frac{1 + \bar{\beta}}{1 -
  \bar{\beta}}\Big) - \frac{1 - 4\,\bar{\beta} + 3 \bar{\beta}^2}{16
  \bar{\beta}^2}\,{\ell n}^2\Big(\frac{1 + \bar{\beta}}{1 -
  \bar{\beta}}\Big) + \frac{4}{\bar{\beta}}\,L\Big(1 - \sqrt{\frac{1 -
    \bar{\beta}}{1 + \bar{\beta}}}\;\Big) \nonumber\\
\hspace{-0.3in}&-& \frac{2}{\bar{\beta}}\,L\Big(\frac{2\bar{\beta}}{1
  + \bar{\beta}}\Big).
\end{eqnarray}
As a result the function $f_B(\bar{E})$ is given by
\begin{eqnarray}\label{labelC.16}
\hspace{-0.3in}&&f_B(\bar{E}) = \frac{3}{2}\,{\ell
  n}\Big(\frac{m_p}{m_e}\Big) + \frac{3}{8} +
\frac{2}{\bar{\beta}^2}\,(1 - \sqrt{1 - \bar{\beta}^2}) +
\frac{4}{\bar{\beta}}\,L\Big(1 - \sqrt{\frac{1 - \bar{\beta}}{1 +
    \bar{\beta}}}\;\Big) - \frac{1 - 4\,\bar{\beta} + 3
  \bar{\beta}^2}{16 \bar{\beta}^2}\,{\ell n}^2\Big(\frac{1 +
  \bar{\beta}}{1 - \bar{\beta}}\Big)\nonumber\\
\hspace{-0.3in}&&- {\ell n}\Big[\Big(\frac{1 +
    \bar{\beta}}{2\bar{\beta}}\Big)\,\frac{\sqrt{1 + \bar{\beta}} +
    \sqrt{1 - \bar{\beta}}}{\sqrt{1 + \bar{\beta}} - \sqrt{1 -
      \bar{\beta}}}\,\Big]\,\Big[\frac{1}{\bar{\beta}}\,{\ell
    n}\Big(\frac{1 + \bar{\beta}}{1 - \bar{\beta}}\Big) - 2\Big] +
\frac{1 - 4\bar{\beta}}{4\bar{\beta}}\,{\ell n}\Big(\frac{1 +
  \bar{\beta}}{1 - \bar{\beta}}\Big),
\end{eqnarray}
where we have used the relations
\begin{eqnarray*}
\hspace{-0.3in}{\ell n}\Big(\frac{1 + \sqrt{1 - \bar{\beta}^2}}{1 -
  \sqrt{1 - \bar{\beta}^2} }\Big) &=& 2\,{\ell n}\Big(\frac{\sqrt{1 +
    \bar{\beta}} + \sqrt{1 - \bar{\beta}}}{\sqrt{1 + \bar{\beta}} -
  \sqrt{1 - \bar{\beta}}}\Big),
\end{eqnarray*}
\begin{eqnarray}\label{labelC.17}
\hspace{-0.3in}{\ell n}\Big(1 + \sqrt{\frac{1 - \bar{\beta}}{1 +
    \bar{\beta}}}\;\Big) &=& \frac{1}{2}\,{\ell n}\Big(\frac{\sqrt{1 +
    \bar{\beta}} + \sqrt{1 - \bar{\beta}}}{\sqrt{1 + \bar{\beta}} -
  \sqrt{1 - \bar{\beta}}}\Big) + \frac{1}{2}\,{\ell
  n}\Big(\frac{2\bar{\beta}}{1 + \bar{\beta}}\Big).
\end{eqnarray}
The radiative corrections, defined by the function
$(\alpha/\pi)\,f_B(\bar{E})$, agrees well with the results, obtained
by Fukugita and Kubota \cite{IBD4} and Raha, Myhrer and Kubodera
\cite{IBD5}.

\section{Appendix D: Amplitude and cross section of radiative inverse 
$\beta$--decay with account for contribution of proton--photon
  interaction} \renewcommand{\theequation}{D-\arabic{equation}}
\setcounter{equation}{0}

In this Appendix we calculate the differential cross section for the
radiative inverse $\beta$--decay by taking into account the
contribution of the proton--photon interaction. Such a contribution is
important for a gauge invariant calculation of the amplitude and the
cross section for the radiative inverse $\beta$--decay.

The amplitude of the radiative inverse $\beta$--decay, taking into
account the contribution of the proton--photon interactions, is
defined by (see Eq.(\ref{labelA.53}))
\begin{eqnarray}\label{labelD.1}
\hspace{-0.3in&&}M(\bar{\nu}_e p \to n e^+ \gamma)_{\lambda'} =
\varepsilon^{*\alpha}_{\lambda'} M_{\alpha}(\bar{\nu}_e p \to n e^+
\gamma) = e\,\frac{G_F}{\sqrt{2}}\,V_{ud} \nonumber\\
\hspace{-0.3in}&&\times \Big\{[\bar{u}_n W^{\mu}u_p]
\Big[\bar{v}_{\bar{\nu}}O_{\mu}\,\frac{1}{m_e + \hat{k} + \hat{q} -
    i0}\,\hat{\varepsilon}^*_{\lambda'}\,v\Big]-
\Big[\bar{u}_nW^{\mu}\,\frac{1}{m_p - \hat{k}_p + \hat{q} -
    i0}\,\hat{\varepsilon}^*_{\lambda'}\,u_p\Big][\bar{v}_{\bar{\nu}}O_{\mu}
  v]\Big\},
\end{eqnarray}
where $q = (\omega, \vec{q} = \omega\,\vec{n}\,)$ and
$\varepsilon^*_{\lambda'}$ are the 4--momentum and 4--polarisation
vector of a photon with $\lambda' = 1,2$. The polarisation vector
$\varepsilon^*_{\lambda'}$ obeys the constraint $q\cdot
\varepsilon^*_{\lambda'} = 0$. The amplitude $M_{\alpha}(\bar{\nu}_e p
\to n e^+ \gamma)$ is gauge invariant. Indeed, replacing
$\varepsilon^{\alpha}_{\lambda'} \to q^{\alpha}$ and using the Dirac
equations $(\hat{k} + m_e)v = 0$ and $(\hat{k}_p - m_p)u_p = 0$ for
the positron and the proton, we obtain
$q^{\alpha}M_{\alpha}(\bar{\nu}_e p \to n e^+ \gamma) = 0$. 

Since the proton and neutron in the radiative inverse $\beta$--decay
are non--relativistic, the amplitude of the radiative inverse
$\beta$--decay can be given by the expression
\begin{eqnarray}\label{labelD.2}
\hspace{-0.3in}M(\bar{\nu}_e p \to n e^+ \gamma)_{\lambda'} = -
\,e\,\frac{G_F}{\sqrt{2}}\,V_{ud}\,\frac{m_p}{\omega}\,\Big(\frac{{\cal
    M}^{(e)}_{\lambda'}}{E - \vec{k}\cdot \vec{n}}- \frac{{\cal
  M}^{(p)}_{\lambda'}}{k_p\cdot n}\Big),
\end{eqnarray}
where $n = q/\omega = (1, \vec{n})$ is a 4--vector, normalised by $n^2
= 0$.  The amplitudes ${\cal M}^{(e)}_{\lambda'}$ and ${\cal
  M}^{(p)}_{\lambda'}$ and their hermitian conjugate ${\cal M}^{(e)
  \dagger}_{\lambda'}$ and ${\cal M}^{(p)\dagger}_{\lambda'}$ are
equal to
\begin{eqnarray}\label{labelD.3}
\hspace{-0.3in}{\cal M}^{(e)}_{\lambda'} &=&
       [\varphi^{\dagger}_n\varphi_p][\bar{v}_{\bar{\nu}}\,\gamma^0(1
         - \gamma^5)\, Q_{\lambda'}\,v] - \lambda\,[\varphi^{\dagger}_n
         \vec{\sigma}\,\varphi_p]\cdot
       [\bar{v}_{\bar{\nu}}\,\vec{\gamma}\,(1 - \gamma^5)\,
         Q\,v],\nonumber\\ {\cal M}^{(p)}_{\lambda'} &=& 2 k_p\cdot \varepsilon^*_{\lambda'}
      \Big( [\varphi^{\dagger}_n\varphi_p][\bar{v}_{\bar{\nu}}\,\gamma^0(1
         - \gamma^5)\,v] - \lambda\,[\varphi^{\dagger}_n
         \vec{\sigma}\,\varphi_p]\cdot
       [\bar{v}_{\bar{\nu}}\,\vec{\gamma}\,(1 - \gamma^5)\, v]\Big),
\end{eqnarray}
and 
\begin{eqnarray}\label{labelD.4}
\hspace{-0.3in}{\cal M}^{(e) \dagger}_{\lambda'} &=&
       [\varphi^{\dagger}_p\varphi_n][\bar{v}\,\bar{Q}_{\lambda'}\,\gamma^0
         (1 - \gamma^5) v_{\bar{\nu}}] - \lambda\,[\varphi^{\dagger}_p
         \vec{\sigma}\,\varphi_n]\cdot [\bar{v}\,
         \bar{Q}\,\vec{\gamma}\,(1 - \gamma^5)
         v_{\bar{\nu}}],\nonumber\\ {\cal M}^{(p) \dagger}_{\lambda'}
       &=& 2 k_p\cdot \varepsilon_{\lambda'}
       \Big([\varphi^{\dagger}_p\varphi_n][\bar{v}\,\gamma^0 (1 -
         \gamma^5) v_{\bar{\nu}}] - \lambda\,[\varphi^{\dagger}_p
         \vec{\sigma}\,\varphi_n]\cdot [\bar{v}\,\vec{\gamma}\,(1 -
         \gamma^5) v_{\bar{\nu}}]\Big),
\end{eqnarray}
where $Q_{\lambda'} = 2 k\cdot \varepsilon^*_{\lambda'} +
\hat{q}\hat{\varepsilon}^*_{\lambda'}$ and $\bar{Q}_{\lambda'} =
\gamma^0 Q^{\dagger}_{\lambda'}\gamma^0 = 2 k\cdot
\varepsilon_{\lambda'} + \hat{\varepsilon}_{\lambda'}\hat{q}$
\cite{Ivanov2013}. For the calculation of the contributions of the
proton--photon interaction we have kept only the leading terms in the
large proton mass expansion.

The squared absolute value of the amplitude Eq.(\ref{labelD.2}),
averaged over polarisations of the proton and summed over
polarisations of the neutron and positron is defined by
\begin{eqnarray}\label{labelD.5}
\hspace{-0.3in}\frac{1}{2}\sum_{\rm pol.}|M(\bar{\nu}_e p \to n e^+
\gamma)_{\lambda'}|^2 &=& \pi
\alpha\,G^2_F|V_{ud}|^2\,\frac{m^2_p}{\omega^2}\,\Bigg\{\frac{1}{(E -
  \vec{k}\cdot \vec{n})^2} \frac{1}{2}\sum_{\rm pol.}|{\cal
  M}^{(e)}_{\lambda'}|^2 + \frac{1}{(k_p\cdot
  n)^2}\frac{1}{2}\sum_{\rm pol.}|{\cal M}^{(p)}_{\lambda'}|^2\nonumber\\
\hspace{-0.3in}&-& \frac{1}{(E - \vec{k}\cdot \vec{n})(k_p\cdot
  n)}\frac{1}{2}\sum_{\rm pol.}\Big({\cal
  M}^{(e)\dagger}_{\lambda'}{\cal M}^{(p)}_{\lambda'} + {\cal
  M}^{(p)\dagger}_{\lambda'}{\cal M}^{(e)}_{\lambda'}\Big)\Bigg\}
\end{eqnarray}
Since the first term has been calculated in Appendix A (see
Eq.(\ref{labelA.57}) and Eq.(\ref{labelA.58})), we calculate below the
second and the third terms only. We get
\begin{eqnarray}\label{labelD.6}
\hspace{-0.3in}&&\frac{1}{2}\sum_{\rm pol}|{\cal M}^{(e)}_{\lambda'}|^2 = 32
E E_{\bar{\nu}} (1 + 3 \lambda^2)\Big\{\Big[(k\cdot
  \varepsilon^*_{\lambda'})(k\cdot \varepsilon_{\lambda'}) \Big(1 +
  \frac{\omega}{E}\Big) - (E - \vec{k}\cdot
  \vec{n}\,)\,\frac{\omega}{E}\,\frac{1}{2}\,\Big((k\cdot
  \varepsilon^*_{\lambda'})\,\varepsilon^0_{\lambda'} + (k\cdot
  \varepsilon_{\lambda'})\,\varepsilon^{0*}_{\lambda'}\Big)\nonumber\\
\hspace{-0.3in}&& - (E - \vec{k}\cdot
\vec{n}\,)\,\frac{1}{2}\,\frac{\omega^2}{E}\,(\varepsilon^*_{\lambda'}\cdot
\varepsilon_{\lambda'})\Big] +
a_0\,\frac{\vec{k}_{\bar{\nu}}}{E_{\bar{\nu}}}\cdot \Big[(k\cdot
  \varepsilon^*_{\lambda'})(k\cdot \varepsilon_{\lambda'})\Big(\vec{\beta} +
  \vec{n}\,\frac{\omega}{E}\Big) - (E - \vec{k}\cdot
  \vec{n}\,)\,\frac{\omega}{E}\,\frac{1}{2}\,\Big((k\cdot
  \varepsilon^*_{\lambda'})\,\vec{\varepsilon}_{\lambda'} + (k\cdot
  \varepsilon_{\lambda'})\,\vec{\varepsilon}^{\,*}_{\lambda'}\Big)\nonumber\\
\hspace{-0.3in}&& - (E - \vec{k}\cdot
\vec{n}\,)\,\frac{1}{2}\,\frac{\omega^2}{E}\,(\varepsilon^*_{\lambda'}\cdot
\varepsilon_{\lambda'})\,\vec{n}\,\Big]\Big\},
\end{eqnarray}
\begin{eqnarray}\label{labelD.7}
\hspace{-0.3in}\frac{1}{2}\sum_{\rm pol.}|{\cal M}^{(p)}_{\lambda'}|^2
= 32 E E_{\bar{\nu}} (1 + 3\lambda^2)\,(k_p\cdot
\varepsilon^*_{\lambda'})(k_p\cdot \varepsilon_{\lambda'})\,\Big(1 + a_0\,\frac{\vec{k}\cdot
  \vec{k}_{\bar{\nu}}}{E E_{\bar{\nu}}}\Big)
\end{eqnarray}
and 
\begin{eqnarray}\label{labelD.8}
\hspace{-0.3in}\frac{1}{2}\sum_{\rm pol.}\Big({\cal
  M}^{(e)\dagger}_{\lambda'}{\cal M}^{(p)}_{\lambda'} + {\cal
  M}^{(p)\dagger}_{\lambda'}{\cal M}^{(e)}_{\lambda'}\Big) = 32 E
E_{\bar{\nu}} (1 + 3\lambda^2)\,\Big((k_p\cdot
\varepsilon^*_{\lambda'})(k_e\cdot \varepsilon_{\lambda'}) + (k_p\cdot
\varepsilon_{\lambda'})(k_e\cdot
\varepsilon^*_{\lambda'})\Big)\,\Big(1 + a_0\,\frac{\vec{k}\cdot
  \vec{k}_{\bar{\nu}}}{E E_{\bar{\nu}}}\Big).
\end{eqnarray}
The contribution of one--virtual photon exchanges is invariant under
gauge transformation of the photon propagator $D_{\alpha\beta}(q) \to
D_{\alpha\beta} + c(q^2)\,q_{\alpha}q_{\beta}$, where $c(q^2)$ is an
arbitrary function and $q^2 \neq 0$. Due to such an invariance one can
calculate the contribution of one--virtual photon exchanges in any
gauge \cite{RC9}. Following \cite{RC9} we have calculated the
contribution of one--virtual photon exchanges in the Feynman gauge
(see also see also Appendix B of \cite{Ivanov2013}).

One may show that the squared absolute value of the amplitude
Eq.(\ref{labelD.2}), averaged over the polarisations of the proton and
summed over the polarisations of the neutron and positron and given by
Eqs.(\ref{labelD.5}) - (\ref{labelD.8}), is invariant under the gauge
transformation $\varepsilon^{\alpha
  *}_{\lambda'}\varepsilon^{\beta}_{\lambda'} \to \varepsilon^{\alpha
  *}_{\lambda'}\varepsilon^{\beta}_{\lambda'} +
c\,q^{\alpha}q^{\beta}$, where $c$ is an arbitrary constant. Indeed,
making such a gauge transformation in Eq.(\ref{labelD.5}) one obtains
an additional contribution, proportional to the constant $c$
\begin{eqnarray}\label{labelD.9}
\hspace{-0.3in}&&\delta\,\frac{1}{2}\sum_{\rm pol.}|M(\bar{\nu}_e p
\to n e^+ \gamma)_{\lambda'}|^2 = c\,32 E E_{\bar{\nu}} (1 + 3 \lambda^2)\,\pi
\alpha\,G^2_F|V_{ud}|^2\,\frac{m^2_p}{\omega^2}\,\Bigg\{\frac{1}{(E -
  \vec{k}\cdot \vec{n})^2}\Bigg((k\cdot q)^2\Big(1 +
\frac{\omega}{E}\Big)\nonumber\\
\hspace{-0.3in}&& - (E - \vec{k}\cdot \vec{n}\,)(k\cdot
q)\,\frac{\omega^2}{E} +
a_0\,\frac{\vec{k}_{\bar{\nu}}}{E_{\bar{\nu}}}\cdot \Big[(k\cdot
  q)^2\Big(\vec{\beta} + \vec{n}\,\frac{\omega}{E}\Big) - (E -
  \vec{k}\cdot \vec{n}\,)\,(k\cdot
  q)\,\vec{n}\,\frac{\omega^2}{E}\Big]\Bigg) + \frac{(k_p\cdot
  q)^2}{(k_p\cdot n)^2}\,\Big(1 +
a_0\,\frac{\vec{k}_{\bar{\nu}}}{E_{\bar{\nu}}}\cdot
\vec{\beta}\,\Big)\nonumber\\
\hspace{-0.3in}&& - \frac{2 (k_p\cdot q)(k\cdot q)}{(E - \vec{k}\cdot
  \vec{n})(k_p\cdot n)}\,\Big(1 +
a_0\,\frac{\vec{k}_{\bar{\nu}}}{E_{\bar{\nu}}}\cdot
\vec{\beta}\,\Big)\Bigg\} = 0,
\end{eqnarray}
where we have used the relations $q = \omega n = \omega (1, \vec{n})$,
$E - \vec{k}\cdot \vec{n} = (k\cdot q)/\omega$ and $(k_p\cdot n) =
(k_p \cdot q)/\omega$. Thus, due to gauge invariance of
Eq.(\ref{labelD.5}) for the summation over polarisation one may use
any gauge. However, it is  obvious that one has to sum over the physical
degrees of freedom of the real photons, which are defined by the
polarisation vector $\varepsilon_{\lambda'} = (0,
\vec{\varepsilon}_{\lambda'})$ \cite{BD1967}. The polarisation vector
$\vec{\varepsilon}_{\lambda'}$ has the following properties \cite{BD1967}
\begin{eqnarray}\label{labelD.10}
\hspace{-0.3in}\vec{q}\cdot \vec{\varepsilon}_{\lambda'} &=&
\vec{q}\cdot \vec{\varepsilon}^{\,*}_{\lambda'} = 0,\nonumber\\
\hspace{-0.3in}\vec{\varepsilon}^{\,*}_{\lambda'}\cdot
\vec{\varepsilon}_{\lambda''} &=&
\delta_{\lambda'\lambda''},\nonumber\\
\hspace{-0.3in}\sum_{\lambda' =
  1,2}\varepsilon^{i\,*}_{\lambda'}\varepsilon^j_{\lambda'} &=&
\delta^{ij} - \frac{q^i q^j}{\omega^2} = \delta^{ij} - n^i n^j.
\end{eqnarray}
Summing over physical degrees of freedom of a real photon we arrive at
the expression
\begin{eqnarray}\label{labelD.11}
\hspace{-0.3in}\frac{1}{2}\sum_{\lambda' = 1,2}\sum_{\rm
  pol.}|M(\bar{\nu}_e p \to n e^+ \gamma)_{\lambda'}|^2 &=& 32 E
E_{\bar{\nu}} (1 + 3 \lambda^2)\,\pi
\alpha\,G^2_F|V_{ud}|^2\,\frac{m^2_p}{\omega^2}\,\Bigg\{\frac{\beta^2
  - (\vec{\beta}\cdot \vec{n}\,)^2}{(1 - \vec{\beta}\cdot
  \vec{n}\,)^2}\Big(1 + \frac{\omega}{E}\Big) + \frac{1}{1 -
  \vec{\beta}\cdot \vec{n}}\,\frac{\omega^2}{E^2}\nonumber\\
\hspace{-0.3in}&+&
a_0\,\frac{\vec{k}_{\bar{\nu}}}{E_{\bar{\nu}}}\cdot \Bigg[\frac{\beta^2
    - (\vec{\beta}\cdot \vec{n}\,)^2}{(1 - \vec{\beta}\cdot
    \vec{n}\,)^2}\Big(\vec{\beta} + \vec{n}\,\frac{\omega}{E}\Big) +
  \frac{\vec{\beta} - \vec{n}\,(\vec{n}\cdot \vec{\beta}\,)}{1 -
    \vec{\beta}\cdot \vec{n}}\,\frac{\omega}{E} + \frac{\vec{n}}{1 -
    \vec{\beta}\cdot \vec{n}}\,\frac{\omega^2}{E^2}\Bigg]\Bigg\}.
\end{eqnarray}
This gives the angular and photon--energy distribution of the
radiative inverse $\beta$--decay, defined by (see Eq.(\ref{labelA.59}))
\begin{eqnarray}\label{labelD.12}
\hspace{-0.3in}\frac{d^2
  \sigma^{(\gamma)}(E_{\bar{\nu}},\cos\theta_{e\bar{\nu}})}{ d\omega\,
  d\cos\theta_{e\bar{\nu}}} &=& (1 +
3\lambda^2)\,\frac{\alpha}{\pi}\,\frac{G^2_F|V_{ud}|^2}{2\pi}
\frac{1}{\omega}\,k\,E
\int\frac{d\Omega_{\vec{n}}}{4\pi}\,\Bigg\{\frac{\beta^2
  -(\vec{\beta}\cdot \vec{n}\,)^2}{(1 - \vec{\beta}\cdot
  \vec{n}\,)^2}\,\Big(1 + \frac{\omega}{E}\Big) + \frac{1}{1 -
  \vec{\beta}\cdot \vec{n}}\,\frac{\omega^2}{E^2}\nonumber\\
\hspace{-0.3in}&+&
a_0\,\frac{\vec{k}_{\bar{\nu}}}{E_{\bar{\nu}}}\cdot\Bigg[\frac{\beta^2
    - (\vec{\beta}\cdot \vec{n}\,)^2}{(1 - \vec{\beta}\cdot
    \vec{n}\,)^2}\,\Big(\vec{\beta} + \vec{n}\,\frac{\omega}{E}\Big) +
  \frac{\vec{\beta} - \vec{n}\,(\vec{\beta}\cdot \vec{n}\,)}{1 -
    \vec{\beta}\cdot \vec{n}}\,\frac{\omega}{E} + \frac{\vec{n}}{1 -
    \vec{\beta}\cdot \vec{n}}\,\frac{\omega^2}{E^2}\Bigg]\Bigg\},
\end{eqnarray}
where $E = \bar{E} - \omega$ and $\beta = k/E = \sqrt{1 - m^2_e/E^2}$.
Up to a common factor the expression in curl brackets coincides with
Eq.(29), obtained by Fukugita and Kubota in Ref.\cite{IBD4}, and
Eq.(9), obtained by Raha, Myhrer and Kudobera in Ref.\cite{IBD5}.

We would like to notice that the contributions of the radiative
inverse $\beta$--decay to the correlation coefficients $A(\bar{E})$
and $B(\bar{E})$, described by the functions $g^{(b1)}(\bar{\beta})$
and $g^{(b2)}(\bar{\beta})$ in the paper by Fukugita and Kubota
\cite{IBD4} (see Eq.(32) in Ref.\cite{IBD4}), in our notation are equal
to
\begin{eqnarray}\label{labelD.13}
\hspace{-0.3in}&&\frac{1}{2}\,g^{(b1)}(\bar{\beta}) =
g^{(1)}_{\beta\gamma}(\bar{E},\mu) = {\ell n}\Big(\frac{m_e}{\mu}\Big)
\,\Big[\frac{1}{\bar{\beta}}\,{\ell n}\Big(\frac{1 + \bar{\beta}}{1 -
    \bar{\beta}}\Big) - 2\Big] + \frac{17}{4} + 2 {\ell
  n}\Big(\frac{2\bar{\beta}}{1 +
  \bar{\beta}}\Big)\,\Big[\frac{1}{\bar{\beta}}\,{\ell n}\Big(\frac{1
    + \bar{\beta}}{1 - \bar{\beta}}\Big) - 2\Big] +
\frac{3}{\bar{\beta}}\,L\Big(\frac{2\bar{\beta}}{1 +
  \bar{\beta}}\Big)\nonumber\\
\hspace{-0.3in}&&+ \frac{1}{4\bar{\beta}}\,{\ell n}^2\Big(\frac{1 + \bar{\beta}}{1 -
  \bar{\beta}}\Big) + \Big(-2 + \frac{7}{8}\,\frac{1}{\bar{\beta}} -
\frac{\bar{\beta}}{8}\Big)\,{\ell n}\Big(\frac{1 + \bar{\beta}}{1 -
  \bar{\beta}}\Big)
\end{eqnarray}
and
\begin{eqnarray}\label{labelD.14}
\hspace{-0.3in}&&\frac{1}{2}\,g^{(b2)}(\bar{\beta}) =
g^{(2)}_{\beta\gamma}(\bar{E},\mu) = {\ell n}\Big(\frac{m_e}{\mu}\Big)
\,\Big[\frac{1}{\bar{\beta}}\,{\ell n}\Big(\frac{1 + \bar{\beta}}{1 -
    \bar{\beta}}\Big) - 2\Big] + \frac{7}{4} + \frac{2}{\bar{\beta}^2}\,(1 - \sqrt{1 -
  \bar{\beta}^2}) - \Big(4 +
\frac{1}{\bar{\beta}}\Big)\,\frac{1}{4}\,{\ell n}\Big(\frac{1 +
  \bar{\beta}}{1 - \bar{\beta}}\Big)\nonumber\\
\hspace{-0.3in}&& + \frac{-1 + 8\bar{\beta} -
  3\bar{\beta}^2}{16\bar{\beta}^2}\,{\ell n}^2\Big(\frac{1 +
  \bar{\beta}}{1 - \bar{\beta}}\Big)-
\frac{2}{\bar{\beta}}\,L\Big(\frac{2\bar{\beta}}{1 + \bar{\beta}}\Big)
+ \frac{4}{\bar{\beta}}\,L\Big(1 - \sqrt{\frac{1 - \bar{\beta}}{1 +
    \bar{\beta}}}\Big) + {\ell n}\Big[\Big(\frac{2\bar{\beta}}{1 +
    \bar{\beta}}\Big)\,\frac{\sqrt{1 + \bar{\beta}} - \sqrt{1 -
      \bar{\beta}}}{\sqrt{1 + \bar{\beta}} + \sqrt{1 -
      \bar{\beta}}}\Big]\nonumber\\
\hspace{-0.3in}&&\times\,\Big[\frac{1}{\bar{\beta}}\,{\ell
    n}\Big(\frac{1 + \bar{\beta}}{1 - \bar{\beta}}\Big) - 2\Big],
\end{eqnarray}
where in the function $g^{(b1)}(\bar{\beta})$, given by Eq.(31) of
Ref.\cite{IBD4}, we have summed up the first two terms. It is easy to
check that our results are in agreement with the results, obtained by
Fukugita and Kubota \cite{IBD4}. For the derivation of
Eqs.(\ref{labelD.13}) and (\ref{labelD.14}) we have used the
definitions of the functions $g^{(1)}_{\beta\gamma}(\bar{E},\mu)$ and
$g^{(2)}_{\beta\gamma}(\bar{E},\mu)$, given by Eqs.(\ref{labelA.68})
and (\ref{labelA.69}), respectively, and the functions
$f^{(\gamma)}_A(\bar{E})$ and $f^{(\gamma)}_B(\bar{E})$, given by
Eqs.(\ref{labelB.19}) and (\ref{labelC.15}), respectively, and the
relations Eqs.(\ref{labelB.8}) and (\ref{labelC.17}).

\end{document}